# Kinetics and mechanism of metal nanoparticle growth *via* optical extinction spectroscopy and computational modeling: the curious case of colloidal gold

M. Reza. Andalibi[*,†,‡], Alexander Wokaun[†], Paul Bowen[‡], Andrea Testino[†]

[†] Energy and Environment Research Division, Paul Scherrer Institute, 5232 PSI Villigen, Switzerland.

[‡] Department of Materials Science and Engineering, École Polytechnique Fedérale de Lausanne (EPFL), Lausanne 1015, Switzerland.

[*] Corresponding author: reza.andalibi@psi.ch

## Abstract

An overarching computational framework unifying several optical theories to describe the temporal evolution of gold nanoparticles (GNPs) during a seeded growth process is presented. To achieve this, we used the inexpensive and widely available optical extinction spectroscopy, to obtain *quantitative* kinetic data. *In situ* spectra collected over a wide set of experimental conditions were regressed using the physical model, calculating light extinction by ensembles of GNPs during the growth process. This model provides temporal information on the size, shape, and concentration of the particles, and any electromagnetic interactions between them. Consequently, we were able to describe the mechanism of GNP growth and divide the process into distinct genesis periods. We provide explanations for several longstanding mysteries, *e.g.*, the phenomena responsible for the purple-greyish hue during the early stages of GNP growth, the complex interactions between nucleation, growth and aggregation events, and a clear distinction between agglomeration and electromagnetic interactions. The presented theoretical formalism has been developed in a generic fashion so that it can readily be adapted to other nanoparticulate formation scenarios such as the genesis of various metal nanoparticles.

## Keywords

Since Faraday's seminal work on the synthesis of colloidal gold nanoparticles (GNPs), there has been a tremendous effort to explore various nanoparticle synthesis routes in order to obtain better control over the properties of the final product.[1,2] Every year numerous papers show up attempting to propose mechanisms for particular synthesis pathways. Most of these studies try to interpret these complicated processes in terms of simplified models and fail to describe the whole process, in particular when dealing with a wide range of operating conditions. This invokes the need for more complete mechanistic theoretical models. By performing global uncertainty/sensitivity analysis on such models one can map out various outcomes of the process *vs.* different possible combinations of the operating conditions.[3] Ideally, this should be the first step in a predictive theory-driven synthesis of nanoparticulate systems.

To develop a theoretical formalism with predictive capacity, extensive *in situ* experimental data over a wide range of conditions has to be collected. In the case of nanoparticles, relatively expensive and laborious characterization techniques such as *in situ* transmission electron microscopy (TEM),[4,5] and synchrotron-based X-ray scattering and X-ray absorption spectroscopy[6–8] are typically employed. This is probably because of the well-developed computational tools available for the analysis of the experimental data (particularly in the case of X-ray based methods).[9–12] Despite this popularity, both the electron beam and synchrotron light can drastically affect the system under investigation, *e.g.*, accelerate the process of GNP formation and growth.[4,5,8,13,14] For synchrotron-based methods, to circumvent this limitation one solution is the application of liquid jets so that a fresh sample is probed throughout the process.[8,15] This method, however, adds to the complexity of the setup and does not allow for real-time measurements.[15] At the same time, optical extinction spectroscopy (OES) is rarely used to follow the formation of nanoparticles in spite of being noninvasive,[13,14] significantly less expensive, and widely available. Indeed, to our knowledge, very few articles have tried to quantitatively employ this characterization method and they are either dealing with relatively simple systems, where the particles behave as optically independent entities, or they simply neglect the interparticle interactions (see below for further discussion).[16,17] For this reason, there is a great need for a general and easily adaptable theoretical framework that utilizes the practical application of OES in the study of nanomaterial formation.

So far, several investigations have addressed the mechanistic aspects during the (seeded) growth of GNPs using various characterization techniques such as atomic force microscopy (AFM),[5,14,18] electrophoretic measurements,[19,20] redox potential/pH measurements,[18–20] dynamic light scattering (DLS),[14,19,21] *ex situ* TEM,[4,5,7,14,19–23] *in situ* TEM,[4,5] and X-ray scattering.[6,7,13,23,24] Many of these studies follow the process also using *ex situ*[5,19–22] or *in situ*[14,23,24] UV-vis spectroscopy but the information is treated merely qualitatively.

From the plethora of research, some of which was summarized above, we know that the processes of seeded growth is typically accompanied by nucleation of new particles.[5,25] This could either be in a homogeneous fashion,[25] or in the close vicinity of the already present seed surface[5] (so called true catalytic secondary nucleation[26] or, equivalently, near surface nucleation followed by particle mediated growth[5]). Additional complications arise from the possibility of agglomeration/aggregation invoked in many studies to describe the transient enhanced extinction in the wavelength range 600-800 nm, namely, the temporary purple-greyish colour of the suspension.[7,18–20,27] Biggs *et al.*[18] and Chow and Zukoski[19] explained this in the light of the reduced colloidal stability in the presence of Au(III) in solution. Later, Rodriguez-Gonzalez and co-workers noticed that a homogeneous Au(III)→Au(I) reduction, followed by Au(I) diffusion toward the seed surface, and its subsequent disproportionation to Au(0) and Au(III) was more consistent with their experimental observations.[20] Along the same lines, several workers have presented strong evidence for the intermediacy of Au(I) in the reduction of Au(III) to metallic Au.[6,21,28,29] Thus, it appears that Au(I) adsorption on the surface of GNPs tampers with their colloidal stability and as it is progressively reduced to Au(0), the surface is repopulated by the abundant citrate anions, and the colloidal stability recovers.[19–21] Another notable observation comes from recent experimental and theoretical findings by Cheng *et al.* who found that the local supersaturation in the region at the seed-solution interface exceeded that in the bulk of the reaction medium.[5] Furthermore, as soon as secondary nuclei form in the vicinity of the seed surface, the gold ions tend to accumulate in the interparticle gap, replacing the citrate ions.[5] This induces even further destabilization and brings the particles closer to each other, giving rise to large numbers of GNPs interacting inside mesoscale superclusters, several times larger than the individual particles.[14,21,24]

Beside the above phenomena, a third and somewhat less addressed complexity concerns the electromagnetic interactions between the particles.[14,23,24,30,31] As we discussed in the previous paragraph, colloidal destabilization can bring particles closer to one another. This in turn provokes electromagnetic coupling between the neighbouring particles.[32,33] Plech *et al.*,[13] and also Förster *et al.*[24] detected evidence for such "correlated assemblies" using small-angle X-ray scattering (SAXS), while Mikhlin *et al.*[14] physically detected such liquid-like mesostructures using AFM, DLS, and TEM. In the literature, this effect has been frequently confused with agglomeration/aggregation wherein the particles are supposed to be in contact with each other (throughout this work we stick to the IUPAC definitions for agglomeration/coagulation and aggregation denoting the formation of physically and chemically bound multi-particle associates, respectively[34]). In our study, however, we explicitly distinguish between these effects in the sense that closely spaced particles are considered to be in electromagnetic interaction (and not agglomerated/aggregated) unless they touch each other (see Supporting Section 2).

Here, we have developed an overarching computational framework, unifying various optical theories, to describe the extinction of light by nanoparticles that may exhibit electromagnetic interactions. We applied the approach to an *in situ* study on citrate-mediated seeded growth of GNPs while the theoretical formalism can be employed in different scenarios such as for other metallic nanoparticles. With this method, we rigorously identified the different mechanistic steps during the evolution of seeds toward final particles, under several experimental conditions. The output was consistent with previous experimental observations. We also provide cogent and comprehensive answers to a number of longstanding questions, such as the phenomena responsible for the purple-greyish hue during the early stages of GNP growth, the complex interactions between nucleation and growth events, and a clear distinction between agglomeration and electromagnetic interactions.

## Computational framework

### Dielectric function of nanosized gold and water (solvent)

The first step in developing an optical model for the extinction behavior of gold suspensions is the modification of the bulk dielectric function so that temperature and finite size effects are properly taken into account. For the bulk data, we used the dielectric function of gold recently measured by Yakubovsky *et al.* on a 25 nm thick film (measured wavelength range 300-2000 nm).[35] A smoothing spline was fitted to these data yielding the dielectric function of gold at room temperature every 1 nm (Fig. 1). Intrinsic size effects for particles smaller than the mean free path of electrons in bulk gold were accounted for in the framework of the extended Drude model assuming size- and temperature-independent contribution from the interband transitions.[32,36,37] Thus, the complex dielectric function of gold reads:

$$\epsilon(\omega, r, T) = \epsilon_{IB}(\omega) + [1 - \frac{\omega_P(T)^2}{\omega(\omega + i\Gamma(r, T))}] \quad (1)$$

where $\epsilon_{IB}$ represents the contribution of the interband transitions by the bound electrons while the term in the square brackets comes from the free-electron contributions. In the latter, $\omega_P$ is the temperature-dependent bulk plasma frequency (rad/s), $\omega = \frac{2\pi c}{\lambda}$ is the vacuum angular frequency of the incident light (rad/s; $c$ and $\lambda$ are the speed of light and wavelength in vacuum, respectively), and $\Gamma$ is the temperature ($T$)- and size (equivalent spherical radius; $r$)-dependent damping frequency (rad/s) for the free electrons. The temperature dependence of $\omega_P$ can be estimated considering the thermal expansion of gold, which in turn reduces the density of the conduction (free) electrons:

$$\omega_P(T) = \frac{\omega_P(T_0)}{\sqrt{1 + \alpha_V(T - T_0)}} \quad (2)$$

Here, $T_0 = 298.15\ K$ is a reference temperature (room temperature), $\omega_P(T_0) = 1.37 \times 10^{16}\ rad/s$,[35] and $\alpha_V = 4.17 \times 10^{-5}\ K^{-1}$ is the volumetric thermal expansion coefficient of gold.[37]

The damping of the free electrons arises from several processes including the interactions with phonons, electrons, lattice defects, grain boundaries, impurities, surface adsorbed species, and surfaces/interfaces themselves.[32,36,38] The temperature dependences of the interactions with phonons ($\Gamma_{e-ph}$) and electrons ($\Gamma_{e-e}$) can be described as:[37]

$$\Gamma_{e-ph}(T) = \Gamma_{e-ph}(T_0) \times [\frac{2}{5} + 4\left(\frac{T}{\theta}\right)^5 \int_0^{\theta/T} \frac{z^4}{e^z - 1} dz] \tag{3}$$

$$\Gamma_{e-e}(\omega, T) = \frac{\pi^3 \Sigma \Delta}{12 \hbar E_F} [(k_B T)^2 + \left(\frac{\hbar \omega}{2\pi}\right)^2] \tag{4}$$

In these equations $\Gamma_{e-ph}(T_0) = 7.09 \times 10^{13}\ rad/s$ (back calculated from[35] $\Gamma_{bulk}(T_0) = 1.2686 \times 10^{14}\ rad/s$), $\theta = 170\ K$ is the Debye temperature of gold, $\Sigma = 0.55$ is the Fermi-surface average of scattering, $\Delta = 0.77$ is the fractional Umklapp scattering, $E_F = 5.51\ eV$ is the Fermi energy, $\hbar$ is the reduced Planck constant, and $k_B$ is the Boltzmann constant.[37] In addition to these two damping mechanisms, the size-dependent damping due to interactions with surfaces/interfaces, grain boundaries, and defects (the free path effect,[32,36] FPE) can be estimated using the following phenomenological approach:

$$\Gamma_{FPE}(r) = A_{eff} \frac{v_f}{L_{eff}} \tag{5}$$

where $v_f = 1.4 \times 10^6\ m/s$ is the Fermi velocity of gold and $L_{eff} = \frac{4 \times volume}{surface\ area}$ ($\frac{4}{3}r\ for\ spheres$) is the effective mean free path of electrons in nanoparticles with arbitrary shape.[39] Further, $A_{eff} \equiv \frac{4}{3}\eta$ where $\eta$ accounts for the details of the scattering process and the additional contributions, e.g., by defects, adsorbed species, and grain boundaries,[38,40,41] and the factor $\frac{4}{3}$ gives a typical proportionality constant 1 for surface-induced scattering as a function of radius in spherical nanoparticles (for which $\eta \equiv 1$).[32,36] Here, the parameter $\eta$ was calibrated using the volume-weighted mean particle radii of seeds and grown GNP estimated from the TEM micrographs (Supporting Section 3).[40] The regression of these two experimental spectra was done by fixing the mean radii and globally fitting $\eta$ along with the aspect ratios of prolate-shaped particles ($\bar{\beta}$) for both seeds and grown GNPs using the Gans formalism in the range 400-800 nm. The value of $\eta$ was constrained in the range [0.2,4].[6-8] The fitted $\eta$ values were used to derive a linear relation $\eta = \eta_0 + \eta_1 \bar{r}_{GNP}$ which was then used for all the intermediate spectra.

Having all the damping mechanisms in place one can calculate the overall damping frequency as[32,36]

$$\Gamma = \Gamma_{e-ph} + \Gamma_{e-e} + \Gamma_{FPE} \tag{6}$$

In practice, we can estimate $\epsilon_{IB}$ by subtracting the bulk Drude expression at room temperature ($r \to \infty \Rightarrow \Gamma_{FPE} \to 0$) from the experimental dielectric function:

$$\epsilon_{IB}(\omega) = \epsilon_{experimental} - [1 - \frac{\omega_P(T_0)^2}{\omega(\omega + i\Gamma(\omega, r \to \infty, T_0))}] \tag{7}$$

Some workers have suggested that correcting only the imaginary part (of the dielectric function) for the intrinsic size effects provides superior fits to experimental data.[42] However, our preliminary simulations proved the opposite in line with the work of Amendola and Meneghetti.[40] For this reason, throughout this work we will correct both the real and imaginary parts of $\epsilon$.

The dielectric function of water at different temperatures was calculated according to the correlation by Fernández-Prini and Dooley.[43]

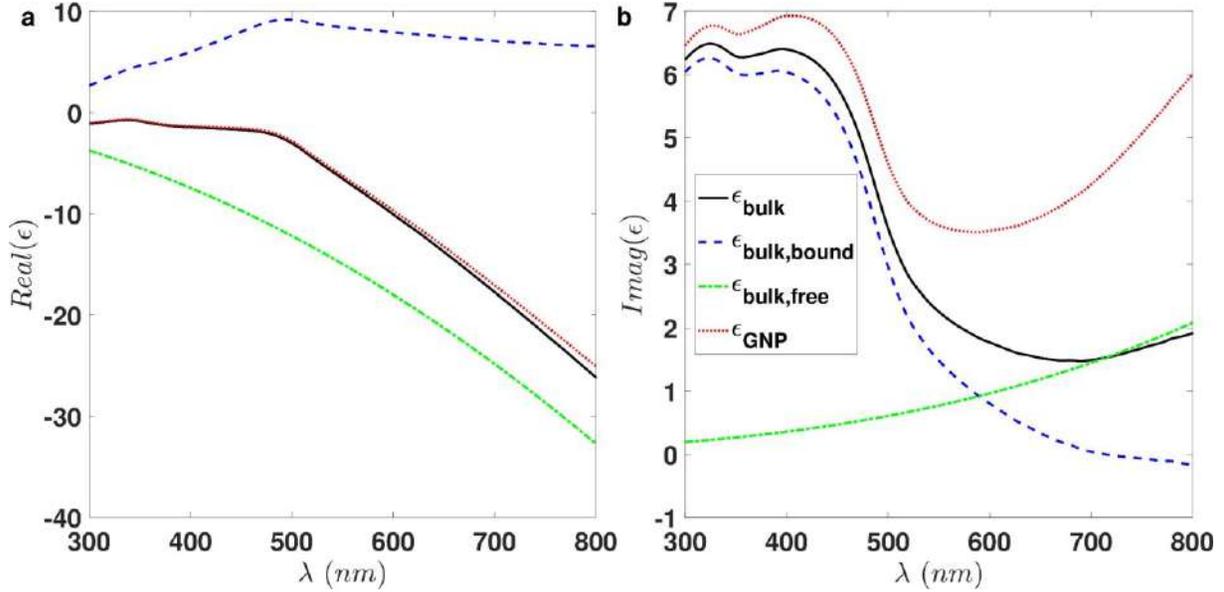

**Fig. 1.** Contributions of the interband transitions (bound electrons) and the free electrons to the real (a) and imaginary (b) parts of the gold's bulk dielectric function taken from Yakubovsky *et al.*[35] For comparison with the bulk data, the size- and temperature-corrected $\epsilon_{GNP}$ are also shown for prolate-shaped GNPs with 5 nm radius and aspect ratio 1.2 at $T = 70°C$ using $A_{eff} = 4/3$.

**Light extinction by ensembles of electromagnetically interacting GNPs**

The extinction of light by an ensemble of noninteracting GNPs is typically described using the Mie theory (in case of spherical particles) or the Gans formalism in the quasi-static regime (spheroidal GNPs with diameter $< \approx 30$ nm).[40,42] In these theories, nanoparticles are assumed to experience only the external field due to the UV-Vis light source. However, more often than not, there are significant electromagnetic interactions between the particles.[30,32,36] In other words, many nanoparticles experience a combination of the external field and the field induced by their neighboring particles (that is, all those inside the mesoscale supercluster embracing the particle). Particularly, in the case of metal nanoparticles synthesis (*e.g.*, growth of GNPs) experimental evidence support the presence of such interactions.[14,24,30] In order to account for such interparticle interactions, there are two alternatives. One can take advantage of a plethora of *numerical* techniques employed to describe the optical extinction by (electromagnetically interacting) particle ensembles. This includes coupled dipole approximation,[44] generalized Mie theory,[45] and boundary element method[46] to name a few. An alternative approach is the application of effective medium theories (EMTs).[36,47–49] While the former applies to any particle size and configuration, the latter is applicable only in the quasi-static regime. Nevertheless, the numerical approaches are computationally prohibitive, particularly for ensembles with large numbers of particles. Therefore, considering the small size of the GNPs in the current study, we chose to employ an EMT formalism to account for the possible electromagnetic interactions between the particles. This allows us to handle the large number of spectra collected during the seeded growth process.

In view of the discussion in the earlier paragraph, the optical response of ensembles composed by interacting GNPs is described using the Maxwell-Garnett effective medium theory, with retarded polarization effects incorporated following Granqvist and Hunderi.[47,48] The interacting GNPs are presumed to form spherical domains wherein their local number concentration exceeds that of the bulk solution (*i.e.*, that averaged over the entire suspension; Fig. 2a inset). The presence of such dense liquid-like superclusters (SCs) has been shown by a variety of characterization techniques (*e.g.*, DLS, AFM, and *in situ* TEM).[4,14] Schatz and colleagues have used this picture to describe the extinction by arrays of plasmonic nanoparticles, *i.e.*, DNA-linked GNP clusters with volume fractions between 1-20%.[44,49]

Granqvist and Hunderi extended the Maxwell-Garnett EMT by introducing a phase factor in the Lorentz local-field equation.[47,48] Doing that, they arrived at the following equation for the effective dielectric function ($\bar{\epsilon}$) of an ensemble of nanoparticles embedded in a surrounding medium with dielectric permittivity $\epsilon_m$:

$$\bar{\epsilon} = \epsilon_m \frac{1 + (1 - \frac{1}{3}e^{i\delta})\sum_j f_j \alpha'_j}{1 - \frac{1}{3}e^{i\delta}\sum_j f_j \alpha'_j} \tag{8}$$

with

$$\alpha'_j \equiv \frac{4\pi\alpha_j}{V_j} = \frac{1}{3}\sum_{i=1}^{3} \frac{\epsilon_j - \epsilon_m}{\epsilon_m + L_{ij}(\epsilon_j - \epsilon_m)} \tag{9}$$

In these equations $f_j$ denotes the fractional filling factors (volume fractions) of various particle classes with $\sum_j f_j = f$, where $f$ is the total volume fraction of particles within the effective medium (individual superclusters). Additionally, $\alpha_j$ denotes the polarizability of particle $j$ with a volume ($V_j$), and $L_{ij}$ represents the triplet of depolarization factors along the different axes of the prolate spheroid $j$.[47,48] Using these quantities, for randomly oriented spheroids we can estimate the average volume-normalized polarizability ($\alpha'_j$).[32,47,48] In Eq. (8), $e^{i\delta}$ is a phase factor that accounts for the retarded nature of interaction with the depolarizing field and can be estimated knowing the mean centre-to-centre distance between the particles:

$$\delta = \frac{2\pi\sqrt{\bar{\epsilon}}\bar{d}_{cc}}{\lambda} \qquad (10)$$

Assuming a uniform distribution of particles throughout the effective medium[24] (Fig. 2a inset), the center-to-center distance between the particles can be estimated as:[24,50]

$$\bar{d}_{cc} = 2\bar{r}_{GNP} \times \sqrt[3]{\frac{\pi}{\sqrt{18}f}} \qquad (11)$$

In this equation, $\bar{r}_{GNP}$ is the average equivalent spherical radius of the GNPs and the factor $\frac{\pi}{\sqrt{18}}$ is the close packing density of the equal spherical domains embracing the individual GNPs (dotted black circles surrounding individual GNPs in Fig. 2a inset). Considering the dependence of $\delta$ on $\bar{\epsilon}$, an iterative procedure is necessary to calculate the $\bar{\epsilon}$, starting with the non-retarded $\bar{\epsilon}$ as an initial guess.[47,48]

A representative effective dielectric permittivity at 70°C for a supercluster of 10 nm GNPs in water with $f = 0.2$ is shown in Fig. 2, calculated using both Maxwell-Garnet theory and its retarded extension by Granqvist and Hunderi.[47,48] Once we have the effective dielectric function of the spherical SCs, we can calculate the extinction cross section of noninteracting SCs using the Mie theory. Schatz *et al.* applied this method to calculate the extinction by DNA-linked nanoparticles and showed that the results match very well with those from the more rigorous coupled dipole approximation.[49]

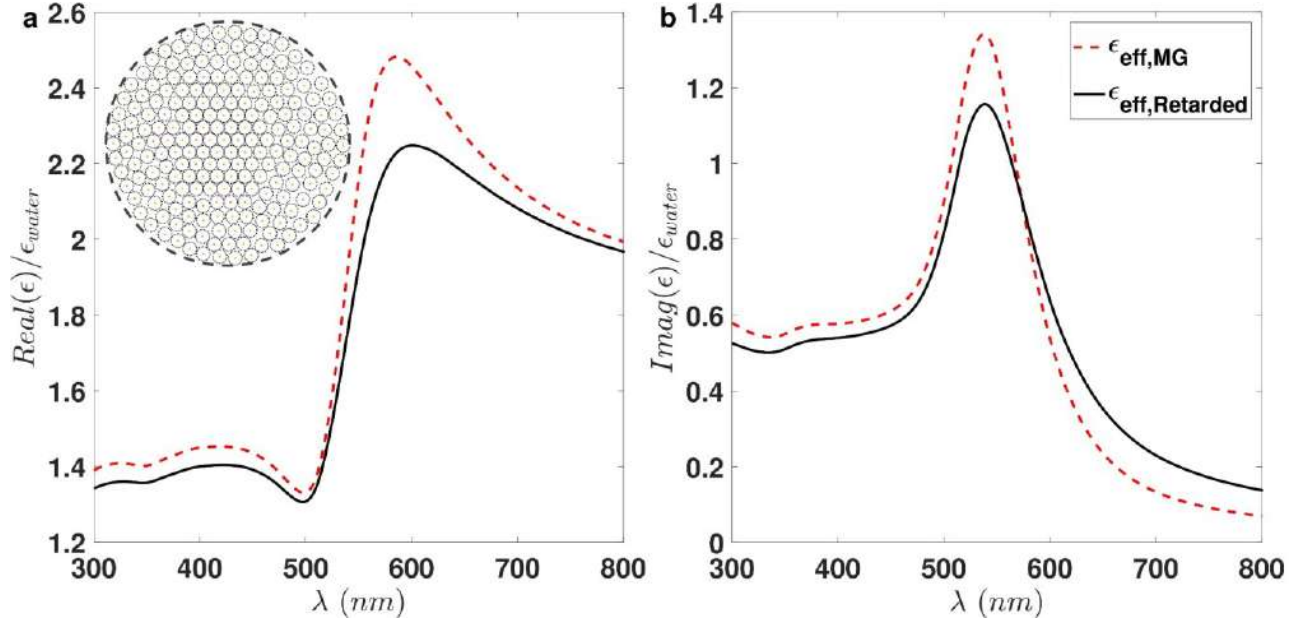

**Fig. 2.** Real (a) and imaginary (b) parts of the dielectric function for an effective medium composed of 20% (by volume) GNPs with $\bar{r}_{GNP} = 5\ nm$ and aspect ratio 1.2 dispersed in water calculated using the Maxwell-Garnett (MG) and retarded effective medium theories at 70°C ($A_{eff} = 4/3$), normalized by the dielectric function of water. The inset in (a) is the schematic representation of an ensemble of electromagnetically interacting gold nanoparticles (yellow dots) composing a dense liquid-like supercluster (SC; large dashed circle in dark gray). Each black dotted circle around individual GNPs represents the portion of SC volume per particle.

## Overall modeling workflow and regression to experimental spectra

Generally, the optical extinction by colloidal metal nanoparticles may arise from two sources. The first is the extinction by well-separated nanoparticles whose interaction with the incoming light is practically independent from one another.[32,40] The second contribution comes from nanoparticles lying close to each other (inside liquid-like superclusters, SCs, but not touching), experiencing a superposition of the external field with the mean field produced by all the neighbouring particles. Accounting for both possibilities in a suspension undergoing kinetic evolution requires the application of the Gans formalism[32,40] for the former and a retarded effective medium theory (EMT)/Mie scattering framework[47–49] for the latter. Hereafter, we refer to this combined Gans + EMT/Mie framework as the GEM model. Using this model, five physically well-defined parameters have to be estimated from regression to experimental spectra: $\bar{r}_{GNP}$ (average equivalent spherical radius), $\bar{\beta}$ (mean major-to-minor axis ratio), $x_{SC}$ (number fraction of SCs out of all the scatterers, *i.e.*, SCs + noninteracting GNPs), $f$ (filling factor, *i.e.*, the volume fraction of GNPs inside SCs), and $\bar{r}_{SC}$ (average radius of spherical SCs).

Neglecting the electromagnetic interactions between the particles one can use the Gans formalism for prolate spheroids with only $\bar{r}_{GNP}$ and $\bar{\beta}$ as the parameters to be determined by regression.[32,40] Using either model (Gans or GEM), a globalized bounded Nelder-Mead optimization scheme was employed for the regression to the experimental data.[51] The 95% confidence intervals on various regression parameters and on different model outputs were obtained by a heuristic approach described by Schwaab et al.[52] As the optical interaction effects disappear the EMT-specific parameters ($x_{SC}$, $f$, and $r_{SC}$) lose their meaning and the common parameters ($\bar{r}_{GNP}$ and $\bar{\beta}$) approach to the same values for regressions with both the GEM and the Gans model. In this case, we selected the Gans formalism as the superior model since it has fewer parameters and provides exactly similar optimization results as the GEM formalism.

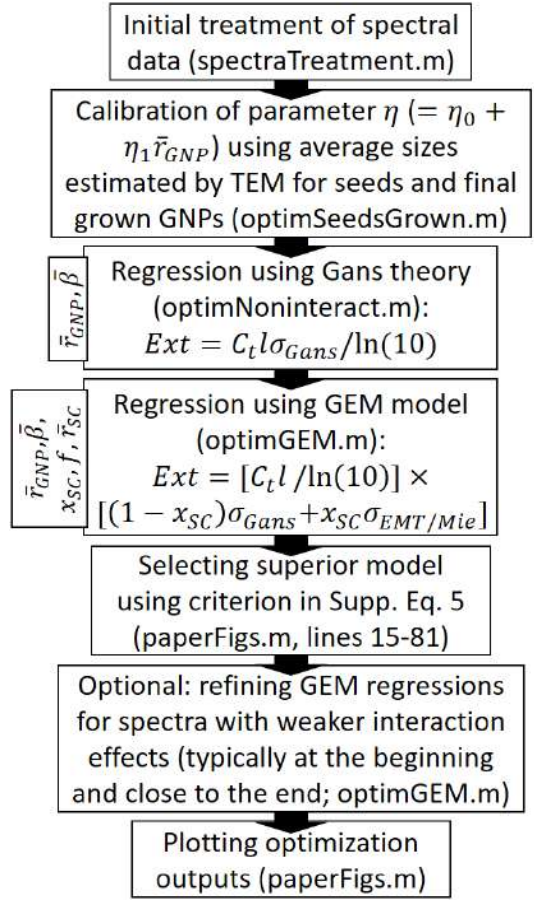

**Fig. 3.** Flowchart summarizing the overall computational framework, including the theory and the regression steps (with the names of the corresponding MATLAB scripts, *.m, wherever applicable). Here, $C_t$ is the total number concentration of the scatterers (noninteracting GNPs + SCs), $l$ is the optical path length, $Ext$ is the overall optical extinction of the sol, and $\sigma_{Gans}$ and $\sigma_{EMT/Mie}$ are the extinction cross sections of the noninteracting GNPs and SCs, respectively.

All the computational codes, both for the calculation of optical extinction and the regression of the experimental spectra, were developed in MATLAB R2017b (available in the Supporting Information). Mie scattering calculations were implemented using the MATLAB/Octave code written by Demers *et al.*[53] Fig. 3 schematically summarizes the overall workflow, including the theory and the regression steps, in a flow chart. See Supporting Section 2 for further details on the theoretical framework and the regression protocol.

## Results and discussion

### Results from the present study

With the summary in the previous section, we are ready to examine the outcome of the *in situ* measurements conducted in the current study. In the rest of this paper, we will primarily focus on the results for the seeds synthesized with 15% citric acid and grown at 70°C (experiment 15:85-70°C; see Section on Computational framework and Supporting Section 1). Results for other experiments, analysed using our overarching model, are either mechanistically similar or simpler and are discussed in Supporting Sections 4-5.

Fig. 4a shows representative regressions for the experiment 15:85-70°C (see Supporting Figs. 4-8 for all the temporal regressions). We see that applying the Gans theory (*i.e.*, neglecting the optical interaction effects) only provides good regressions at early and later times (Fig. 4a spectra at 0, 20, 80, 750, 1130, 1500 s and Fig. 4b). During these periods, the GEM model describes the data equally well and yields the same optimal values, as those obtained from the Gans theory, for the common parameters $\bar{r}_{GNP}$, $\bar{\beta}$ (Supporting Fig. 30). Nevertheless, the regression becomes insensitive to the EMT-specific parameters ($x_{SC}$, $f$, and $r_{SC}$; Supporting Section 8) and thus, the Gans formalism should be selected to avoid overfitting. During the intermediate period (spectra at 150 to 670 s), however, it is crucial to account for the electromagnetic interactions and the GEM framework has to be chosen (Fig. 4b; see the Section on Computational framework and Supporting Section 2 for further details on model selection). With these considerations in mind, our theoretical framework provides excellent regressions to the experimental spectra throughout the kinetic process. Quantitatively, the range-normalized root mean square errors (NRMSE) are invariably less than 2% for all the regressions while the deviations from experimental data at individual wavelengths and time points are never more than 5% of the extinction range (Fig. 4b and c, respectively).

Fig. 5 presents the temporal evolution of the optimized model parameters (a-e) and the particle number concentration (f) for the experiment 15:85-70°C. In this figure, the red open circles represent the outputs of the GEM model regression whereas the black dots are those corresponding to the selected model along with their heuristically estimated[52] 95% confidence bounds (see the Section on Computation framework and Supporting Section 8 for details). Fig. 6 also provides some additional kinetic outputs obtained from regressing the experimental spectra (a-c), representative recorded spectra (d), the evolution of peak width at half-maximum (PWHM) calculated from the experimental spectra according to Natan *et al.*[54] (e; twice the difference between $\lambda_{LSPR}$ and the wavelength to its red with an extinction half the value at LSPR), and the *in situ* measured solution pH and reduction potential (f; Eh *vs.* Ag/AgCl/c(KCl) = 3 mol/L). The grey-to-white shaded regions in Figs. 3b, 4, and 5 (denoted by Roman) characterize the various mechanistic steps as discussed below.

Fig. 5a,b,f show that the GEM model approaches the Gans formalism when the interaction effects are negligible (that is, when % GNP in SCs approaches zero; 5b). We can see this self-consistent behaviour more clearly in Supporting Fig. 30 where the common parameters/outputs of the two models ($\bar{r}_{GNP}$, $\bar{\beta}$, GNP number concentration, and effective density) are plotted on the same axes. As expected, the discrepancy grows when the interaction effects become more prominent and it reaches a maximum around the same time the percentage of GNPs inside SCs is the highest. With the self-consistency checked, we can now scrutinize the temporal evolution during the seeded growth of GNPs by dividing the overall process time into five regions, as discussed below.

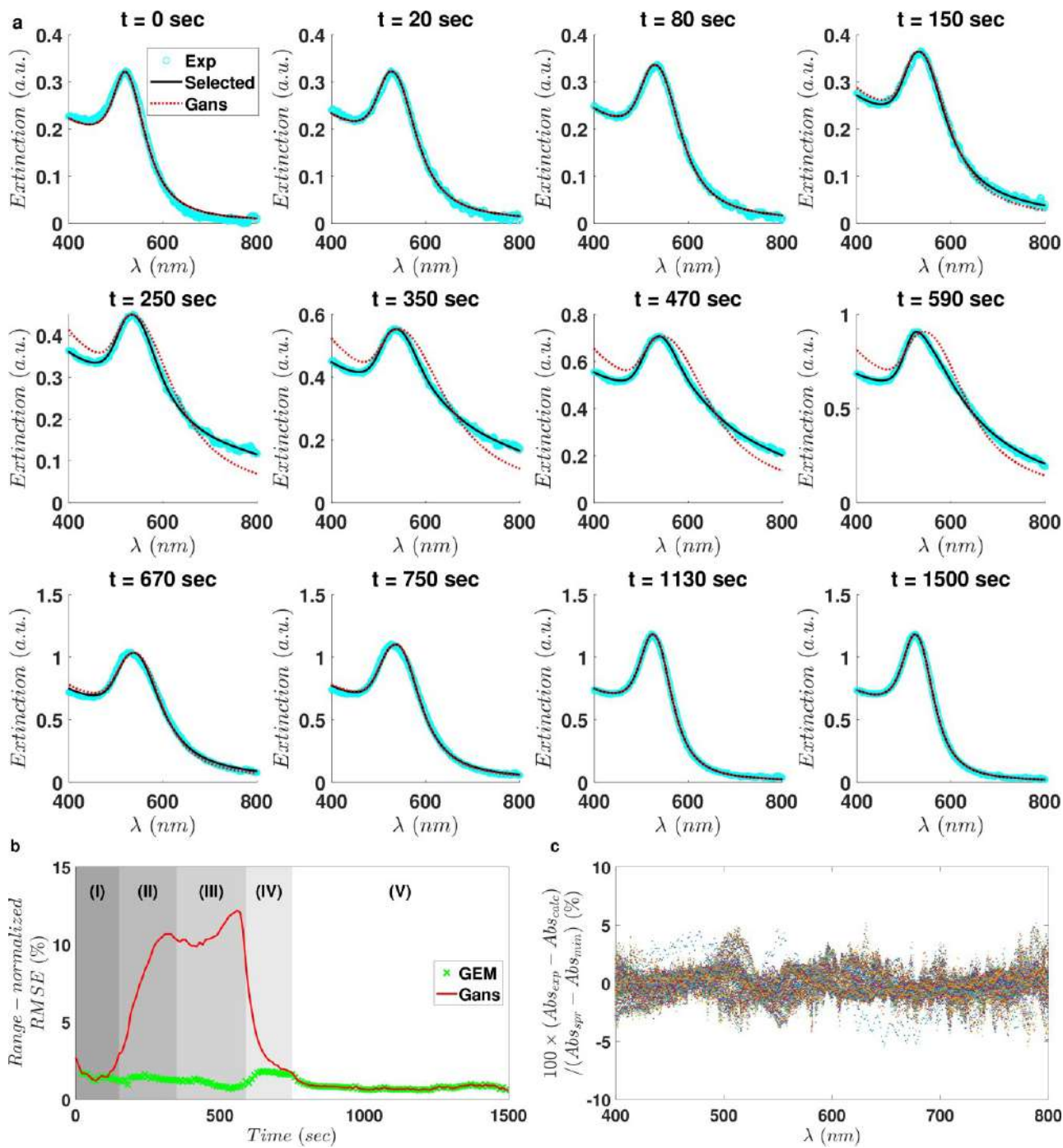

**Fig. 4.** Summary of spectral regressions and their quality (experiment 15:85-70°C). (a) Representative experimental UV-vis spectra (cyan curves composed of several open circles) and their corresponding theoretical fits using the Gans theory (red dashed lines) and the selected model (black lines). (b) Range-normalized root mean square error (NRMSE) for the temporal regressions using the two models. The grey-to-white shaded regions characterize various mechanistic steps (denoted by Roman numerals) as discussed in the main text. (c) Range-normalized deviations from experimental data for individual wavelengths and temporal points (obtained from regressions by the selected model).

In region (I) (0-150 s), as soon as the concentrated Au(III) solution is injected into the seed suspension (Fig. 6d; the spectrum at 20 s) $\bar{r}_{GNP}$ drops from 7.4 to 6.1 nm (Fig. 5a) while the number concentration of particles almost doubles (Fig. 5f). This implies the occurrence of a nucleation event which quickly decelerates as a result of pH change (1.7 in the 20 mM Au(III) solution to 5.5 in the reaction medium at 70°C) converting the highly reactive [AuCl$_4$]$^-$ species into its less reactive, hydrolysed counterparts ([AuCl$_{4-x}$(OH)$_x$]$^-$; x = 1-4).[7,55] The sudden leap in PWHM and its subsequent stabilization correlates with this observation (Fig. 6e). Presumably, the new particles are generated by primary and secondary nucleation (*i.e.*, nucleation independent from or induced by the presence of already formed gold nanoparticles, respectively[26]), as both mechanisms are possible under this operating condition. More specifically, the temperature, pH and number concentration of the seeds lie within a range where both nucleation types are viable (note that at the point of injection the high local supersaturation could induce an even faster nucleation).[5,25] Alternatively, by lowering the temperature to 20°C (in a suspension prepared by only Na$_3$Cit) one can minimize primary nucleation (*i.e.*, preserve the particle number concentration).[5] In the latter case, all the injected Au(III) would be spent to enlarge the seeds through catalytic secondary nucleation and molecular growth (note that compared to primary nucleation, both processes have smaller free energy barriers[26]).[5] Indeed, Wuithschick *et al.* employed this notion to determine the reaction time at which primary nucleation stops during the method of Turkevich.[7]

Another notable feature in region (I) is the formation of more anisotropic nanoparticles (increase in $\bar{\beta}$ from 1.25 to 1.5; Fig. 5b) which is also documented in TEM observations by other workers.[5,22,25,56] Such anisotropic secondary particles can arise due to either slow agglomeration/aggregation,[56] or secondary nucleation.[5] In this case, considering the short duration of region (I) and the near-bulk effective density of the population (Fig. 6a), a secondary nucleation followed by integration to the seeds, *via* interparticle molecular growth, seems to be responsible for the anisotropy (see Supporting Section 2 for the protocol to estimate effective density).[5] A detailed discussion regarding the origins of anisotropy in secondary particles is presented in the classic paper by Enustun and Turkevich.[56]

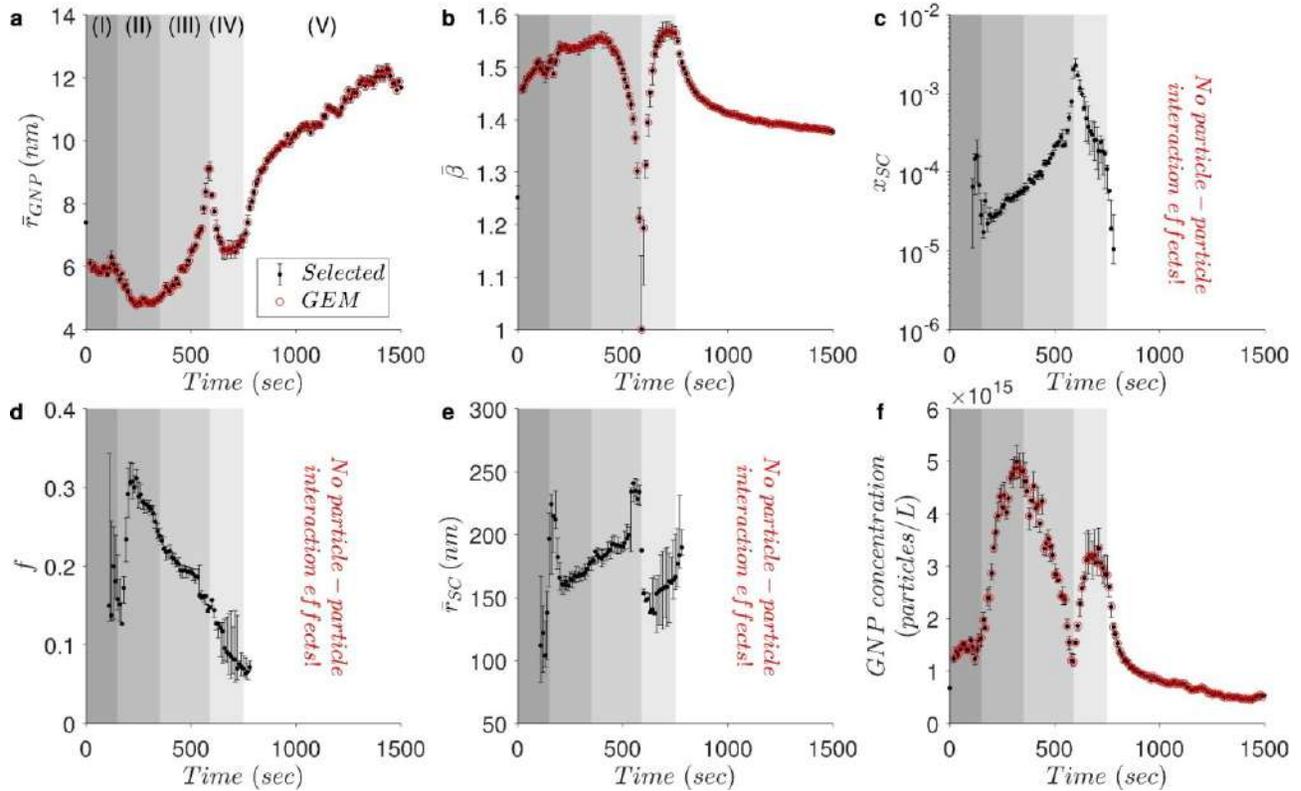

**Fig. 5.** Kinetics of gold nanoparticle growth (experiment 15:85-70°C) from *in situ* UV-vis spectroscopy and model regression. (a-e) Regressed physical parameters (and their corresponding 95% confidence bounds found by a heuristic search method[52]) including mean particle radius (a), mean particle aspect ratio (b), number fraction of liquid-like superclusters (c), volume fraction of particles in superclusters (d), and mean supercluster radius (e). (f) Temporal particle number concentration (along with 95% confidence bounds[52]) obtained from scaling the calculated extinction cross sections with the experimental spectra. Legends are common in all the plots and the grey-to-white shaded regions characterize various mechanistic steps (denoted by Roman numerals in the first plot) as discussed in the main text.

After the nucleation event at the beginning of region (I), ensuing the reduced colloidal stability (due to Au(I) adsorption on the surface) and increased number concentration, there is a quiescent period during which particles start to gradually approach each other. This gives rise to electromagnetic interaction effects and the appearance of mesoscale liquid-like superclusters (Fig. 5c-e and Fig. 6b,c; Fig. 6d,e enhanced extinction in the wavelength range 600-800 nm and peak broadening).[14] Looking at the potential measurements (Fig. 6f), we see that the pH continuously drops which is due to the injection of chloroauric acid solution and $[AuCl_4]^- \rightarrow Au(I)$ reduction by citrate.[21,55] The latter, which dominates after the mixing period (probably less than a second), provides growth units/monomers for the densification of secondary particles (*via* molecular growth and neck formation[5]) in addition to a subsequent nucleation event which happens in region (II) (see below). Looking at the Eh data (Fig. 6f) in region (I), and knowing that the Au(I)/Au(0) couple mainly determines the potential of the gold electrode,[20,28] we understand that the quick initial drop in potential is related to the consumption of Au(I) by nucleation and molecular growth. Following this shallow drop, Eh rises back up which indicates the build-up of Au(I) consistent with the pH behaviour discussed earlier.

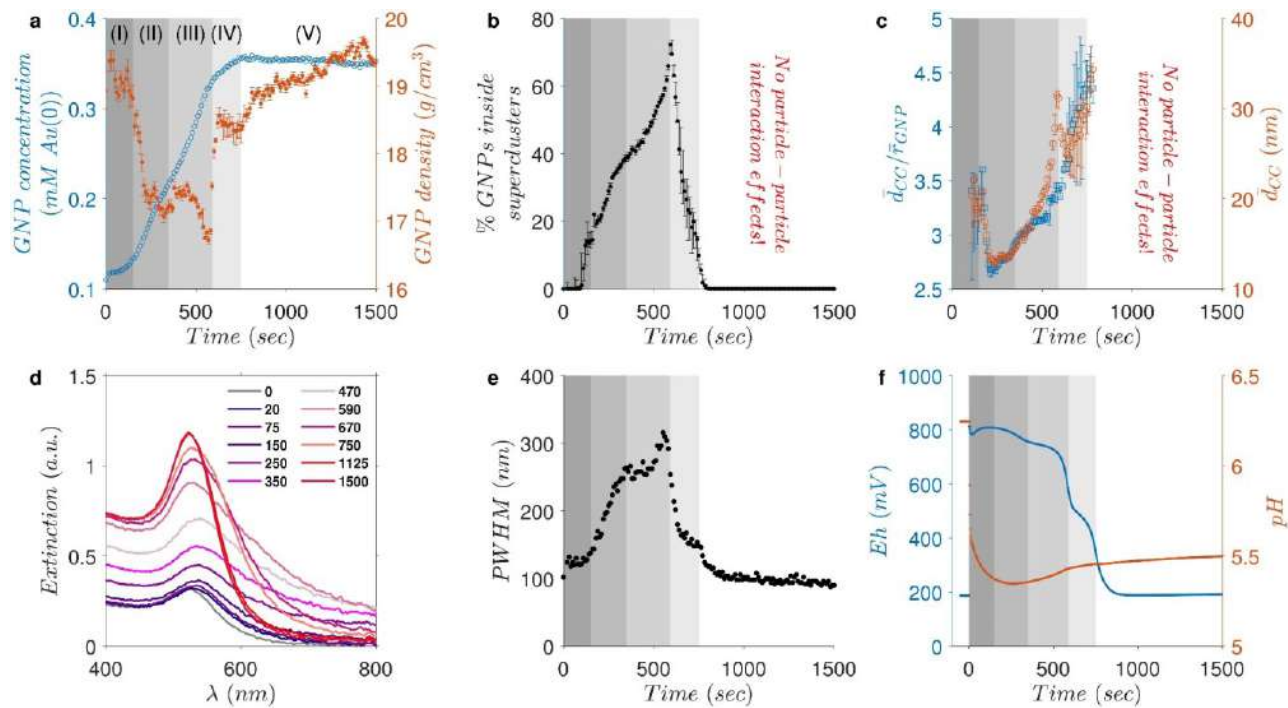

**Fig. 6.** Complementary kinetic data (experiment 15:85-70°C) from *in situ* UV-vis spectroscopy and electrochemical potential measurements. (a) Gold nanoparticle concentration (mM Au(0); obtained from extinction at 400 nm[27]) and the estimated volumetric mass density (effective) of nanoparticles (with 95% confidence bounds found by a heuristic search method[52]). (b) Estimated percentage of electromagnetically interacting particles out of the overall population (*i.e.*, interacting + noninteracting) along with the respective 95% confidence bounds.[52] (c) Mean centre-to-centre distance between electromagnetically interacting nanoparticles both normalized to the average particle radius and in absolute units (with 95% confidence bounds[52]). (d) Experimental extinction spectra at selected temporal points. (e) Temporal evolution of peak width at half-maximum obtained from experimentally measured spectra.[54] (f) *In situ* measured solution pH and reduction potential (Eh *vs.* Ag/AgCl/c(KCl) = 3 mol/L). The grey-to-white shaded regions (all the plots except for (d)) characterize various mechanistic steps (denoted by Roman numerals in the first plot) as discussed in the main text.

In region (II) (150-350 s) the formation of Au(0) accelerates (Fig. 6a) due to a burst of new particles increasing the number concentration (Fig. 5f) and lowering $\bar{r}_{GNP}$ (Fig. 5a). This corresponds to a second nucleation event (both primary and secondary nucleation). One major difference compared to region (I) is the rapidly dropping effective density of GNPs (Fig. 6a) which indicates that most of the Au(III) precursor is consumed in nucleation rather than molecular growth. This way, inside the resultant anisotropic secondary particles, mainly physical bonds hold primary particles together. We may identify these secondary particles as agglomerates but we should note that the mechanism behind their formation is true catalytic secondary nucleation[5,26] and not agglomeration.

As we mentioned earlier, in region (I) the concentration of Au(I) builds up,[21,29] which in turn provides a high supersaturation level that allows for significant primary nucleation alongside secondary nucleation in region (II). Concurrent with the rapid increase in number concentration and in the presence of surface-adsorbed Au(I), colloidal stability decreases further and more GNPs approach each other to form liquid-like superclusters (Fig. 5c-e and Fig. 6b,c; Fig. 6d enhanced extinction in 600-800 nm). Nanoparticles assemble into mesoscale structures ~ 350 nm in diameter (Fig. 5e) with an average centre-to-centre distance ~ 15 nm ($\cong 3\ \bar{r}_{GNP}$) between the GNPs (Fig. 6c). The presence of both significant electromagnetic interactions and elongated nanoparticle agglomerates induces a two-fold rise in the experimentally measured PWHM (Fig. 6e).

Consulting Fig. 6f in region (II), the pH remains almost constant although the reduction of Au(III) by citrate releases protons.[21,55] To explain this, we note that most of the highly reactive [AuCl$_4$]$^-$ species has already converted either into Au(I) or [AuCl$_{4-x}$(OH)$_x$]$^-$. For this reason, Au(III) speciation is shifted toward [AuCl$_4$]$^-$ (Le Chatelier's principle), releasing OH$^-$:

$$[AuCl_4]^- + x\ OH^- \rightleftharpoons [AuCl_{4-x}(OH)_x]^- + x\ Cl^-;\ x=1\text{-}4 \quad (12)$$

This shift counterbalances the pH decreasing effect of Au(III) reduction and remains until the end of the process, as witnessed by the continuous slow increase in pH. Finally, with Au(I)/Au(0) as the chief potential determining couple,[20,28] the gradual decrease in Eh (Fig. 6f) implies a continuous $Au(I) \rightarrow Au(0)$ reaction in region (II).

Region (III) (350-590 s) is accompanied by a precipitous drop in the number concentration of GNPs (Fig. 5f) which come together and produce larger, secondary particles around 9 nm in radius (Fig. 5a). These secondary particles are predicted to have a quite spherical shape (Fig. 5b) and are of lower effective
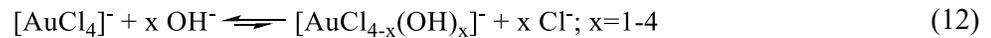

density compared to bulk gold (Fig. 6a). Therefore, they appear to be randomly packed fluffy agglomerates, whose morphology results from fast agglomeration under reduced colloidal stability.[19,20,56] The high number concentration at the beginning of this period supports such a fast coagulation process.[7,18–20,56] Concurrently, growth continues to consume ionic Au as evidenced by the continuous increase in Au(0) concentration (Fig. 6a).

Looking at the state of particle-particle interaction, it continues to grow (Fig. 6b) and reaches a maximum at the end of this period (more than 70% of GNPs experiencing electromagnetic interactions with neighbouring particles). This is consistent with the continued enhancement of extinction in the wavelength range 600-800 (the transient purple-greyish hue observed in the evolving suspension; Fig. 6d). Consulting Fig. 6e, initially PWHM does not significantly change, but toward the end of this period, it rises toward a maximum of more than 300 nm (more than threefold the value in the seed spectrum). To explain this, we note that at the beginning the formation of more spherical agglomerates counteracts the peak broadening caused by enhanced electromagnetic interactions. Yet, eventually interaction effects dominate and broaden the localized surface plasmon resonance (LSPR) band (Fig. 6e).

In region (III), the majority of Au(III) is in the hydrolysed form (considering the pH level, Fig. 6f) and they gradually reduce to Au(I), either directly or *via* [AuCl$_4$]$^-$.[55] Hence, unlike region (I) the level of supersaturation (Au(I) activity) is comparatively low. Consequently, contrary to the dense secondary particles in region (I), the agglomerates forming in region (III) do not have as extensive chemical bonds and they are less dense (Fig. 6a). Towards the end of this region, Eh decreases quickly indicating the onset of another nucleation event, manifesting in region (IV) (Fig. 6f).

In region (IV) (590-750 s) the number concentration rises once again (Fig. 5f) and $\bar{r}_{GNP}$ drops (Fig. 5a). Meanwhile, $\bar{\beta}$ rapidly increases due to the reappearance of anisotropic secondary particles. We can describe these observations by another nucleation event. Here, considering the relatively low supersaturation degrees (Fig. 6f), the energetically more favourable secondary nucleation is probably more significant than primary nucleation. A second mechanism that can account for the increased number concentration is the (partial) peptization (*i.e.*, redispersion of agglomerated particles back to colloidally stable primary particles[34]) of the agglomerates formed in region (III). Chemical bond formation between primary particles inside these agglomerates gives rise to denser secondary particles (Fig. 6a). During this compaction, the loosely bound primary particles would have a chance to repeptize. This is assisted by the recovered colloidal stability due to the relatively low particle number concentration at the end of region (III) (Fig. 5f) and the lower level of Au(I) (Fig. 6f). Complementary evidence for the regained colloidal stability comes from the rapidly disappearing interaction effects (Fig. 5c,d and Fig. 6b,c). As the interaction effects vanish, PWHM decreases as well, with a deceleration towards the end of the period (Fig. 6d), the latter being due to the reappearance of anisotropic entities (Fig. 5b).

In the last temporal region (750-1500 s), there is an abrupt drop in the number concentration (Fig. 5f), and a fast rise in $\bar{r}_{GNP}$ (Fig. 5a) and the average effective density (Fig. 6a). Notably, Au(0) concentration does not perceptibly increase anymore (Fig. 6a) implying an almost complete depletion of supersaturation. Eh being very close to its final equilibrium value further corroborates this hypothesis (Fig. 6f). Hence, this period is dominated by an Ostwald ripening event (the amorphous nature of newly formed GNPs, as documented by Loh *et al.*,[4] can help the dissolution of small nanoparticles). Thus, larger particles grow at the expense of smaller particles and agglomerates coalesce to structures that are denser (Fig. 6a). In this period, coalescence along with intra-particle ripening render the particles more spherical.[5,22] Finally, in the absence of electromagnetic interactions, the two optical theories (GEM and Gans) predict similar outputs (Supporting Fig. 30).

Fig. 7 schematically summarizes the mechanistic picture presented earlier, enumerating the most prominent events in each temporal region. In this scheme, the degree of overlap between primary particles inside secondary particles is proportionate to the effective density of the latter (*i.e.*, the more the overlap, the higher the effective density). Note that molecular growth, defined as particle enlargement by the addition of molecularly sized entities, can happen as long as the system is supersaturated.[26] In region (V), this could be in the form of Ostwald/interparticle ripening.

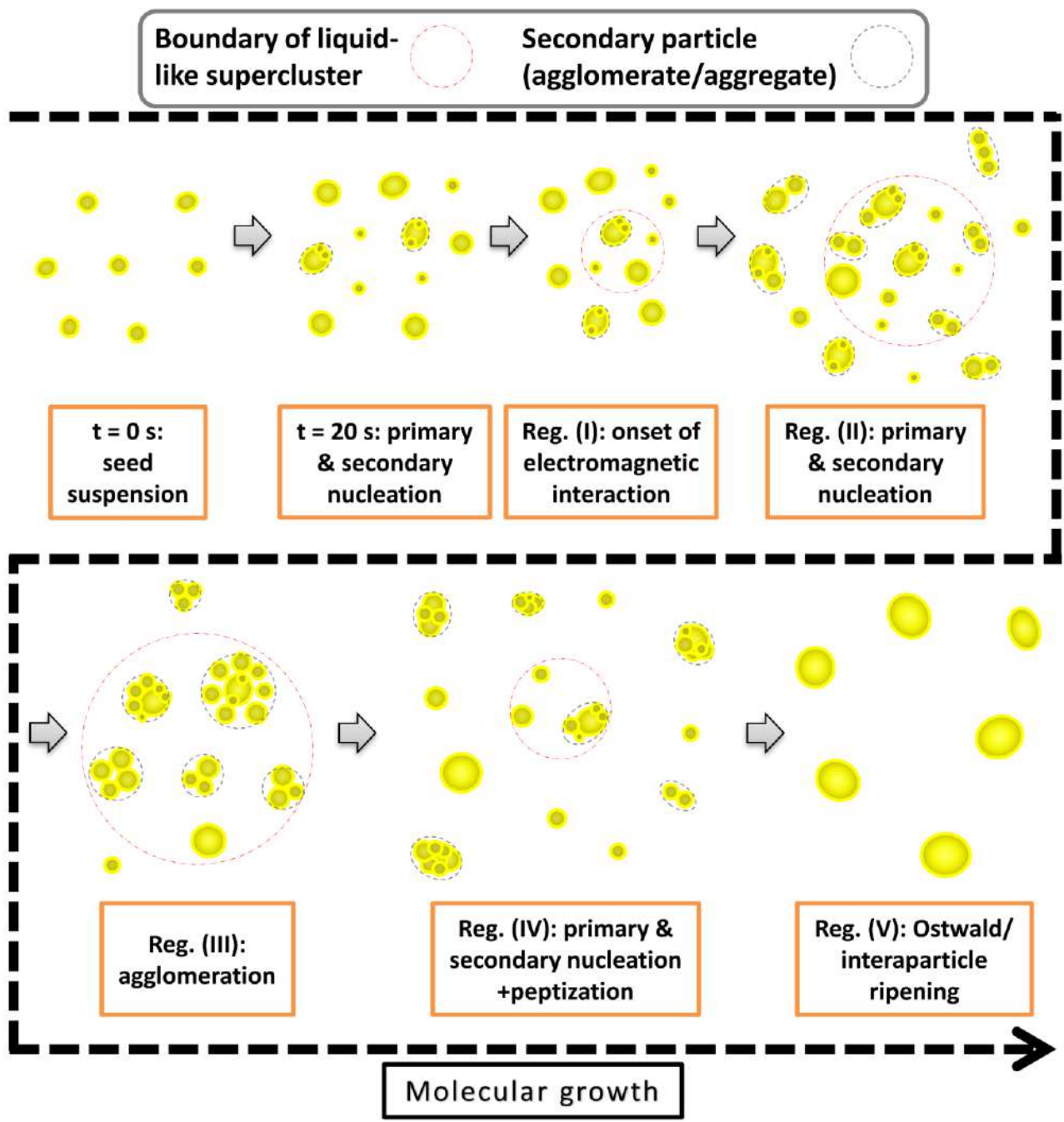

**Fig. 7.** Proposed mechanism for the seeded growth of gold nanoparticles with a schematic representation portraying the sequence of processes encountered during the seeded growth of GNPs in the experiment 15:85-70°C. The degree of overlap between primary particles inside secondary particles correlates with the effective density of the latter (*i.e.*, more overlap means a higher effective density). Seeds and final grown GNPs are shown as individual particles to denote coalescence into single entities. Molecular growth happens throughout the process as represented by the continuous dashed arrow. To save space, the relative size of superclusters is not drawn to scale (they should be much larger than the individual secondary particles).

**Corroborating evidence from principal component analysis**

Principal component analysis (PCA) is a useful and accessible tool to identify the major contributors to an overall signal.[57,58] For an evolving colloidal suspension, the optical response can be decomposed into linearly uncorrelated variables, called principal components (PCs), giving coefficients (or loadings) with their corresponding temporal scores. In simple words, each coefficient represents the spectral signature of a PC while the score correlates with the weight (significance) of that PC at a specific time. PCA has a limitation in the sense that each spectrum is assumed to be a linear combination of several pure spectra. A tacit assumption in the previous statement is the constancy of the pure spectra (over time), which does not hold for a growing colloidal metal suspension. Therefore, PCA would detect artificial compounds, the linear combinations of which give the temporal spectra. Nevertheless, it is not possible to assign the coefficients to particular species.[57] In spite of these limitations, PCA allows for a model-free estimation of the number of contributing phenomena and provides rough optical signatures for each of them.

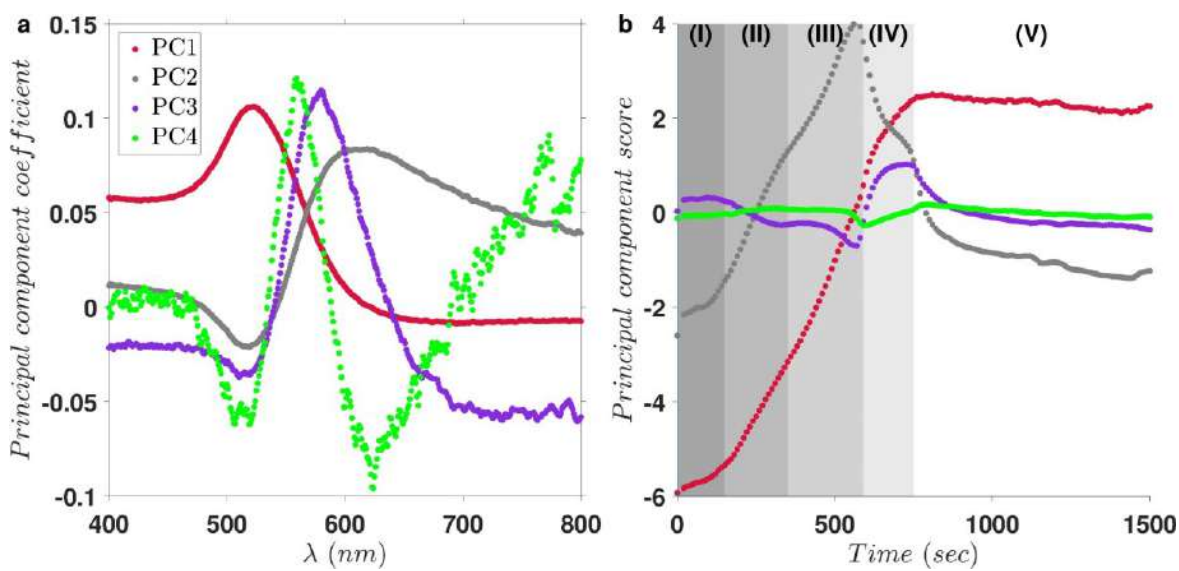

**Fig. 8.** Coefficients (a) and scores (b) from principal component analysis on temporal UV-vis spectra (15:85-70°C experiment). Legends are common in both plots. The grey-to-white shaded regions in part b characterize various mechanistic steps (denoted by Roman numerals) as discussed in the main text.

With the introduction above in mind, PCA over a spectral range 400-800 nm using the centered extinction data from the experiment 15:85-70°C (Fig. 8) indicates that there are three main contributors while the fourth PC is practically capturing noise (as implied by its noisy coefficient and low score throughout the process). Note that there is no contribution from the dissolved ionic gold species nor from the organic compounds in the selected spectral range.[7,28] Therefore, only three PCs, all originating from the colloidal nanoparticles, capture > 99.8% of the temporal variation in the spectra with individual contributions 75.9, 22.8, 1.3% for PCs 1 to 3, respectively.

Taking a closer look at the coefficient plot (Fig. 8a) reveals that the first PC (red full circles) is mainly related to individual GNPs with their LSPR band around 520 nm. The corresponding score plot (Fig. 8b) indicates a monotonic evolution toward the final GNPs which bear more Au(0). This is consistent with the continuously increasing concentration of atomic gold shown in Fig. 6a.

The second PC (Fig. 8a; gray full circles) exhibits an inverse behavior close to the LSPR band (plasmon damping) while it correlates with the enhanced extinction in the range 600-800 nm. The corresponding score grows to a maximum and drops subsequently (Fig. 8b). Thus, this PC majorly represents the electromagnetic interactions between the particles that appear early on, grow as the particles approach each other, and fade out as they peptize back to noninteracting scatterers. Mikhlin *et al.*[14] identified a very similar temporal evolution during the one-pot synthesis of GNPs, where large (100-500 nm) three-dimensional networks of GNPs suspended in aqueous droplets appeared early on and disintegrated as the process proceeded to the end.

Finally, the third PC (Fig. 8a; purple full circles) demonstrates a similar coefficient behavior as PC2 but with a narrower peak around 580 nm and a sharper contrast between this peak and the LSPR-damping band. The contrast signifies the distinction between longitudinal and transversal extinction cross sections in prolate-shaped particles. Therefore, this PC can mainly be ascribed to the particle shape evolution that dampens the LSPR band and enhances the extinction in the red region. Looking at Fig. 8b, the temporal variation in the corresponding score plot correlates quite closely with the evolution observed in $\bar{\beta}$ (Fig. 5b).

To summarize, the collective nonlinear combination of the three effects identified by PCA (*i.e.*, the extinction by individual GNPs as related to their Au(0) content, the particle aspect ratio, and the electromagnetic interactions between the GNPs) dictates the temporal optical behavior of the particle ensemble. This nonlinear (and correlated) dependence does not allow for a one-to-one quantitative assignment between the PCs and the physical characteristics. That is why we tackled the problem directly adopting the nonlinear regression of a physicochemical model to the experimental data. Nevertheless, as we demonstrated earlier PCA does provide qualitative insight about the physics of the problem and provides support for the proposed theoretical framework. Results from PCA on other datasets exhibit similar resemblance to the output of the optical modeling framework and they are presented in Supporting Section 7.

## Conclusions

In summary, we developed a comprehensive theoretical framework to extract kinetic and mechanistic information about the seeded growth of gold nanoparticles from time-resolved *in situ* optical extinction spectroscopy (OES). When compared to more sophisticated and expensive characterization techniques such as SAXS and *in situ* TEM, OES is appealing due to its availability, low cost, and relatively straightforward data collection procedure under various operating conditions. So far, the quantitative application of this technique was hampered by the lack of a complete theoretical formalism capable of extracting key temporal kinetic information like particle size, shape, and concentration, as well as the state of particle-particle interaction (mesoscale aspects). In this article, we drew on the extensive previous research in the field and developed a physical model to describe the optical response of a growing gold sol. We demystified the complexity of the process, which involves several temporally overlapping subproceses, *e.g.* primary and secondary nucleation, molecular growth, agglomeration, and Ostwald ripening. We observed similar events for a wide range of experiments (*i.e.*, six sets of data at different T and pH) and presented mechanistic pathways to account for the various trends observed in the outputs of the spectral regression. In this respect, we identified the role of nucleation events accompanying the particle enlargement process. Moreover, we showed that the particle enlargement occurs *via* a combination of molecular growth (in the form of Ostwald ripening, particularly close to the end of the process), secondary nucleation, and agglomeration. We further identified instances of peptization consistent with the previous experimental evidence reported by other workers.[18–20] We also addressed the longstanding question about the nature of moieties responsible for the purple-greyish color during the growth of gold nanoparticles. This was determined by distinguishing between the optical signatures of electromagnetic interactions between the particles inside liquid-like superclusters, from that of relatively small fluffy agglomerates. Finally, our theoretical formalism was developed in a generic form so that it can also be applied to other nanoparticulate systems such as various metal nanoparticles.

## Materials and methods

### Materials

Metallic gold foil (99.99% pure) was purchased from Nuova Franco Suisse Italia. Hydrogen chloride (37 wt. % in $H_2O$, 99.999% trace metals basis), nitric acid (70% $HNO_3$, ≥99.999% trace metals basis), sodium citrate tribasic dihydrate ($Na_3Cit.2H_2O$, ACS reagent ≥99.0%), and citric acid ($H_3Cit$, ACS reagent ≥99.5%) were purchased from Sigma Aldrich. Ultrapure water (18.2 MΩ.cm, MilliQ) was used throughout the experiments. All the glassware and PTFE-coated magnetic stirring bars were cleaned with freshly prepared *aqua regia* and rinsed thoroughly with MilliQ water before each experiment. A stock solution of $HAuCl_4$ in water was prepared according to the procedure by Gross[59] and used in all the syntheses. Fresh solutions of $HAuCl_4$ and citrate buffer (with the desired concentrations) were prepared before each experiment and stored in tightly closed glass bottles.

### Synthesis of GNP seeds and subsequent growth experiments

Gold nanoparticle seeds were prepared by injecting 1 mL of a 50 mM $HAuCl_4$ aqueous solution into 499 mL boiling solution of 2 mM citrate buffer under intensive stirring. The buffer's pH was tuned by adjusting the molar ratio citric acid/trisodium citrate (0%:100%, 15%:85%, and 25%:75%). We then used these particles in a seeded growth process, the kinetics of which was of interest. For this purpose, 0.75 mL 20 mM $HAuCl_4$ solution was rapidly injected into 60 mL of a seed suspension (under vigorous stirring) at different temperatures (40, 60, 65, 70°C). The temporal evolution of the system was followed by UV-vis spectroscopy (one spectrum every 20 s for the experiments 25:75-40°C and 0:100-70°C, and every 10 s for the rest of the experiments), reduction potential (Eh) measurement, and pH measurement (Eh and pH data points collected every two seconds). Six sets of experimental data were collected under various conditions (T and pH) using different seed suspensions. This includes experiments at 60, 65, and 70°C using the 15%:85%-buffer seed, experiments at 70°C using the 25%:75%- and 0%:100%-buffer seeds, and an experiment at 40°C using the 25%:75%-buffer seed suspension. Reference to each experiment denotes the buffer ratio followed by the experimental temperature, *e.g.*, 15:85-70°C (experiment using 15%:85%-buffer seed at 70°C). See Supporting Section 1 for further details on the kinetic experiments.

**Characterization**

The volume-averaged equivalent spherical radius for the seeds and grown GNP were found by TEM (JEOL-2010), measuring the semi-major and semi-minor axes of at least 100 particles using Fiji package and assuming prolate shaped particles (the axis perpendicular to the image is taken as a minor axis).[54,60,61] The TEM samples were prepared following the protocol recommended by Parak *et al.*.[61] *In situ* UV-vis spectra were collected using an Ocean Optics PEEK immersion probe with 1 cm optical path length (pure water was used as the reference). The probe was connected to a DT-mini-2-GS UV-Vis-NIR light source and a USB2000+XR1 spectrometer, both from Ocean Optics. Each spectrum was an average of four 2.5-second (5 s for the data in 0:100-70°C and 25:75-40°C experiments) acquisitions smoothed with a boxcar width two. Measurements of pH and Eh (*vs.* Ag/AgCl/c(KCl) = 3 mol/L) were done using a pH glass electrode (Unitrode Pt 1000) and a combined gold ring electrode, respectively, provided by Metrohm AG.

## Acknowledgements

This work was supported by the Energy and Environment Research Division (ENE), Paul Scherrer Institute, Switzerland.

The authors thank Mr. Joerg Schneebeli (PSI) for providing the UV-Vis spectrophotometer, Prof. Christian Ludwig (PSI and EPFL) for valuable discussions, Dr. Ivo Alxneit (PSI) and Dr. Ralf Kaegi (EMPA) for their help with TEM sample preparation and imaging, and Dr. Agnese Carino and Dr. Mohamed Tarik for helping with the experiments.

## Associated content

**Supporting Information Available:** Additional details on experimental and computational procedures, TEM micrographs of all the samples along with their particle size distributions, outputs and discussion for datasets not explored in the main text, convergence of the two optical models to the same output, discussion on confidence bounds obtained during regression on experimental data, and fully-commented MATLAB codes for the entire computational workflow. This material is available free of charge *via* the Internet at http://pubs.acs.org.

## References

(1) Faraday, M. The Bakerian Lecture: Experimental Relations of Gold (and Other Metals) to Light. *Philos. Trans. R. Soc. London* **1857**, *147*, 145–181.

(2) Zhao, P.; Li, N.; Astruc, D. State of the Art in Gold Nanoparticle Synthesis. *Coord. Chem. Rev.* **2013**, *257*, 638–665.

(3) Saltelli, A.; Ratto, M.; Tarantola, S.; Campolongo, F. Update 1 of: Sensitivity Analysis for Chemical Models. *Chem. Rev.* **2012**, *112*, PR1–PR21.

(4) Loh, N. D.; Sen, S.; Bosman, M.; Tan, S. F.; Zhong, J.; Nijhuis, C. A.; Král, P.; Matsudaira, P.;

Mirsaidov, U. Multistep Nucleation of Nanocrystals in Aqueous Solution. *Nat. Chem.* **2016**, *9*, 77.

(5) Cheng, Y.; Tao, J.; Zhu, G.; Soltis, J. A.; Legg, B. A.; Nakouzi, E.; De Yoreo, J. J.; Sushko, M. L.; Liu, J. Near Surface Nucleation and Particle Mediated Growth of Colloidal Au Nanocrystals. *Nanoscale* **2018**, *10*, 11907–11912.

(6) Abécassis, B.; Testard, F.; Kong, Q.; Francois, B.; Spalla, O. Influence of Monomer Feeding on a Fast Gold Nanoparticles Synthesis: Time-Resolved XANES and SAXS Experiments. *Langmuir* **2010**, *26*, 13847–13854.

(7) Wuithschick, M.; Birnbaum, A.; Witte, S.; Sztucki, M.; Vainio, U.; Pinna, N.; Rademann, K.; Emmerling, F.; Kraehnert, R.; Polte, J. Turkevich in New Robes: Key Questions Answered for the Most Common Gold Nanoparticle Synthesis. *ACS Nano* **2015**, *9*, 7052–7071.

(8) Li, T.; Senesi, A. J.; Lee, B. Small Angle X-Ray Scattering for Nanoparticle Research. *Chem. Rev.* **2016**, *116*, 11128–11180.

(9) Benfatto, M.; Congiu-Castellano, A.; Daniele, A.; Della Longa, S. *MXAN* : A New Software Procedure to Perform Geometrical Fitting of Experimental XANES Spectra. *J. Synchrotron Radiat.* **2001**, *8*, 267–269.

(10) Ravel, B.; Newville, M.; IUCr. *ATHENA* , *ARTEMIS* , *HEPHAESTUS* : Data Analysis for X-Ray Absorption Spectroscopy Using *IFEFFIT*. *J. Synchrotron Radiat.* **2005**, *12*, 537–541.

(11) Ilavsky, J.; Jemian, P. R.; IUCr. *Irena* : Tool Suite for Modeling and Analysis of Small-Angle Scattering. *J. Appl. Crystallogr.* **2009**, *42*, 347–353.

(12) Breßler, I.; Kohlbrecher, J.; Thünemann, A. F.; IUCr. *SASfit* : A Tool for Small-Angle Scattering Data Analysis Using a Library of Analytical Expressions. *J. Appl. Crystallogr.* **2015**, *48*, 1587–1598.

(13) Plech, A.; Kotaidis, V.; Siems, A.; Sztucki, M. Kinetics of the X-Ray Induced Gold Nanoparticle Synthesis. *Phys. Chem. Chem. Phys.* **2008**, *10*, 3888.

(14) Mikhlin, Y.; Karacharov, A.; Likhatski, M.; Podlipskaya, T.; Zubavichus, Y.; Veligzhanin, A.; Zaikovski, V. Submicrometer Intermediates in the Citrate Synthesis of Gold Nanoparticles: New Insights into the Nucleation and Crystal Growth Mechanisms. *J. Colloid Interface Sci.* **2011**, *362*, 330–336.

(15) Mohammed, A. S. A.; Carino, A.; Testino, A.; Andalibi, M. R.; Cervellino, A. In Situ Liquid SAXS Studies on the Early Stage of Calcium Carbonate Formation. *Part. Part. Syst. Charact.* **2018**, *36*, 1800482.

(16) Mizutani, T.; Ogawa, S.; Murai, T.; Nameki, H.; Yoshida, T.; Yagi, S. In Situ UV-Vis Investigation of Growth of Gold Nanoparticles Prepared by Solution Plasma Sputtering in NaCl Solution. *Appl. Surf. Sci.* **2015**, *354*, 397–400.

(17) Khoury, R. A.; Ranasinghe, J. C.; Dikkumbura, A. S.; Hamal, P.; Kumal, R. R.; Karam, T. E.; Smith, H. T.; Haber, L. H. Monitoring the Seed-Mediated Growth of Gold Nanoparticles Using in Situ Second Harmonic Generation and Extinction Spectroscopy. *J. Phys. Chem. C* **2018**, *122*, 24400–24406.

(18) Biggs, S.; Mulvaney, P.; Zukoski, C. F.; Grieser, F. Study of Anion Adsorption at the Gold-Aqueous Solution Interface by Atomic Force Microscopy. *J. Am. Chem. Soc.* **1994**, *116*, 9150–9157.

(19) Chow, M. K.; Zukoski, C. F. Gold Sol Formation Mechanisms: Role of Colloidal Stability. *J. Colloid Interface Sci.* **1994**, *165*, 97–109.

(20) Rodríguez-González, B.; Mulvaney, P.; Liz-Marzán, L. M. An Electrochemical Model for Gold Colloid Formation *via* Citrate Reduction. *Zeitschrift fur Phys. Chemie* **2007**, *221*, 415–426.

(21) Leng, W.; Pati, P.; Vikesland, P. J. Room Temperature Seed Mediated Growth of Gold Nanoparticles: Mechanistic Investigations and Life Cycle Assesment. *Environ. Sci. Nano* **2015**, *2*, 440–453.

(22) Ji, X.; Song, X.; Li, J.; Bai, Y.; Yang, W.; Peng, X. Size Control of Gold Nanocrystals in Citrate Reduction: The Third Role of Citrate. *J. Am. Chem. Soc.* **2007**, *129*, 13939–13948.

(23) Koerner, H.; MacCuspie, R. I.; Park, K.; Vaia, R. A. In Situ UV/Vis, SAXS, and TEM Study of Single-Phase Gold Nanoparticle Growth. *Chem. Mater.* **2012**, *24*, 981–995.

(24) Chen, X.; Schröder, J.; Hauschild, S.; Rosenfeldt, S.; Dulle, M.; Förster, S. Simultaneous SAXS/WAXS/UV-Vis Study of the Nucleation and Growth of Nanoparticles: A Test of Classical Nucleation Theory. *Langmuir* **2015**, *31*, 11678–11691.

(25) Bastús, N. G.; Comenge, J.; Puntes, V. Kinetically Controlled Seeded Growth Synthesis of Citrate-Stabilized Gold Nanoparticles of up to 200 Nm: Size Focusing *versus* Ostwald Ripening. *Langmuir* **2011**, *27*, 11098–11105.

(26) Andalibi, M. R.; Kumar, A.; Srinivasan, B.; Bowen, P.; Scrivener, K.; Ludwig, C.; Testino, A. On the Mesoscale Mechanism of Synthetic Calcium–Silicate–Hydrate Precipitation: A Population Balance Modeling Approach. *J. Mater. Chem. A* **2018**, *6*, 363–373.

(27) Hendel, T.; Wuithschick, M.; Kettemann, F.; Birnbaum, A.; Rademann, K.; Polte, J. In Situ Determination of Colloidal Gold Concentrations with Uv-Vis Spectroscopy: Limitations and Perspectives. *Anal. Chem.* **2014**, *86*, 11115–11124.

(28) Lingane, J. J. Standard Potentials of Half-Reactions Involving + 1 and + 3 Gold in Chloride Medium. *J. Electroanal. Chem.* **1962**, *4*, 332–342.

(29) Fong, Y. Y.; Visser, B. R.; Gascooke, J. R.; Cowie, B. C. C.; Thomsen, L.; Metha, G. F.; Buntine,

M. A.; Harris, H. H. Photoreduction Kinetics of Sodium Tetrachloroaurate under Synchrotron Soft X-Ray Exposure. *Langmuir* **2011**, *27*, 8099–8104.

(30) Ghosh, S. K.; Pal, T. Interparticle Coupling Effect on the Surface Plasmon Resonance of Gold Nanoparticles: From Theory to Applications. *Chem. Rev.* **2007**, *107*, 4797–4862.

(31) Likhatski, M.; Karacharov, A.; Kondrasenko, A.; Mikhlin, Y. On a Role of Liquid Intermediates in Nucleation of Gold Sulfide Nanoparticles in Aqueous Media. *Faraday Discuss.* **2015**, *179*, 235–245.

(32) Quinten, M. *Optical Properties of Nanoparticle Systems: Mie and Beyond*; Wiley–VCH, Weinheim, Germany: Weinheim, Germany, 2011.

(33) Carl, N.; Prévost, S.; Fitzgerald, J. P. S.; Karg, M. Salt-Induced Cluster Formation of Gold Nanoparticles Followed by Stopped-Flow SAXS, DLS and Extinction Spectroscopy. *Phys. Chem. Chem. Phys.* **2017**, *19*, 16348–16357.

(34) Alemán, J. V.; Chadwick, A. V.; He, J.; Hess, M.; Horie, K.; Jones, R. G.; Kratochvíl, P.; Meisel, I.; Mita, I.; Moad, G.; *et al.* Definitions of Terms Relating to the Structure and Processing of Sols, Gels, Networks, and Inorganic-Organic Hybrid Materials (IUPAC Recommendations 2007). *Pure Appl. Chem.* **2007**, *79*.

(35) Yakubovsky, D. I.; Arsenin, A. V.; Stebunov, Y. V.; Fedyanin, D. Y.; Volkov, V. S. Optical Constants and Structural Properties of Thin Gold Films. *Opt. Express* **2017**, *25*, 25574.

(36) Kreibig, U.; Vollmer, M. *Optical Properties of Metal Clusters*; Springer Series in Materials Science; Springer Berlin Heidelberg: Berlin, Heidelberg, 1995; Vol. 25.

(37) Bouillard, J.-S. G.; Dickson, W.; O'Connor, D. P.; Wurtz, G. A.; Zayats, A. V. Low-Temperature Plasmonics of Metallic Nanostructures. *Nano Lett.* **2012**, *12*, 1561–1565.

(38)  Foerster, B.; Joplin, A.; Kaefer, K.; Celiksoy, S.; Link, S.; Sönnichsen, C. Chemical Interface Damping Depends on Electrons Reaching the Surface. *ACS Nano* **2017**, *11*, 2886–2893.

(39)  Coronado, E. A.; Schatz, G. C. Surface Plasmon Broadening for Arbitrary Shape Nanoparticles: A Geometrical Probability Approach. *J. Chem. Phys.* **2003**, *119*, 3926–3934.

(40)  Amendola, V.; Meneghetti, M. Size Evaluation of Gold Nanoparticles by UV−vis Spectroscopy. *J. Phys. Chem. C* **2009**, *113*, 4277–4285.

(41)  Kildishev, A. V.; Shalaev, V. M.; Chen, K.-P.; Borneman, J. D.; Drachev, V. P. Drude Relaxation Rate in Grained Gold Nanoantennas. *Nano Lett.* **2010**, *10*, 916–922.

(42)  Khlebtsov, N. G. Determination of Size and Concentration of Gold Nanoparticles from Extinction Spectra. *Anal. Chem.* **2008**, *80*, 6620–6625.

(43)  Fernández-Prini, R.; Dooley, R. B. Release on the Refractive Index of Ordinary Water Substance as a Function of Wavelength, Temperature and Pressure. *Int. Assoc. Prop. Water Steam* **1997**, 1–7.

(44)  Ross, M. B.; Mirkin, C. A.; Schatz, G. C. Optical Properties of One-, Two-, and Three-Dimensional Arrays of Plasmonic Nanostructures. *J. Phys. Chem. C* **2016**, *120*, 816–830.

(45)  Pellegrini, G.; Mattei, G.; Bello, V.; Mazzoldi, P. Interacting Metal Nanoparticles: Optical Properties from Nanoparticle Dimers to Core-Satellite Systems. *Mater. Sci. Eng. C* **2007**, *27*, 1347–1350.

(46)  Hohenester, U.; Trügler, A. MNPBEM – A Matlab Toolbox for the Simulation of Plasmonic Nanoparticles. *Comput. Phys. Commun.* **2012**, *183*, 370–381.

(47)  Granqvist, C.; Hunderi, O. Retardation Effects on the Optical Properties of Ultrafine Particles. *Phys. Rev. B* **1977**, *16*, 1353–1358.

(48)  Granqvist, C. G.; Hunderi, O. Optical Properties of Ag-SiO2 Cermet Films: A Comparison of

Effective-Medium Theories. *Phys. Rev. B* **1978**, *18*, 2897–2906.

(49) Lazarides, A. A.; Lance Kelly, K.; Jensen, T. R.; Schatz, G. C. Optical Properties of Metal Nanoparticles and Nanoparticle Aggregates Important in Biosensors. *J. Mol. Struct. THEOCHEM* **2000**, *529*, 59–63.

(50) Corti, C. W.; Cotterill, P.; Fitzpatrick, G. A. The Evaluation of the Interparticle Spacing in Dispersion Alloys. *Int. Metall. Rev.* **1974**, *19*, 77–88.

(51) Oldenhuis, R. MATLAB Minimization Algorithm "Minimize" Version 1.7. Https://Github.Com/Rodyo/FEX-Minimize. 2017.

(52) Schwaab, M.; Biscaia, Jr., E. C.; Monteiro, J. L.; Pinto, J. C. Nonlinear Parameter Estimation through Particle Swarm Optimization. *Chem. Eng. Sci.* **2008**, *63*, 1542–1552.

(53) Demers, S. M. E.; Hsieh, L. J. H.; Shirazinejad, C. R.; Garcia, J. L. A.; Matthews, J. R.; Hafner, J. H. Ultraviolet Analysis of Gold Nanorod and Nanosphere Solutions. *J. Phys. Chem. C* **2017**, *121*, 5201–5207.

(54) Brown, K. R.; Walter, D. G.; Natan, M. J. Seeding of Colloidal Au Nanoparticle Solutions. 2. Improved Control of Particle Size and Shape. *Chem. Mater.* **2000**, *12*, 306–313.

(55) Ojea-Jiménez, I.; Campanera, J. M. Molecular Modeling of the Reduction Mechanism in the Citrate-Mediated Synthesis of Gold Nanoparticles. *J. Phys. Chem. C* **2012**, *116*, 23682–23691.

(56) Enustun, B. V.; Turkevich, J. Coagulation of Colloidal Gold. *J. Am. Chem. Soc.* **1963**, *85*, 3317–3328.

(57) Wang, J.; Boelens, H. F. M.; Thathagar, M. B.; Rothenberg, G. In Situ Spectroscopic Analysis of Nanocluster Formation. *ChemPhysChem* **2004**, *5*, 93–98.

(58) Vogt, F.; Booksh, K. Chemometrics. In *Kirk-Othmer Encyclopedia of Chemical Technology*; John

Wiley & Sons, Inc.: Hoboken, NJ, USA, 2004.

(59) Gross, S. Colloidal Dispersion of Gold Nanoparticles. In *Materials Syntheses: A Practical Guide*; Springer Vienna: Vienna, 2008; pp 155–161.

(60) Schindelin, J.; Arganda-Carreras, I.; Frise, E.; Kaynig, V.; Longair, M.; Pietzsch, T.; Preibisch, S.; Rueden, C.; Saalfeld, S.; Schmid, B.; *et al.* Fiji: An Open-Source Platform for Biological-Image Analysis. *Nat. Methods* **2012**, *9*, 676–682.

(61) Hühn, J.; Carrillo-Carrion, C.; Soliman, M. G.; Pfeiffer, C.; Valdeperez, D.; Masood, A.; Chakraborty, I.; Zhu, L.; Gallego, M.; Yue, Z.; et al. Selected Standard Protocols for the Synthesis, Phase Transfer, and Characterization of Inorganic Colloidal Nanoparticles. *Chem. Mater.* **2017**, *29*, 399–461.

# Supporting Information

# Kinetics and Mechanism of Metal Nanoparticle Growth *via* Optical Extinction Spectroscopy and Computational Modeling: The Curious Case of Colloidal Gold

M. Reza. Andalibi[*a,b], Alexander Wokaun[c], Paul Bowen[b], Andrea Testino[a]

[a] Paul Scherrer Institute (PSI), ENE-LBK-CPM, Villigen-PSI, Switzerland.

[b] École polytechnique fédérale de Lausanne (EPFL), STI-IMX-LMC, Lausanne, Switzerland.

[c] Paul Scherrer Institute (PSI), Energy and Environment Division, Villigen-PSI, Switzerland.

[*] Corresponding author: reza.andalibi@psi.ch

# Contents

## Supporting Section 1. Additional details on kinetic experiments

Among several seed synthesis protocols investigated, those synthesized in buffers with 15% citric acid provide a very good model system for kinetic studies because of the slow enough temporal evolution providing the chance to collect enough data with a reasonable time resolution of 10 seconds. On the other hand, the process is not excessively slowed down such that the particles interact for too long and aggregate (*i.e.*, form chemically cemented secondary particles[1]) in an uncontrolled manner. This situation, which gives rise to large fractal-like structures, has been thoroughly studied by Karg *et al.*[2] The latter happens, for instance, for seeds synthesized using 100% $Na_3Cit$ and grown at T < 60°C. Both the higher pH and the lower temperature in this sample hindered the reduction of Au(III), the presence of which favors attractive interactions between the particles.[3,4] Throughout the main text, we have primarily focused on the results for seeds synthesized with 15% citric acid and grown at 70°C (named as 15:85-70°C). Results for the other experiments (including those for the same seeds grown at 60 and 65°C) are either mechanistically similar or simpler and we have discussed them in Supporting Sections 4-6.

## Supporting Section 2. Overall modeling workflow and regression to experimental spectra

We started the modeling procedure by converting the raw data, *i.e.*, dark, reference, and temporal sample intensities, to extinction spectra. According to Parak *et al.* absorption by small noninteracting GNPs converges to zero at wavelength range 800-1200 nm.[5] Therefore, the possibly drifted spectra of the seeds and grown particles were background corrected by shifting them to 0.01 and 0.03 absorption units at λ = 800 nm, respectively (these values are estimated from preliminary extinction calculations using the TEM characterized size and aspect ratio). The intermediate spectra were background-corrected using the shift factors estimated by linear interpolation between that of the seeds and the grown GNP. Several spectra collected on seeds before the injection of Au(III) solution were averaged to yield the seed data with improved signal to noise ratio. Background-corrected spectra were then fitted using smoothing splines with automatically selected smoothing parameters using the MATLAB's curve fitting app and the data were saved with a 1 nm resolution.

We then started the regression procedure by calibrating the parameter $\eta$ using the volume-weighted mean particle radii of seeds and grown GNP estimated from TEM micrographs (Supporting Section 3).[6] The regression of these two experimental spectra was done by fixing the mean radii and globally fitting $\eta$ along with $\bar{\beta}$ for both seeds and grown GNPs using the Gans formalism in the range 400-800 nm. The value of $\eta$ was constrained in the range [0.2,4].[6–8] The fitted $\eta$ values were used to derive a linear relation $\eta = \eta_0 + \eta_1 \bar{r}_{GNP}$ which was then used for all the intermediate spectra.

In the next step, we fitted all the intermediate spectra using the Gans theory with only two parameters ($\bar{r}_{GNP}$ and $\bar{\beta}$). The same was also done using an overarching framework in which we considered possible contributions from both noninteracting GNPs as well as SCs composed of electromagnetically interacting particles (namely, GEM model ≡ Gans theory for noninteracting GNPs + retarded EMT/Mie for interacting GNPs). Assuming similar average size and aspect ratio for GNPs both inside and outside the SCs, we have five parameters to regress: $\bar{r}_{GNP}$, $\bar{\beta}$, $x_{SC}$ (the number fraction of SCs out of all the scatterers, *i.e.*, SCs + noninteracting GNPs), $f$, and $\bar{r}_{SC}$ (the average radius of SCs). All the spectra were regressed searching over a wide space for $\bar{\beta}$ and the EMT-specific parameters ($x_{SC}$, $f$, and $\bar{r}_{SC}$) while constraining $\bar{r}_{GNP}$ to within 10% (50% for the second spectrum) of the previous optimized value (obtained from Gans theory for the first spectrum and GEM for the rest of the intermediate spectra):

$$\max(0.9\bar{r}_{GNP,j-1}, 2\ nm) \leq \bar{r}_{GNP} \leq \min(1.1\bar{r}_{GNP,j-1}, 15\ nm)$$

$$1 \leq \bar{\beta} \leq 3$$

$$-8 \leq \log_{10}(x_{SC}) \leq 0 \quad \text{(Supp. Eq. 1)}$$

$$-3 \leq \log_{10}(f) \leq \log_{10}(0.4)$$

$$25 \leq \bar{r}_{SC} \leq 250\ nm$$

where $j$ represents the index of the current spectrum being fitted. There are three physical constraints to be considered when performing this regression:

$$N_{GNP/SC} = f\left(\frac{\bar{r}_{SC}}{\bar{r}_{GNP}}\right)^3 \geq 1000 \quad \text{(Supp. Eq. 2)}$$

$$\frac{f}{\frac{4}{3}\pi\bar{r}_{GNP}^3} > (1-x_{SC})C_t + x_{SC}C_t f\left(\frac{\bar{r}_{SC}}{\bar{r}_{GNP}}\right)^3 \quad \text{(Supp. Eq. 3)}$$

$$\frac{4}{3}\pi C_t [x_{SC}\bar{r}_{SC}^3 + (1-x_{SC})\bar{r}_{GNP}^3] \leq \frac{\pi}{\sqrt{18}} \quad \text{(Supp. Eq. 4)}$$

with $N_{GNP/SC}$ denoting the average number of GNPs interacting inside each SC and $C_t$ being the total number concentration of the scatterers (noninteracting GNPs + SCs).

The first constraint (Supp. Eq. 2) specifies that there have to be at least 1000 particles for the interaction to be meaningful. This is a somewhat arbitrary lower bound considering the loosely defined stipulation of *"large"* number of particles ($N_{GNP/SC} \gg 1$) necessary for the mean-field averaging in the EMT derivation to be meaningful.[9] This number was selected based on our preliminary regressions using a lower bound of two (the smallest physically meaningful value) which revealed that generally during the interaction period more than 1000 GNPs are involved inside each SC. Nevertheless, this would not introduce any problems because smaller $N_{GNP/SC}$ values happen in cases where the particle ensemble can adequately be described by a noninteracting Gans formalism. In these conditions, the size and aspect ratio fitted by the GEM model approaches that obtained by the Gans model (see Supporting Section 6).

The second constraint (Supp. Eq. 3), coming directly from our definition of superclusters as effective media for the interacting particles, requires that the local GNP number concentration inside an SC be higher than the bulk (*i.e.*, averaged over all the suspension) concentration.

The third constraint (Supp. Eq. 4) finally states that the suspension volume cannot be filled with GNPs and SCs beyond the close sphere packing density. This constraint is rarely violated during an optimization problem, as it requires an extremely high concentration of scatterers.

Being only dependent on the optimization variables, Supp. Eq. 2 can be introduced as a nonlinear constraint in the optimization scheme. On the contrary, the other two constraints have the parameter $C_t$, which is calculated by scaling the computed extinction cross section with the experimental spectrum in the vicinity of the surface-plasmon resonance band ($\lambda_{spr} \pm 1\ nm$). Therefore, to exert the constraints in Supp. Eqs. 3 and 4, the objective function is written so that it returns infinity ("Inf") in case either of them is violated.

Having the constraints in place, a globalized bounded Nelder-Mead optimization,[10] run from 20 randomly generated start points (complying with the first constraint), was used to find the set of model parameters that fit the intermediate spectra optimally. Since the search space for the parameters $x_{SC}$ and $f$ span several orders of magnitude, the random sample was generated on their $\log_{10}$ transformation. Moreover, in order to bring all the variables in the order of unity and make the optimization more robust the $\log_{10}$-transformation of the three EMT parameters ($x_{SC}$, $f$, and $\bar{r}_{SC}$) were optimized and all the parameters (including the GNP parameters, $\bar{r}_{GNP}$ and $\bar{\beta}$) were centered and scaled in the range [-1,1].[11] The objective function ($fval \equiv \chi^2 = \sum_i (A_{experimental} - A_{caluclated})^2$) was also scaled to be in the order of unity close to the optimal point. Throughout the optimization, we saved the function evaluations whose $fval$ were within the 95% confidence bounds of the optimized point. This information was then used in order to estimate the corresponding confidence regions on the optimized parameters as well as the other ancillary model outputs (*e.g.*, $N_{GNP/SC}$). Details of this heuristic approach to the construction of 95% confidence bounds are presented in the work of Schwaab *et al.*[12]

When all the spectra are fitted using both the noninteracting GNP model as well as the GEM model, the former is selected as the superior one whenever the following conditions hold:

$$\frac{|\bar{r}_{GNP,GEM} - \bar{r}_{GNP,Gans}|}{\bar{r}_{GNP,GEM}} \leq 0.01$$

$$\frac{|\bar{\beta}_{GEM} - \bar{\beta}_{Gans}|}{\bar{\beta}_{GEM}} \leq 0.01$$

(Supp. Eq. 5)

This condition coincides with % GNP in SCs (obtained from the GEM regressions) approaching zero. This practically corresponds to a situation where there are no tangible electromagnetic interactions between the particles. For periods during which relatively weak interaction effects are present, typically at the beginning and close to the end of the seeded growth process, regressions are repeated over a narrower search space that is compatible with the intermediate data points (*i.e.*, when the interaction effects are typically stronger). This is done either by setting a more stringent bound (*e.g.*, setting $lb_{\log_{10}(f)} = 0.1$) or by constraining the parameter(s) to a neighborhood of the previous optimal output (*e.g.*, $\max(0.8 \times \bar{r}_{SC,j-1}, 25\ nm) \leq \bar{r}_{SC} \leq \min(1.2 \times \bar{r}_{SC,j-1}, 250\ nm)$). This new regression is retained as long as it does not deteriorate the quality of the fit substantially. This would help to obtain more reliable estimates for various parameters, especially the EMT variables.

One of the ancillary outputs of the current theoretical framework is the volumetric mass density of nanoparticles. Estimating the concentration of GNPs (in mM Au(0)) from the linear correlation with absorbance at $\lambda = 400\ nm$ (Abs$_{400}$),[13] one can calculate the density of nanoparticles by comparing the previous value with that computed using the bulk gold density (19.3 g.cm$^{-3}$), the regressed $\bar{r}_{GNP}$, and the number concentration. Practically, the seeds and the grown particles have an apparent mass density similar to the bulk value (bear in mind a ~ 5% uncertainty in the concentration from Abs$_{400}$,[13] as well as possible uncertainties in the sizes and concentrations of the initial and final GNPs).

As discussed in the main text, in our theoretical framework we explicitly distinguish between agglomeration/aggregation and electromagnetic interactions between the particles. To this end, closely spaced particles are considered to be in electromagnetic interaction (and not agglomerated/aggregated) unless they touch each other. This is straightforwardly implemented by setting an upper bound of 0.4 for the filling factor, which is equivalent to a center-to-center distance of $\sim 2.5\bar{r}_{GNP}$ (the minimum filling factor at which particles may be touching each other corresponds to random loose packing of spheres which has a filling factor of 0.55[14]).

## Supporting Section 3. TEM images and particle size distributions

In this section, we have summarized the number-based particles size distributions (PSD) and the representative TEM micrographs for different seed and grown GNP suspensions. For each PSD the normal distribution with the same mean and standard deviation is plotted as well. Note, however, that the volume-weighted mean particle radii are used throughout the spectroscopic analyses as the extinction signal is weighted by particle volumes.[6]

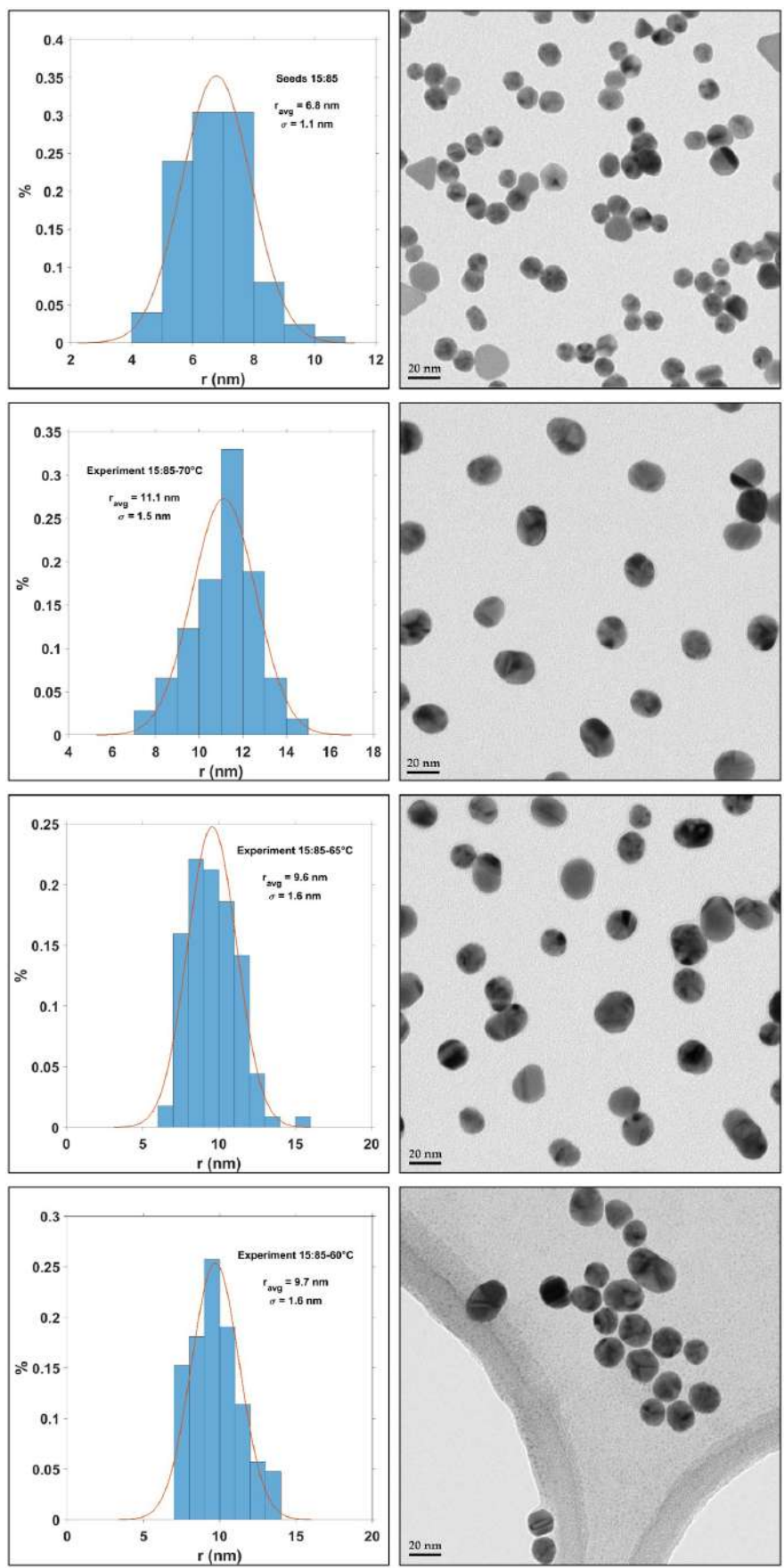

**Supporting Fig. 1.** Particle size distributions (left panel) for the seeds and grown GNPs prepared with a 15%:85% citric acid/trisodium citrate buffer along with their representative TEM images (right panel).

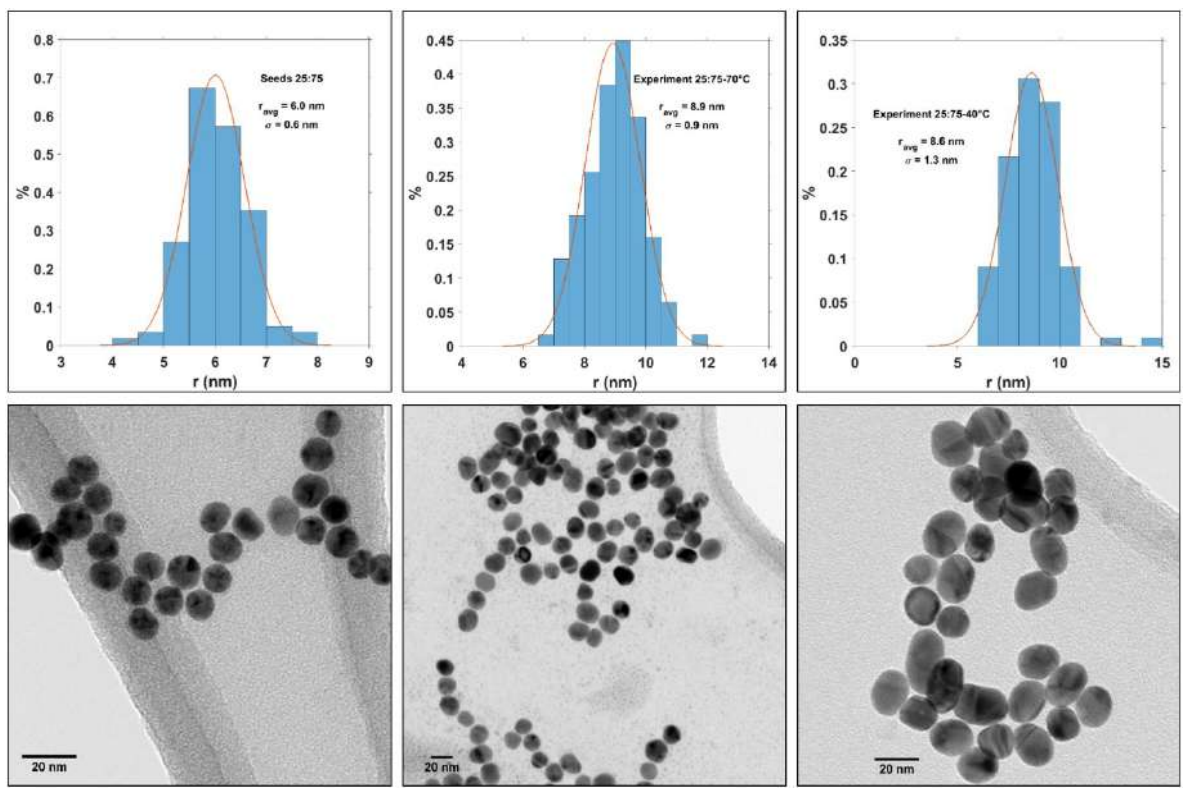

**Supporting Fig. 2.** Particle size distributions (top panel) for the seeds and grown GNPs prepared with a 25%:75% citric acid/trisodium citrate buffer along with their representative TEM images (bottom panel).

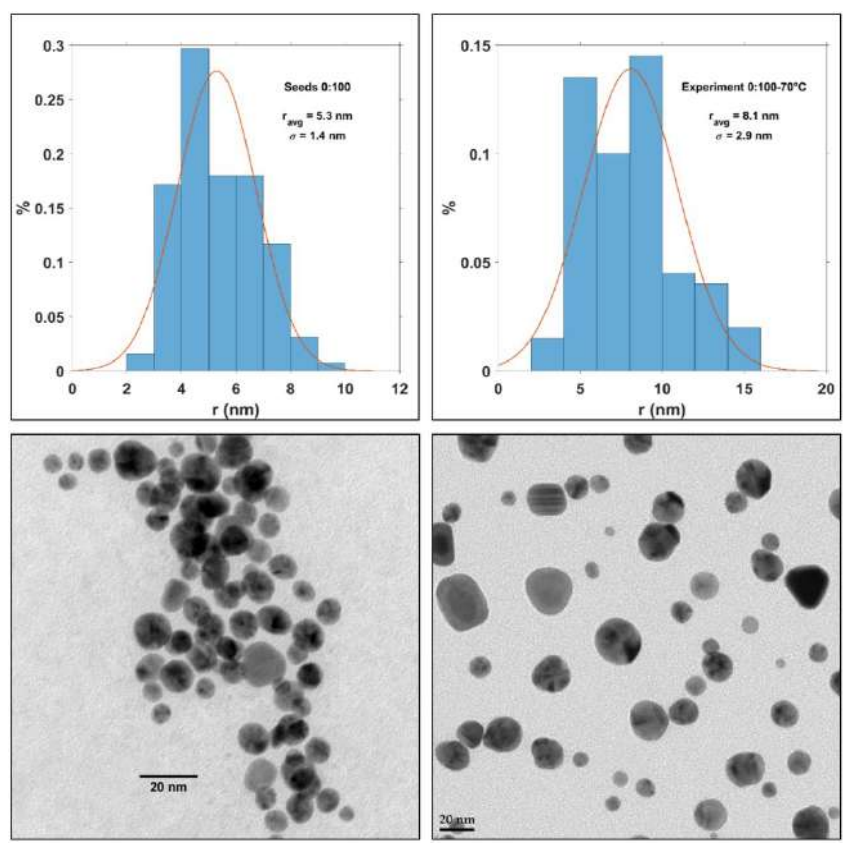

**Supporting Fig. 3.** Particle size distributions (top panel) for the seeds and grown GNPs prepared with 100% trisodium citrate along with their representative TEM images (bottom panel).

# Supporting Section 4. Regression plots and fit qualities

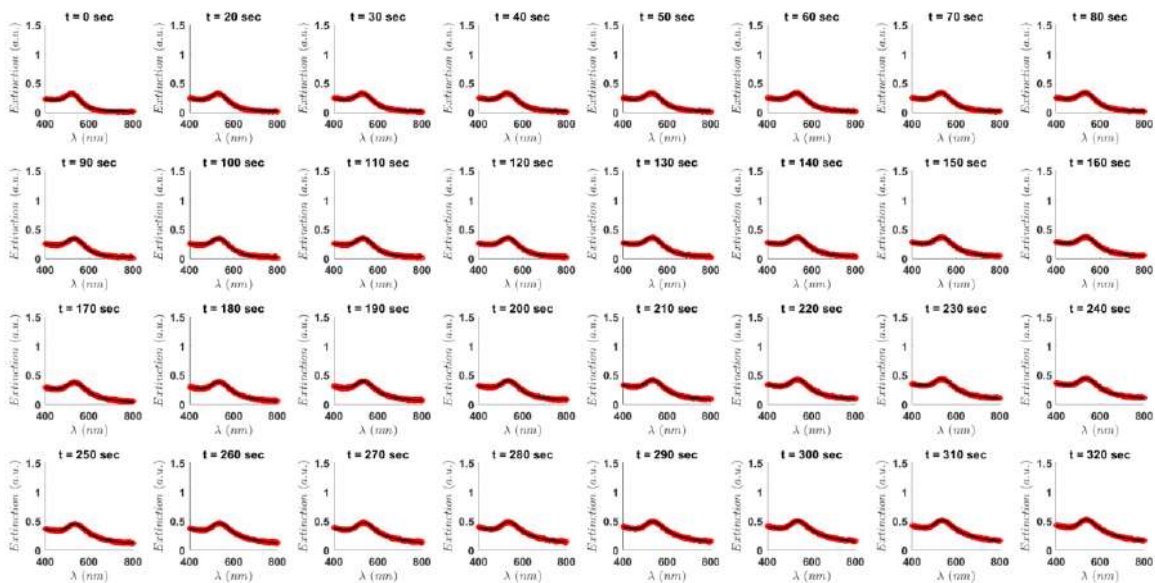

**Supporting Fig. 4.** Experimental spectra (red curves) and their corresponding theoretical fits (black lines) for the 15:85-70°C experiment (t = 0-320 s).

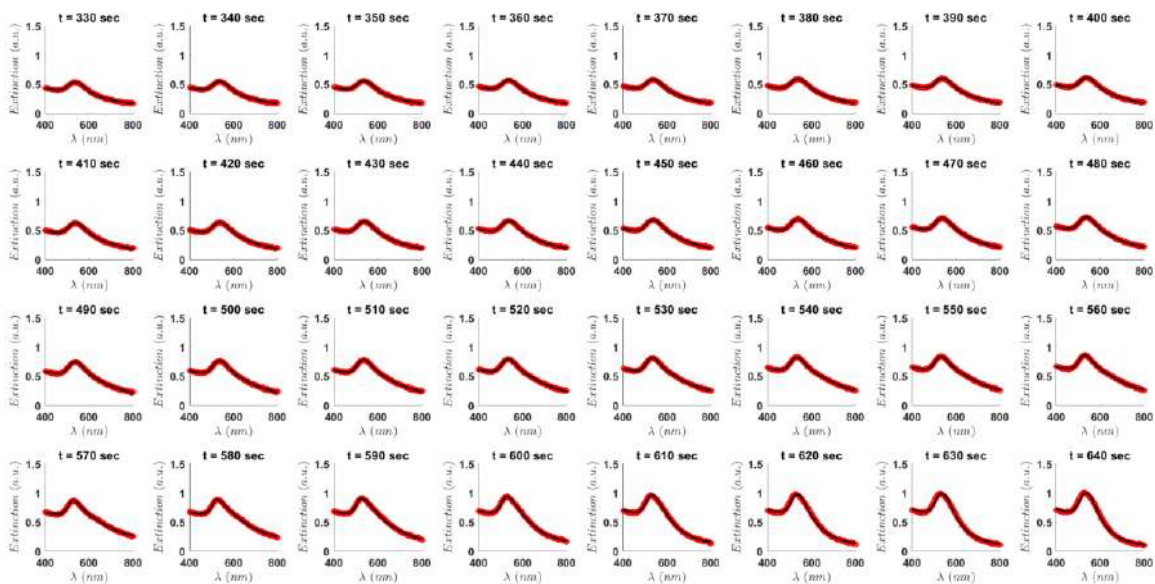

**Supporting Fig. 5.** Experimental spectra (red curves) and their corresponding theoretical fits (black lines) for the 15:85-70°C experiment (t = 330-640 s).

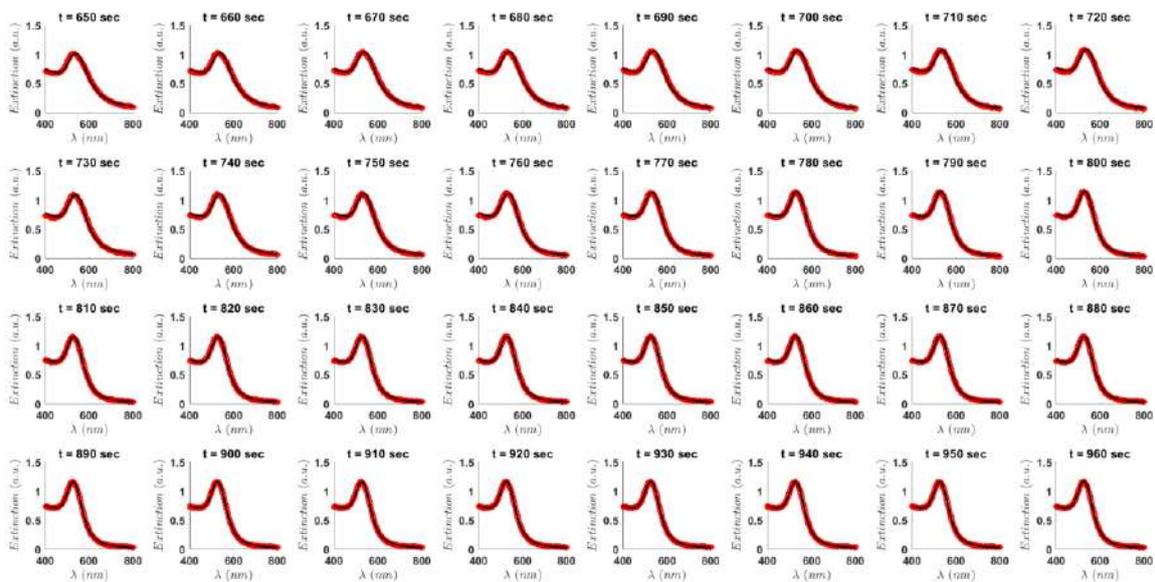

**Supporting Fig. 6.** Experimental spectra (red curves) and their corresponding theoretical fits (black lines) for the 15:85-70°C experiment (t = 650-960 s).

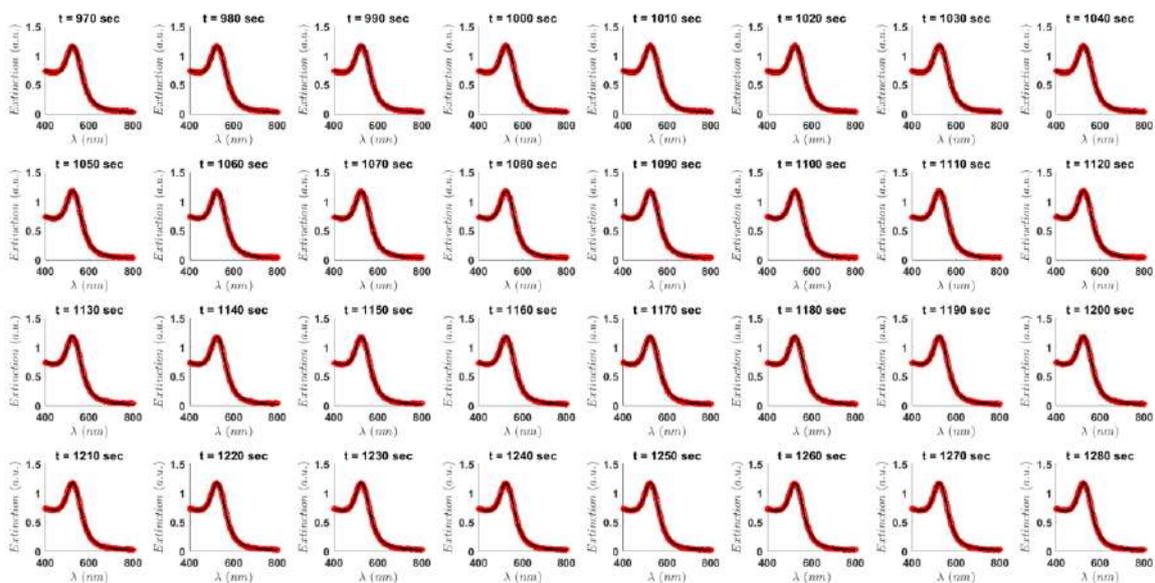

**Supporting Fig. 7.** Experimental spectra (red curves) and their corresponding theoretical fits (black lines) for the 15:85-70°C experiment (t = 970-1280 s).

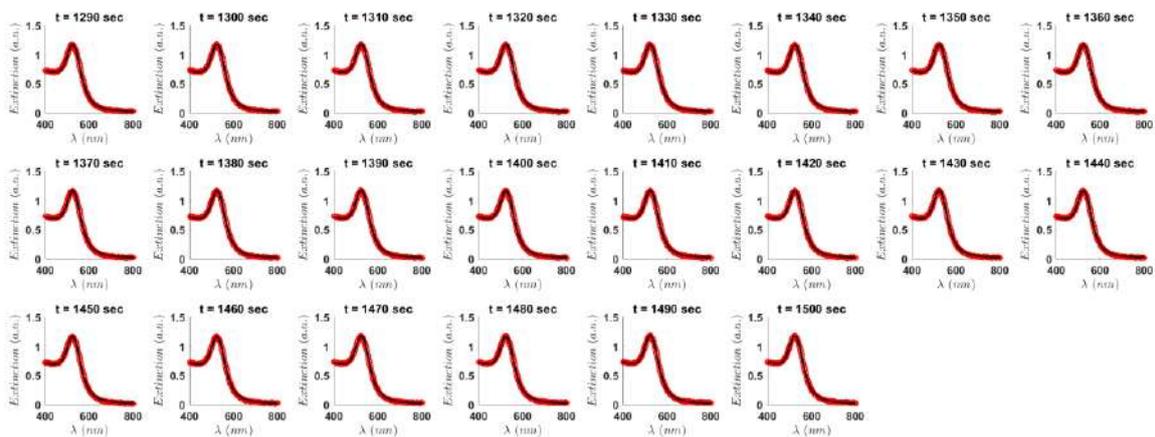

**Supporting Fig. 8.** Experimental spectra (red curves) and their corresponding theoretical fits (black lines) for the 15:85-70°C experiment (t = 1290-1500 s).

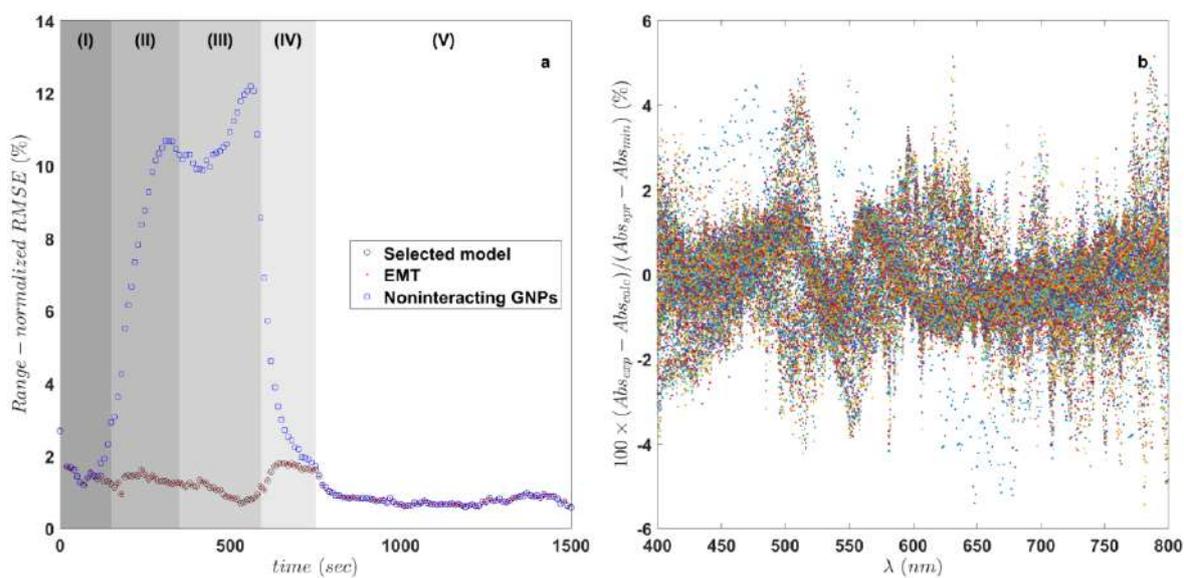

**Supporting Fig. 9.** Range-normalized root mean square error (NRMSE; **a**) and range-normalized errors for individual wavelengths and temporal points (**b**) for the fits on spectra in the 15:85-70°C experiment.

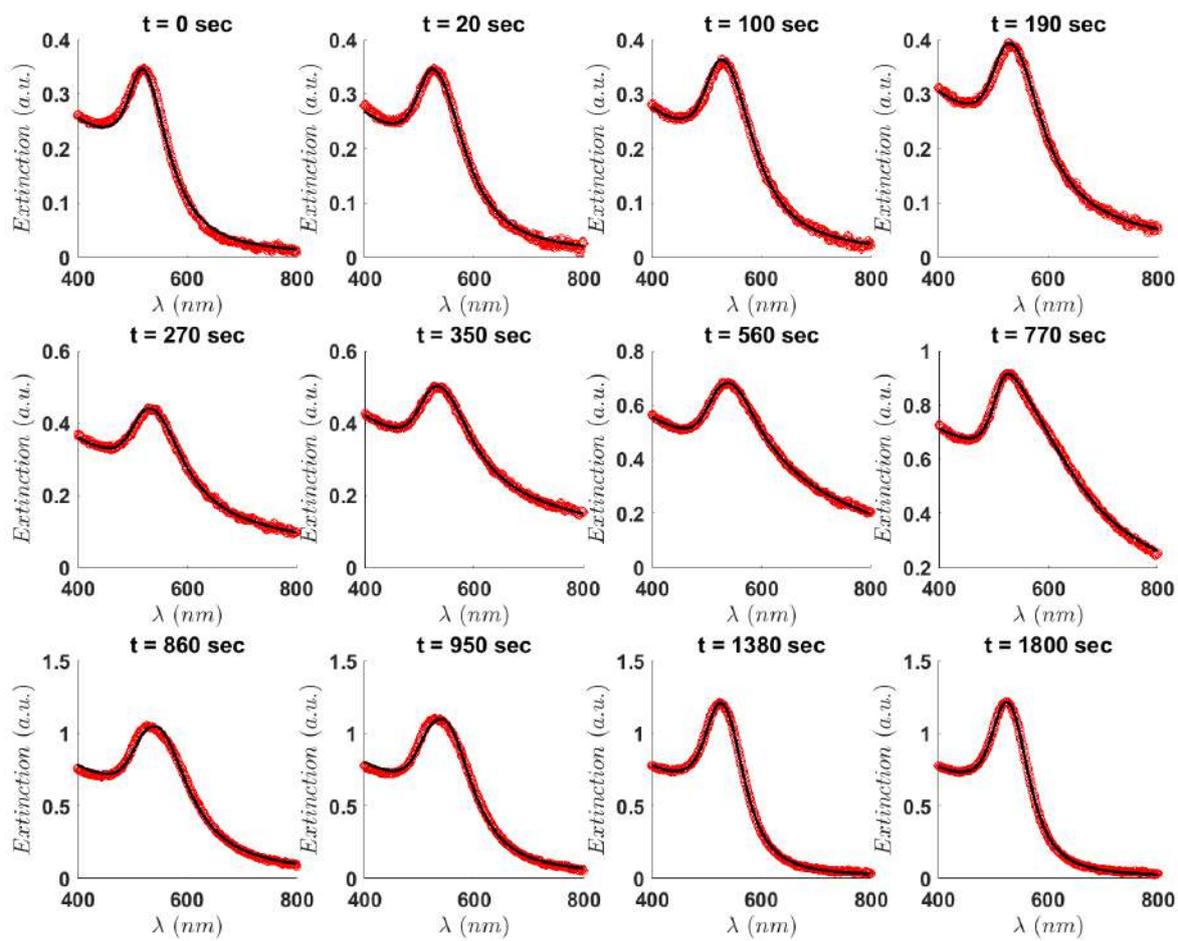

**Supporting Fig. 10.** Representative experimental UV-Vis spectra (red curves composed of several open circles) and their corresponding theoretical fits (black lines) for the 15:85-65°C experiment.

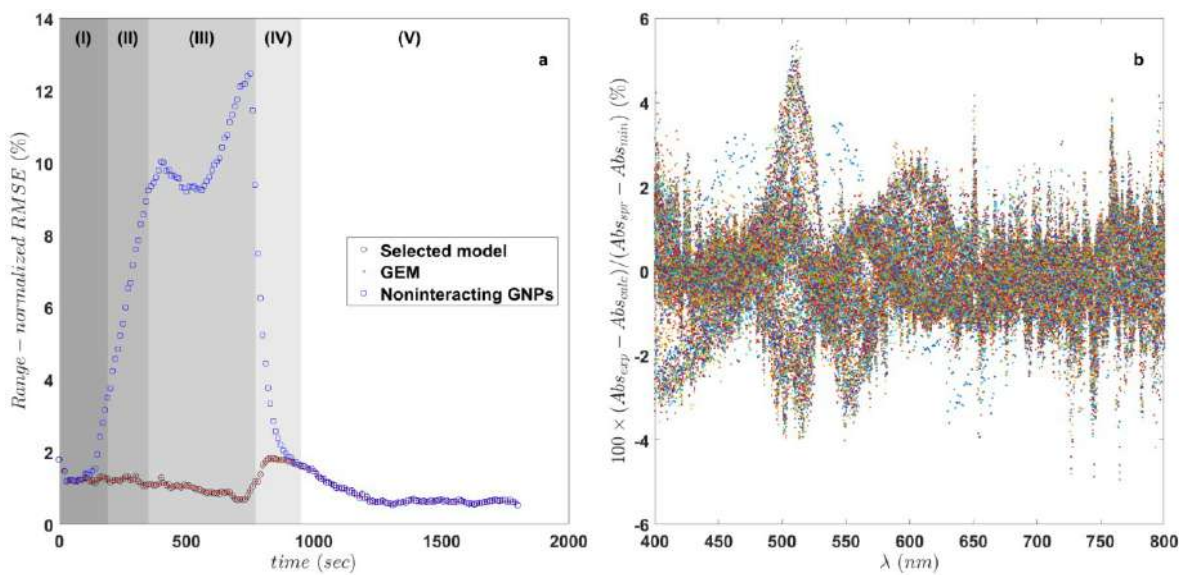

**Supporting Fig. 11.** Range-normalized root mean square error (NRMSE; **a**) and range-normalized errors for individual wavelengths and temporal points (**b**) for the fits on spectra in the 15:85-65°C experiment.

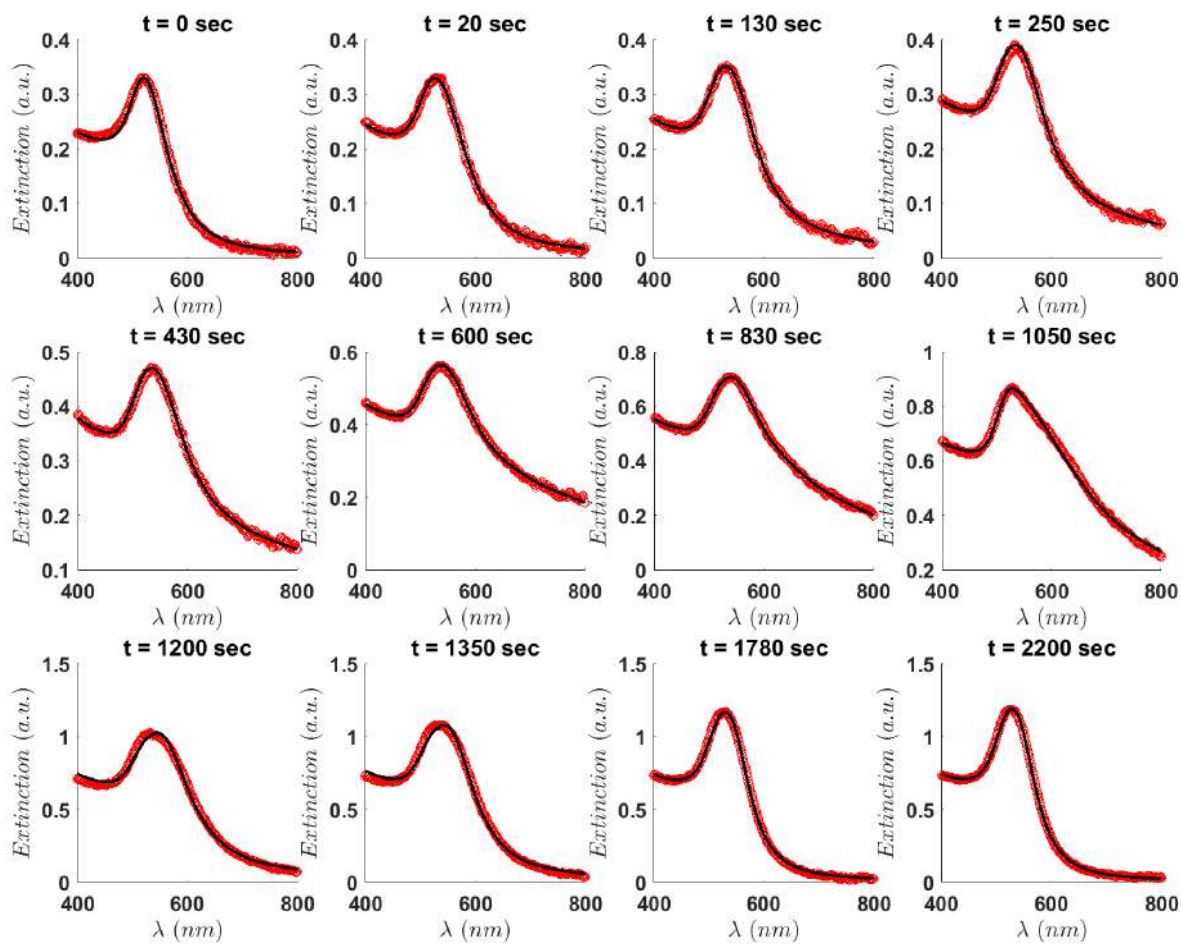

**Supporting Fig. 12.** Representative experimental UV-Vis spectra (red curves composed of several open circles) and their corresponding theoretical fits (black lines) for the 15:85-60°C experiment.

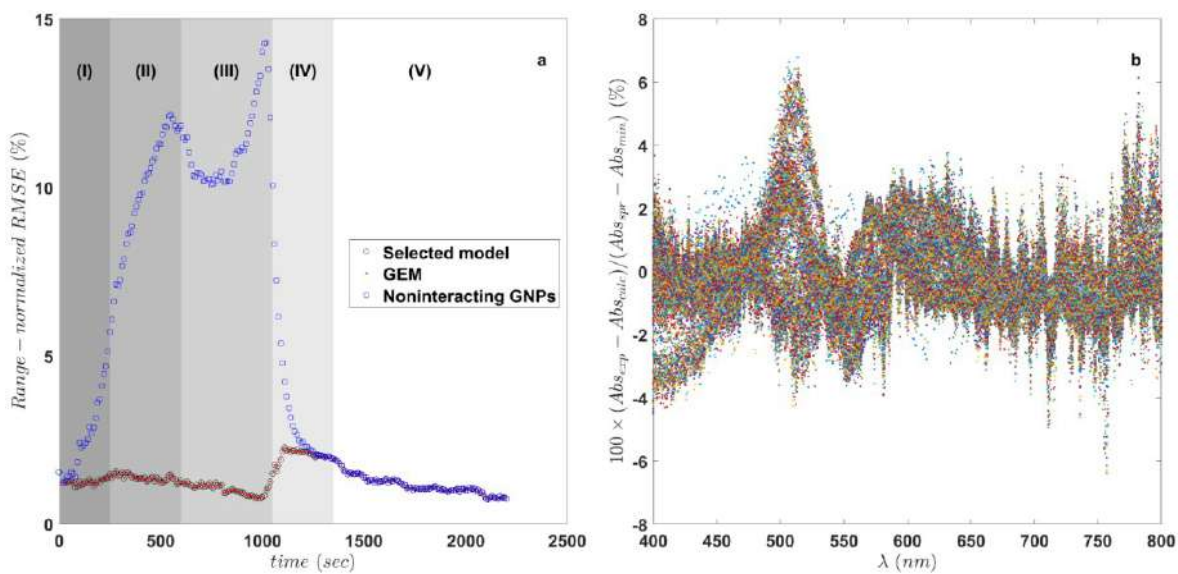

**Supporting Fig. 13.** Range-normalized root mean square error (NRMSE; **a**) and range-normalized errors for individual wavelengths and temporal points (**b**) for the fits on spectra in the 15:85-60°C experiment.

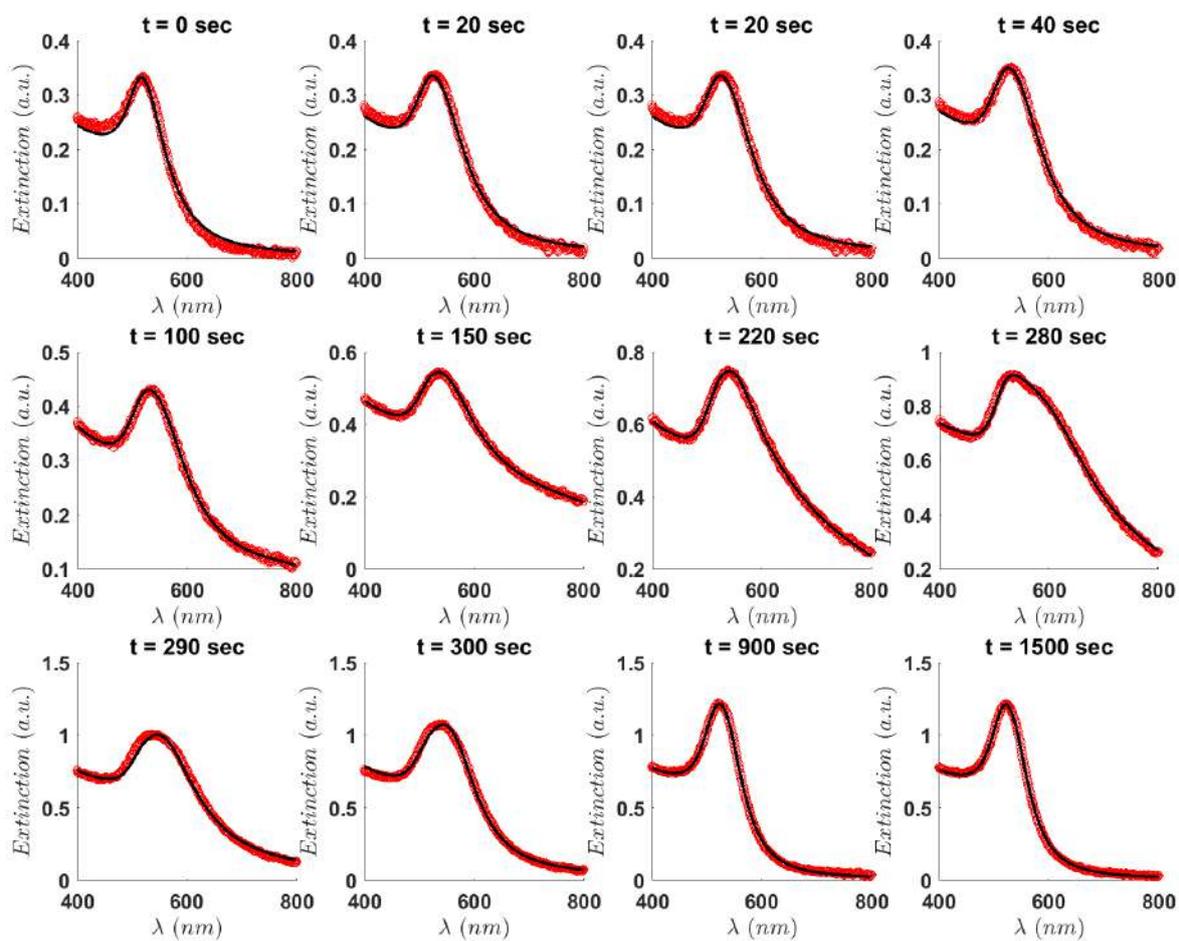

**Supporting Fig. 14.** Representative experimental UV-Vis spectra (red curves composed of several open circles) and their corresponding theoretical fits (black lines) for the 25:75-70°C experiment.

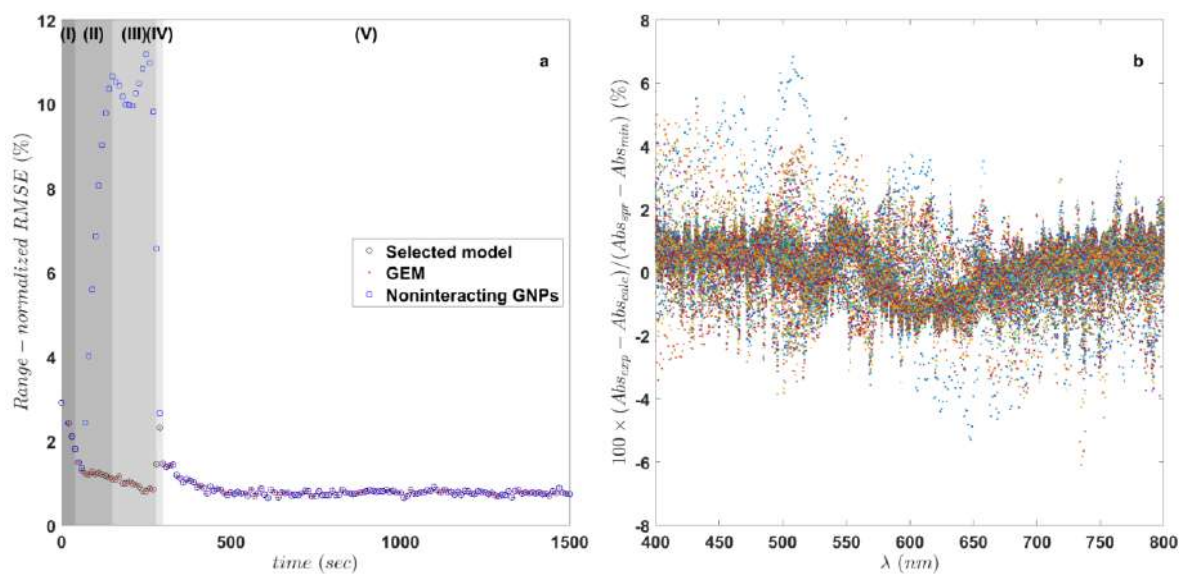

**Supporting Fig. 15.** Range-normalized root mean square error (NRMSE; **a**) and range-normalized errors for individual wavelengths and temporal points (**b**) for the fits on spectra in the 25:75-70°C experiment.

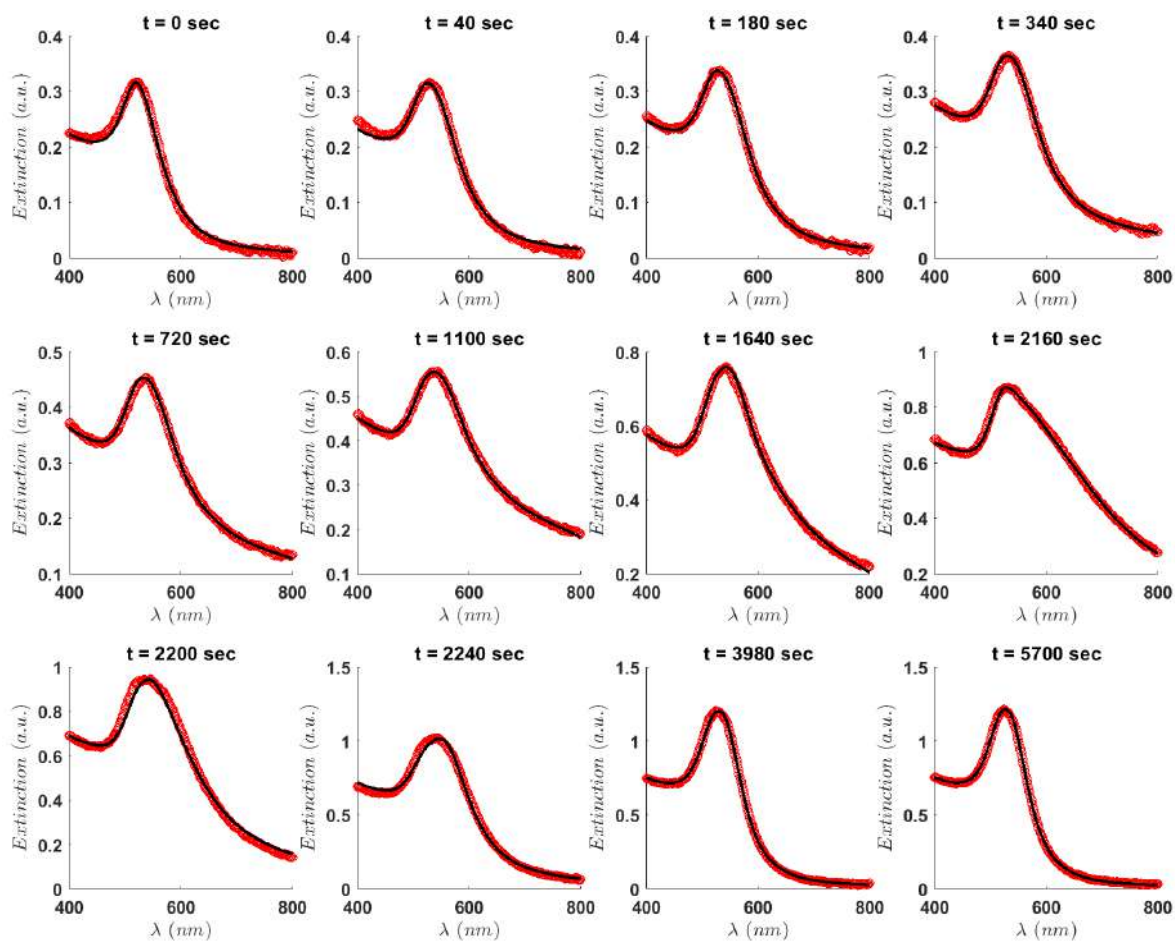

**Supporting Fig. 16.** Representative experimental UV-Vis spectra (red curves composed of several open circles) and their corresponding theoretical fits (black lines) for the 25:75-40°C experiment.

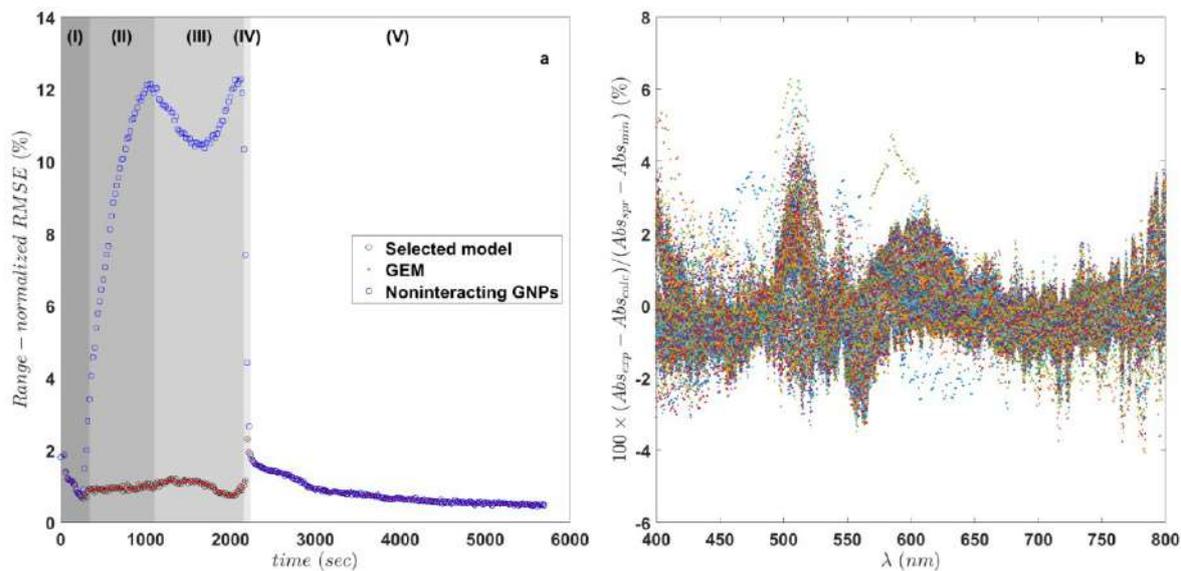

**Supporting Fig. 17.** Range-normalized root mean square error (NRMSE; **a**) and range-normalized errors for individual wavelengths and temporal points (**b**) for the fits on spectra in the 25:75-40°C experiment.

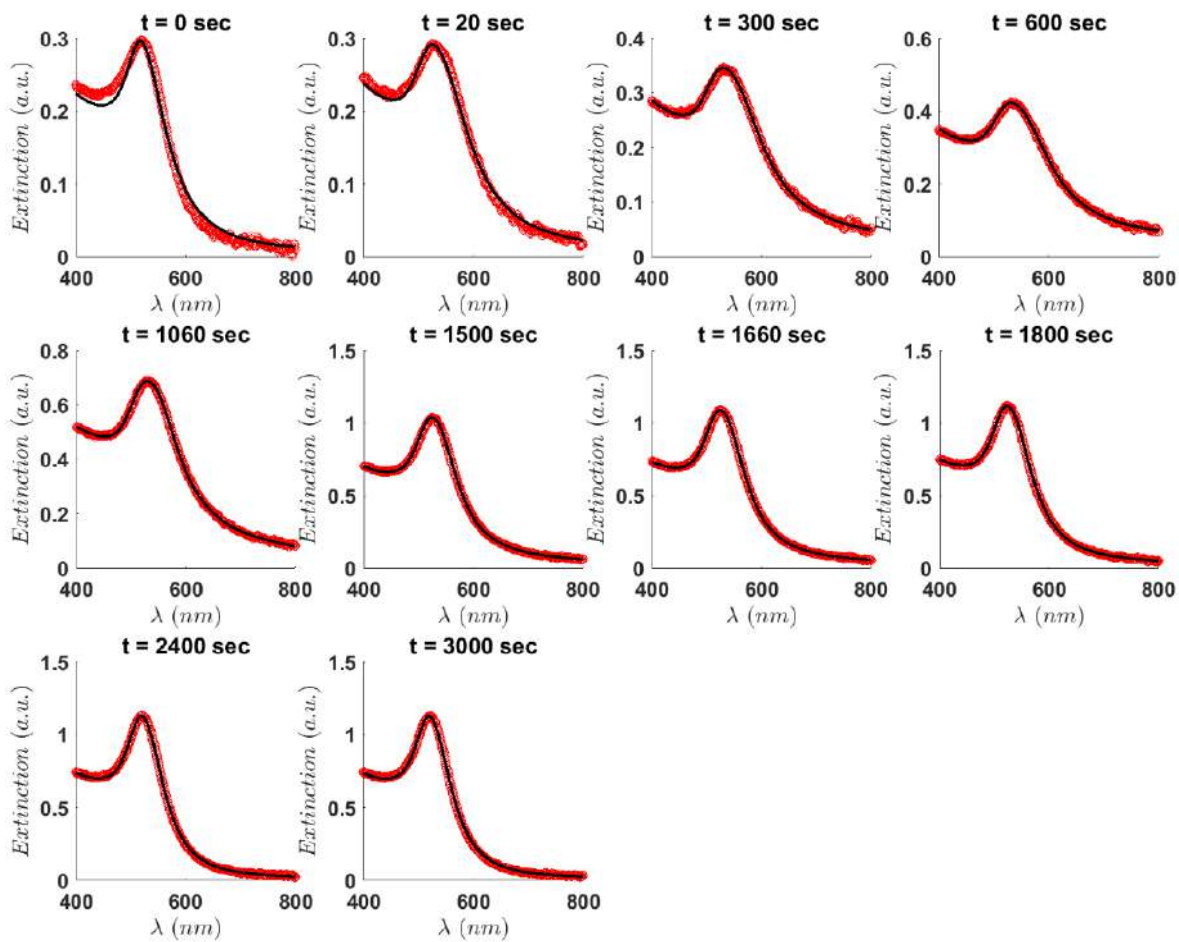

**Supporting Fig. 18.** Representative experimental UV-Vis spectra (red curves composed of several open circles) and their corresponding theoretical fits (black lines) for the 0:100-70°C experiment.

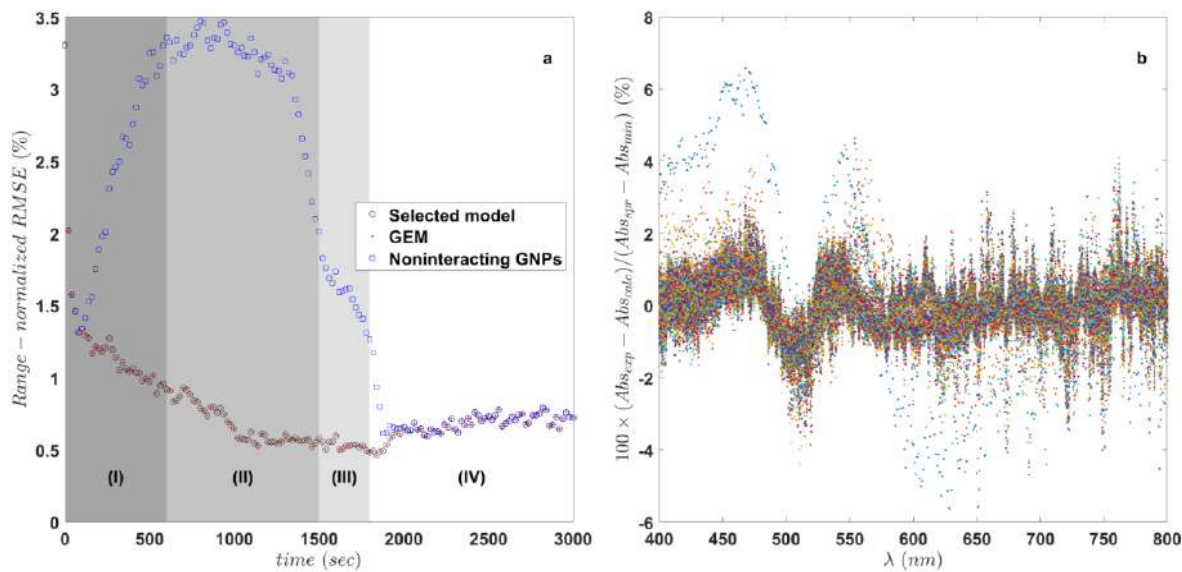

**Supporting Fig. 19.** Range-normalized root mean square error (NRMSE; **a**) and range-normalized errors for individual wavelengths and temporal points (**b**) for the fits on spectra in the 0:100-70°C experiment.

## Supporting Section 5. Regression outputs for experiments other than 15:85-70°C

In this section, we have presented the model outputs for all the datasets not discussed in the main text. The major experimental conditions varied from one set to another are the pH of the seed suspension, tuned by the citric acid/trisodium citrate ratio, and the temperature at which the growth process happened. Both lower pH and higher temperature favor a more facile reduction of Au(III) to Au(I) which in turn accelerates the whole process. In all the experiments performed with buffer (that is, all the experiments but 0:100-70°C) the overall mechanism of the seeded growth process is very similar to that of the experiment 15:85-70°C discussed in the main text. The essential difference from one experiment to another is the time scale corresponding to each mechanistic region (Supporting Figs. 20-27). On the other hand, in the experiment conducted with 100% trisodium citrate ($Na_3Cit$) the mechanism appears to be slightly different (Supporting Figs. 28 and 29). In this experiment, the electromagnetic interaction effects are generally much weaker than the rest of the experiments (Supporting Fig. 29b). The significantly smaller enhancement of the extinction in the range 600-800 nm (Supporting Fig. 29d), as well as the invariably lower PWHM values (almost half the values in other experiments, close to the maximum; Supporting Fig. 29e) support this hypothesis.

For the experiment 0:100-70°C, the overall seeded growth process can be divided into four mechanistic regions (grey-to-white shaded regions in Supporting Figures 28 and 29 denoted by Roman numerals). As soon as the Au(III) solution is injected into the seed suspension, a rapid jump happens in Eh (Supporting Fig. 29f) denoting the formation of Au(I) which couples with Au(0) to control the potential.[15,16] Following that, there is an immediate drop in Eh indicating the conversion of Au(I) to Au(0). This provides the necessary monomers for a nucleation event, both primary and secondary, at the beginning of region (I) (0-600 s). The sudden drop in the average GNP radius and the leap in number concentration notify such an event. At the same time, similar to the experiment 15:85-70°C, true catalytic secondary nucleation gives rise to relatively anisotropic scatterers (Supporting Fig. 28b). In the presence of Au(I) and enhanced number concentration,[3,4,16] the colloidal destabilization of nanoparticles bring them closer to each other inducing electromagnetic interaction effects (Supporting Fig. 28c-e; Supporting Fig. 29b,c; Supporting Fig. 29d,e enhanced extinction in the wavelength range 600-800 nm and peak broadening). Here, a notable difference is that the average GNP density starts to drop below the bulk value early on (Supporting Fig. 29a). Furthermore, the molar concentration of Au(0) exhibits no quiescent period in its temporal evolution. This means that in region (I) a continuous but slow primary nucleation competes with secondary nucleation and molecular growth. As a result, the number concentration keeps increasing slowly while the former competition preserves a more or less constant particle size (Supporting Fig. 28a,f). In the meantime, secondary nucleation keeps increasing the mean aspect ratio of the GNP scatterers (Supporting Fig. 28b). From what we described here, region (I) in this experiment resembles a combination of regions (I) and (II) of the typical mechanism presented in the main text.

In region (II) (600-1500 s), following a deceleration in nucleation, colloidal destabilization induces an agglomeration event. Consequently, the mean particle size increases while the GNP number concentration goes down. Slow but continuous provision of Au(I) (Supporting Fig. 29a,f) allows for the densification of these agglomerates into aggregates (Supporting Fig. 29a). Here, the agglomeration being slower than in the typical mechanism, the rate of decrease in $\bar{\beta}$ is smaller. Close to the end of this region, simultaneous with a quick drop in Eh (Supporting Fig. 29f), the decrease of $\bar{\beta}$ accelerates. The could be due to an accelerated coalescence of the agglomerates into more compact morphologies.[17]

In region (III) (1500-1800 s), the previous aggregation process decelerates (Supporting Fig. 28a,f) while slow molecular growth continues to consume ionic gold (Supporting Fig. 29a). By the end of this region, the suspension is practically depleted from ionic Au (Supporting Fig. 29a).

Finally, with no more ionic gold available in region (IV) (1800-3000 s) Ostwald ripening/interparticle ripening induces further enlargement in the average size and reduction in the average aspect ratio (Supporting Fig. 28a,b and Supporting Fig. 29f).[17–19] During this period, as the electromagnetic interaction effects disappear, the two optical theories (GEM and Gans) approach to similar outputs (Supporting Fig. 35).

As one can see from the discussion above, the overall mechanism during the experiment 0:100-70°C is very similar to the general picture explained in the main text. In this respect, the decelerated rate of Au(I) production, due to the higher pH levels, renders the process more gradual and continuous when compared to the experiment 15:85-70°C. Furthermore, the slower Au(I) formation and the higher pH level reduce the electromagnetic interaction effects which consequently introduces much less deviation from the predictions of the Gans theory (compare Supporting Fig. 35 with Supporting Fig. 30-34). Additional support for the proposed mechanism is presented in Supporting Section 7 (using principal component analysis).

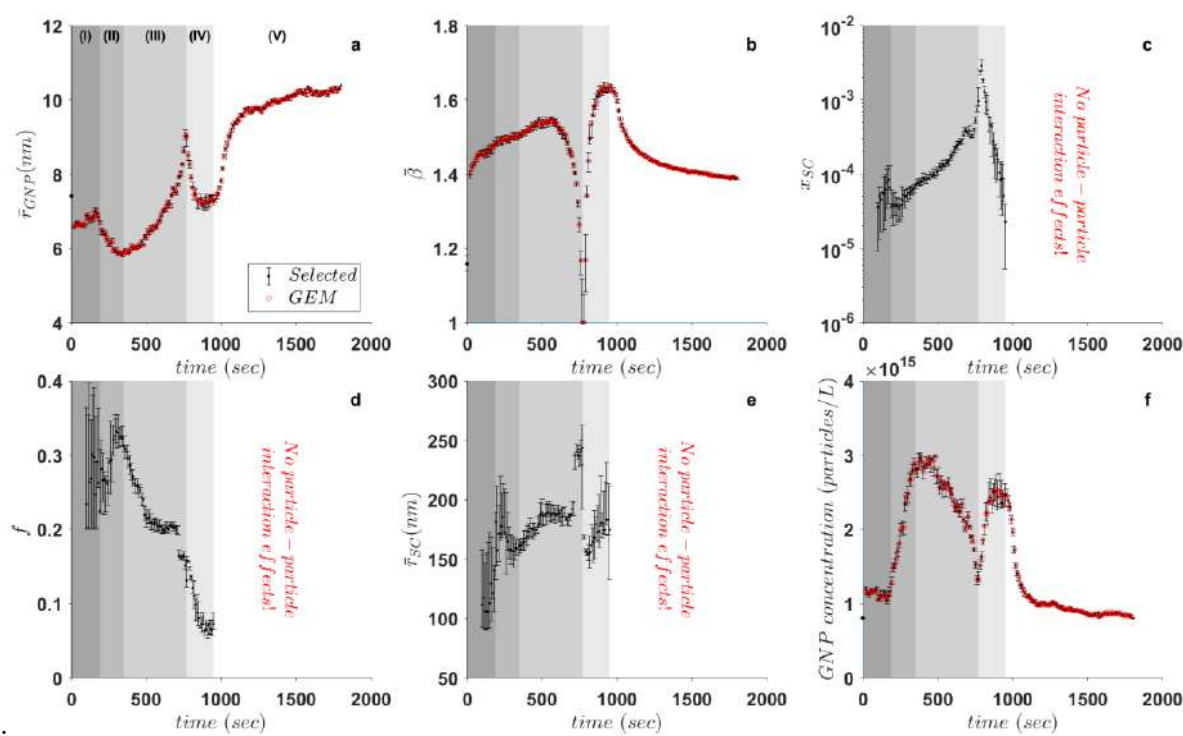

**Supporting Fig. 20.** Kinetics of gold nanoparticle growth (15:85-65°C experiment) from *in situ* UV-Vis spectroscopy and model regression. (a-e) Regressed physical parameters (and their corresponding 95% confidence bounds found by a heuristic search method) including mean particle radius (a), mean particle aspect ratio (b), number fraction of liquid-like superclusters (c), volume fraction of particles in superclusters (d), and mean supercluster radius (e). f, Temporal particle number concentration (along with the respective 95% confidence bounds[46]) obtained from scaling the calculated extinction cross sections with the experimental spectra. Legends are common in all the plots and the grey-to-white shaded regions denote various mechanistic steps (denoted by Roman numerals in the first plot).

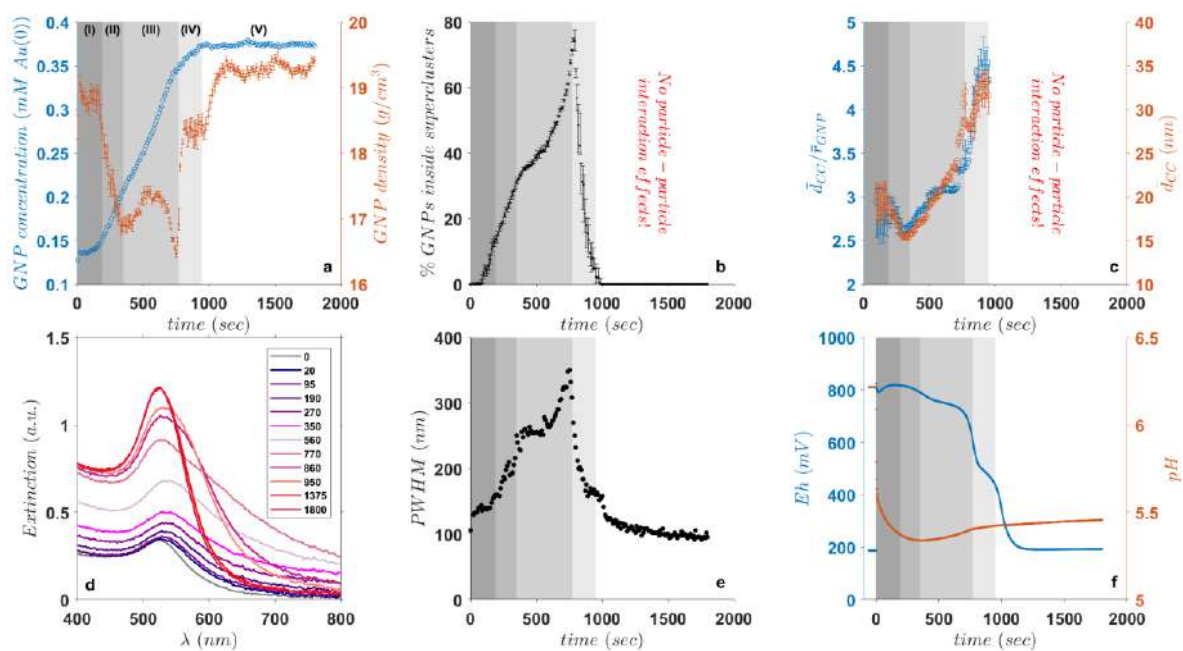

**Supporting Fig. 21.** Complementary kinetic data (15:85-65°C experiment) from *in situ* UV-Vis spectroscopy and electrochemical potential measurements. a, Gold nanoparticle concentration (as mM Au(0); obtained from extinction at 400 nm[47]) and the estimated volumetric mass density of nanoparticles (with 95% confidence bounds found by a heuristic search method). b, Estimated percentage of electromagnetically interacting particles out of the overall population (*i.e.*, interacting + noninteracting) along with the respective 95% confidence bounds.[46] c, Mean center-to-center distance between electromagnetically interacting nanoparticles both normalized to the average particle radius and in absolute units (with 95% confidence bounds[46]). d, Experimental extinction spectra at selected temporal points. e, Temporal evolution of peak width at half-maximum obtained from experimentally measured spectra.[6] f, *In situ* measured solution pH and reduction potential (Eh) (*vs.* Ag/AgCl/c(KCl) = 3 mol/L). The grey-to-white shaded regions (all the plots except for (d)) denote various mechanistic steps (denoted by Roman numerals in the first plot).

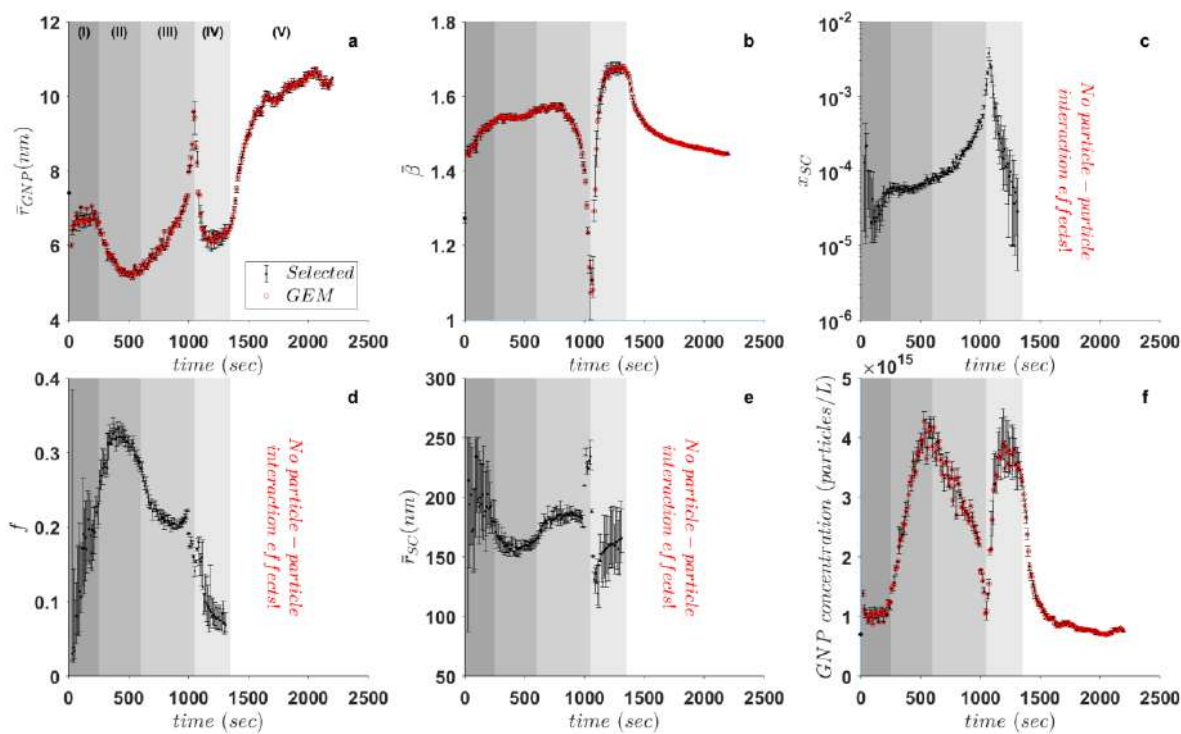

**Supporting Fig. 22.** Kinetics of gold nanoparticle growth (15:85-60°C experiment) from *in situ* UV-Vis spectroscopy and model regression. (a-e) Regressed physical parameters (and their corresponding 95% confidence bounds found by a heuristic search method) including mean particle radius (a), mean particle aspect ratio (b), number fraction of liquid-like superclusters (c), volume fraction of particles in superclusters (d), and mean supercluster radius (e). f, Temporal particle number concentration (along with the respective 95% confidence bounds[46]) obtained from scaling the calculated extinction cross sections with the experimental spectra. Legends are common in all the plots and the grey-to-white shaded regions denote various mechanistic steps (denoted by Roman numerals in the first plot).

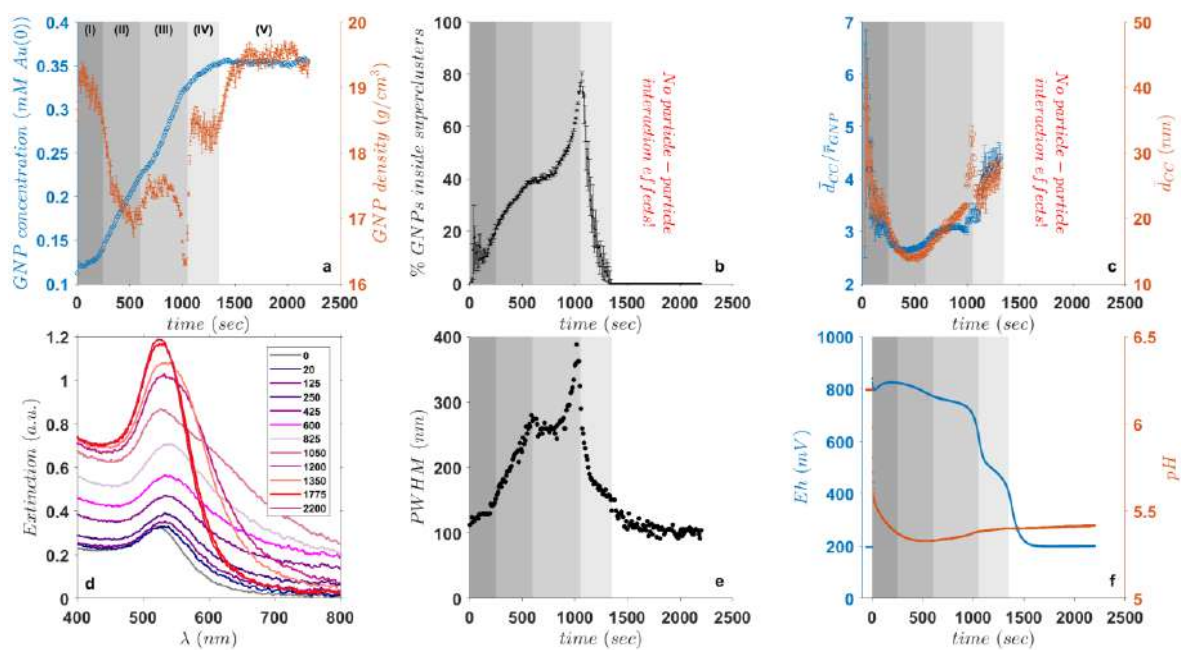

**Supporting Fig. 23.** Complementary kinetic data (15:85-60°C experiment) from *in situ* UV-Vis spectroscopy and electrochemical potential measurements. a, Gold nanoparticle concentration (as mM Au(0); obtained from extinction at 400 nm[47]) and the estimated volumetric mass density of nanoparticles (with 95% confidence bounds found by a heuristic search method). b, Estimated percentage of electromagnetically interacting particles out of the overall population (*i.e.*, interacting + noninteracting) along with the respective 95% confidence bounds.[46] c, Mean center-to-center distance between electromagnetically interacting nanoparticles both normalized to the average particle radius and in absolute units (with 95% confidence bounds[46]). d, Experimental extinction spectra at selected temporal points. e, Temporal evolution of peak width at half-maximum obtained from experimentally measured spectra.[6] f, *In situ* measured solution pH and reduction potential (Eh) (*vs.* Ag/AgCl/c(KCl) = 3 mol/L). The grey-to-white shaded regions (all the plots except for (d)) denote various mechanistic steps (denoted by Roman numerals in the first plot).

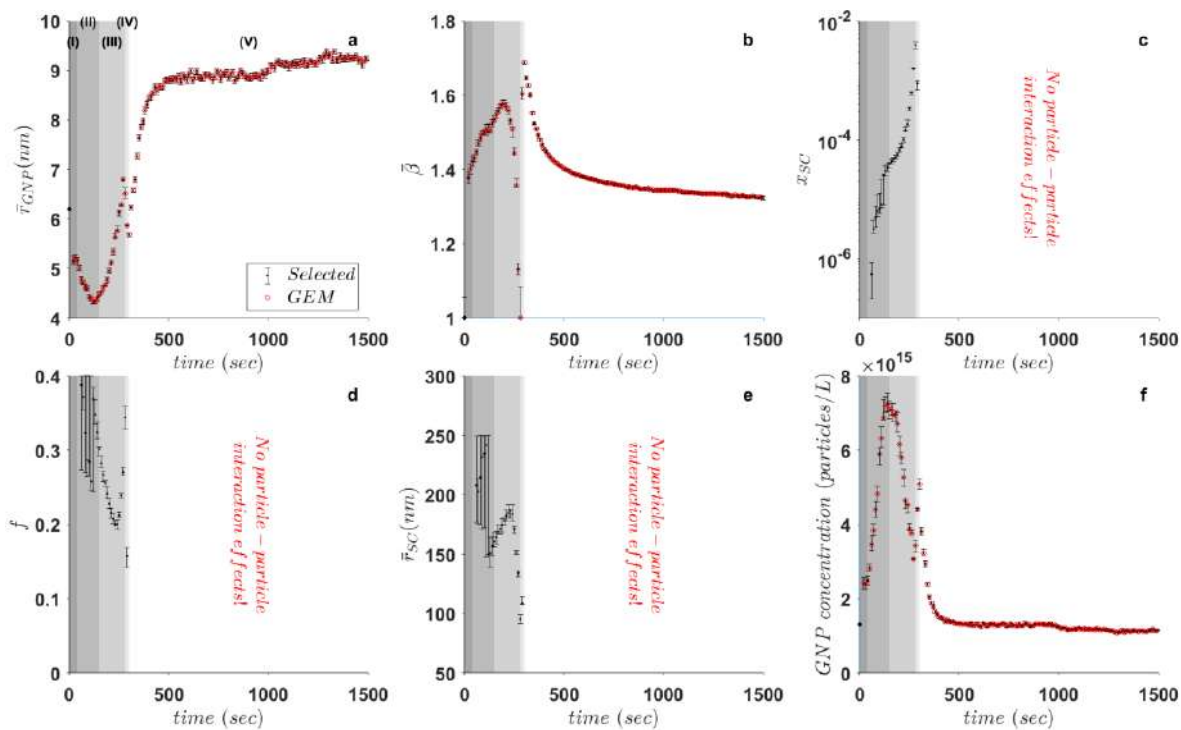

**Supporting Fig. 24.** Kinetics of gold nanoparticle growth (25:75-70°C experiment) from *in situ* UV-Vis spectroscopy and model regression. (a-e) Regressed physical parameters (and their corresponding 95% confidence bounds found by a heuristic search method) including mean particle radius (a), mean particle aspect ratio (b), number fraction of liquid-like superclusters (c), volume fraction of particles in superclusters (d), and mean supercluster radius (e). f, Temporal particle number concentration (along with the respective 95% confidence bounds[46]) obtained from scaling the calculated extinction cross sections with the experimental spectra. Legends are common in all the plots and the grey-to-white shaded regions denote various mechanistic steps (denoted by Roman numerals in the first plot).

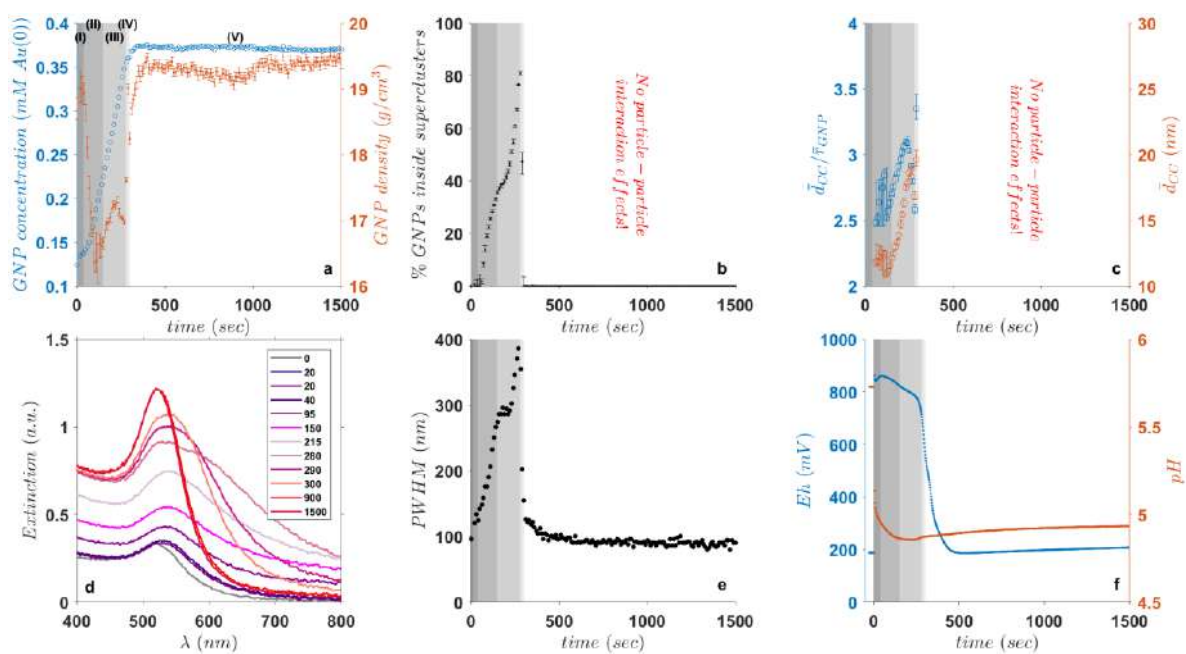

**Supporting Fig. 25.** Complementary kinetic data (25:75-70°C experiment) from *in situ* UV-Vis spectroscopy and electrochemical potential measurements. a, Gold nanoparticle concentration (as mM Au(0); obtained from extinction at 400 nm[47]) and the estimated volumetric mass density of nanoparticles (with 95% confidence bounds found by a heuristic search method). b, Estimated percentage of electromagnetically interacting particles out of the overall population (*i.e.*, interacting + noninteracting) along with the respective 95% confidence bounds.[46] c, Mean center-to-center distance between electromagnetically interacting nanoparticles both normalized to the average particle radius and in absolute units (with 95% confidence bounds[46]). d, Experimental extinction spectra at selected temporal points. e, Temporal evolution of peak width at half-maximum obtained from experimentally measured spectra.[6] f, *In situ* measured solution pH and reduction potential (Eh) (*vs.* Ag/AgCl/c(KCl) = 3 mol/L). The grey-to-white shaded regions (all the plots except for (d)) denote various mechanistic steps (denoted by Roman numerals in the first plot).

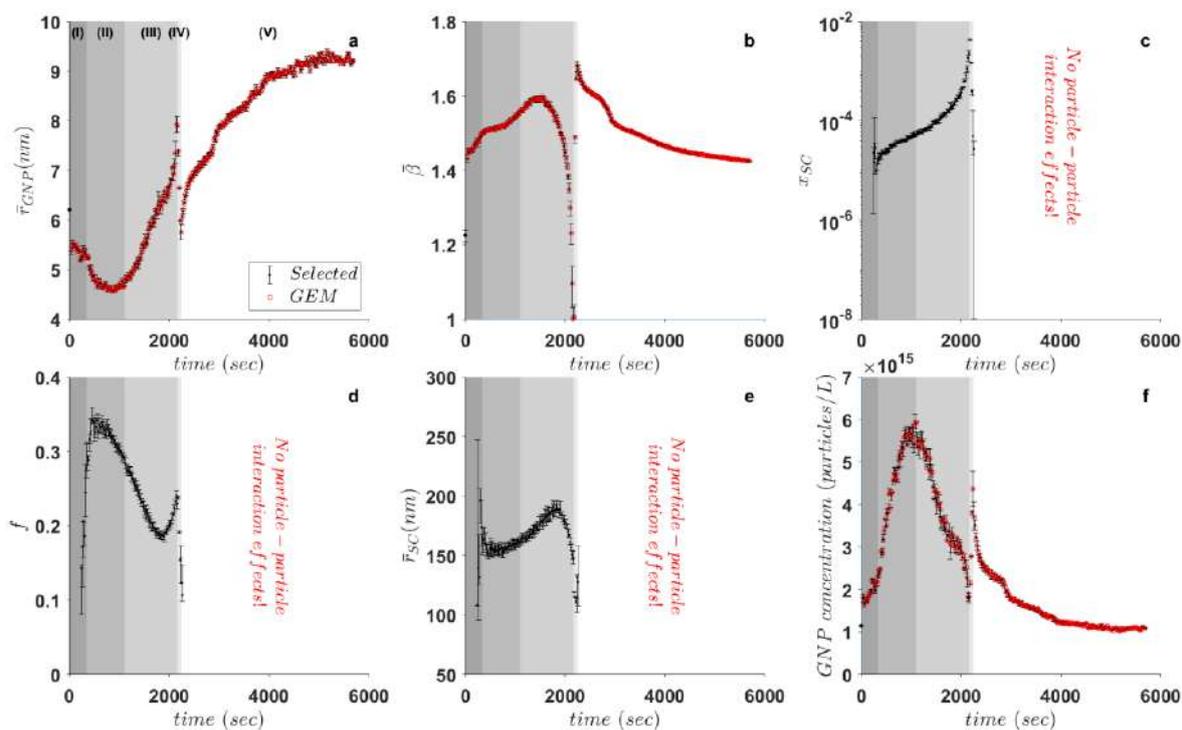

**Supporting Fig. 26.** Kinetics of gold nanoparticle growth (25:75-40°C experiment) from *in situ* UV-Vis spectroscopy and model regression. (a-e) Regressed physical parameters (and their corresponding 95% confidence bounds found by a heuristic search method) including mean particle radius (a), mean particle aspect ratio (b), number fraction of liquid-like superclusters (c), volume fraction of particles in superclusters (d), and mean supercluster radius (e). f, Temporal particle number concentration (along with the respective 95% confidence bounds[46]) obtained from scaling the calculated extinction cross sections with the experimental spectra. Legends are common in all the plots and the grey-to-white shaded regions denote various mechanistic steps (denoted by Roman numerals in the first plot).

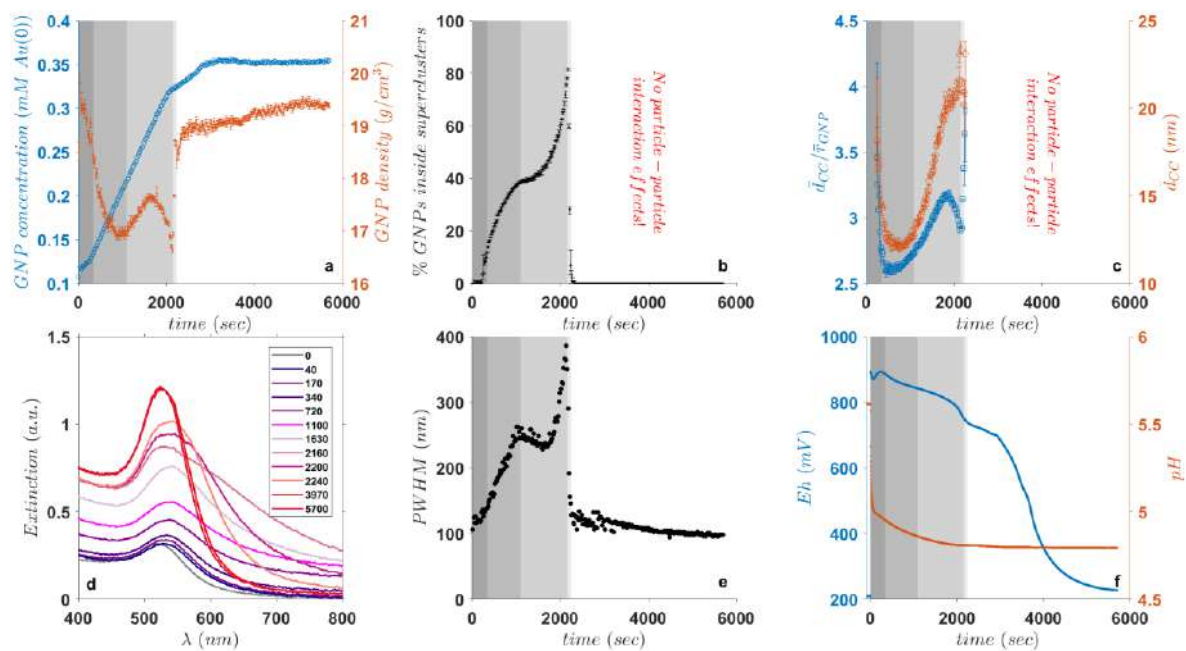

**Supporting Fig. 27.** Complementary kinetic data (25:75-40°C experiment) from *in situ* UV-Vis spectroscopy and electrochemical potential measurements. a, Gold nanoparticle concentration (as mM Au(0); obtained from extinction at 400 nm[47]) and the estimated volumetric mass density of nanoparticles (with 95% confidence bounds found by a heuristic search method). b, Estimated percentage of electromagnetically interacting particles out of the overall population (*i.e.*, interacting + noninteracting) along with the respective 95% confidence bounds.[46] c, Mean center-to-center distance between electromagnetically interacting nanoparticles both normalized to the average particle radius and in absolute units (with 95% confidence bounds[46]). d, Experimental extinction spectra at selected temporal points. e, Temporal evolution of peak width at half-maximum obtained from experimentally measured spectra.[6] f, *In situ* measured solution pH and reduction potential (Eh) (*vs.* Ag/AgCl/c(KCl) = 3 mol/L). The grey-to-white shaded regions (all the plots except for (d)) denote various mechanistic steps (denoted by Roman numerals in the first plot).

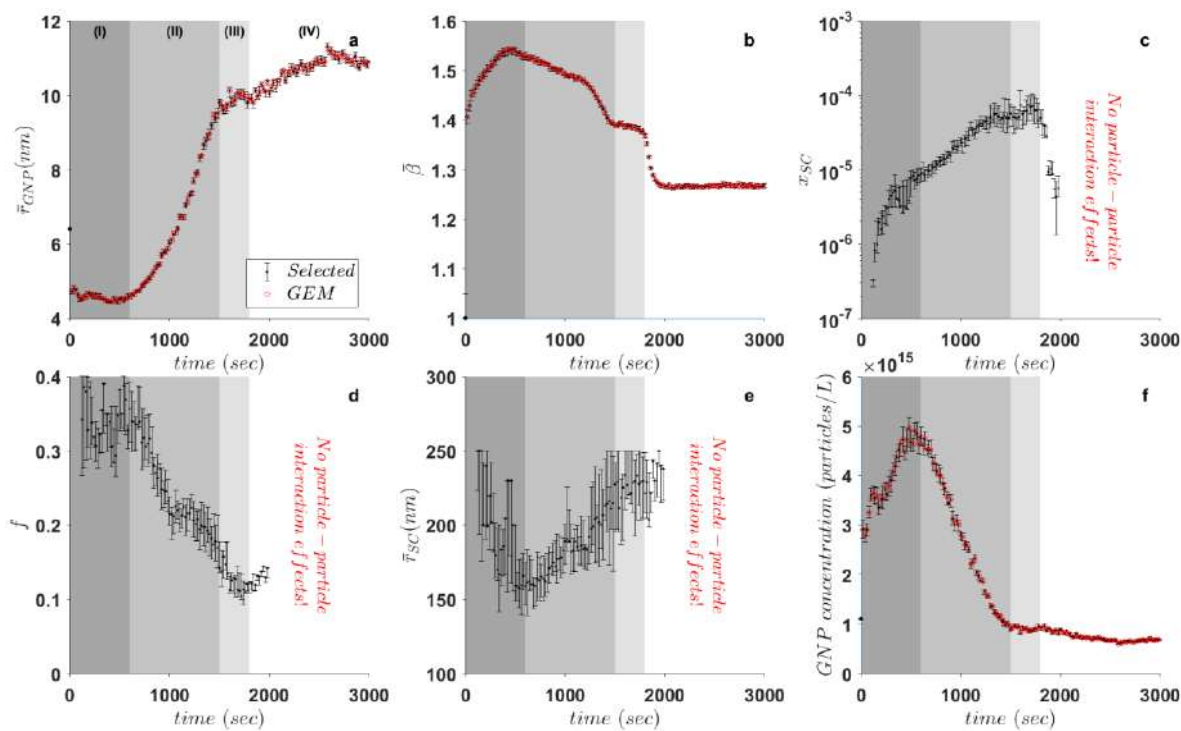

**Supporting Fig. 28.** Kinetics of gold nanoparticle growth (0:100-70°C experiment) from *in situ* UV-Vis spectroscopy and model regression. (a-e) Regressed physical parameters (and their corresponding 95% confidence bounds found by a heuristic search method) including mean particle radius (a), mean particle aspect ratio (b), number fraction of liquid-like superclusters (c), volume fraction of particles in superclusters (d), and mean supercluster radius (e). f, Temporal particle number concentration (along with the respective 95% confidence bounds[46]) obtained from scaling the calculated extinction cross sections with the experimental spectra. Legends are common in all the plots and the grey-to-white shaded regions denote various mechanistic steps (denoted by Roman numerals in the first plot).

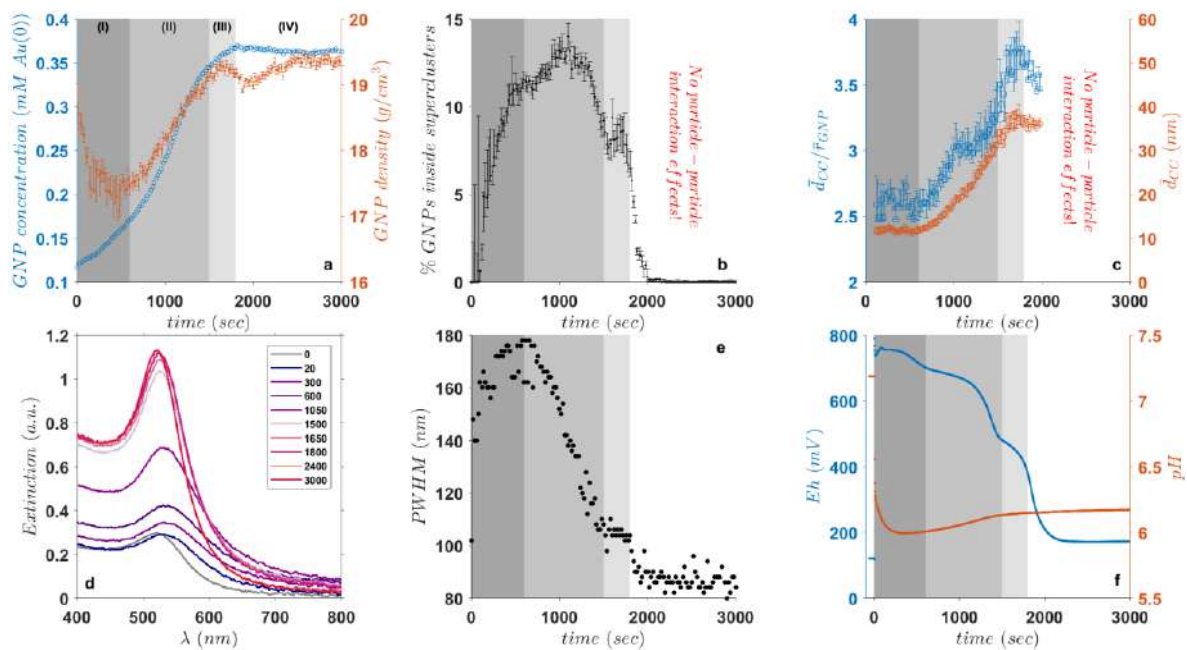

**Supporting Fig. 29.** Complementary kinetic data (0:100-70°C experiment) from *in situ* UV-Vis spectroscopy and electrochemical potential measurements. a, Gold nanoparticle concentration (as mM Au(0); obtained from extinction at 400 nm[47]) and the estimated volumetric mass density of nanoparticles (with 95% confidence bounds found by a heuristic search method). b, Estimated percentage of electromagnetically interacting particles out of the overall population (*i.e.*, interacting + noninteracting) along with the respective 95% confidence bounds.[46] c, Mean center-to-center distance between electromagnetically interacting nanoparticles both normalized to the average particle radius and in absolute units (with 95% confidence bounds[46]). d, Experimental extinction spectra at selected temporal points. e, Temporal evolution of peak width at half-maximum obtained from experimentally measured spectra.[6] f, *In situ* measured solution pH and reduction potential (Eh) (*vs.* Ag/AgCl/c(KCl) = 3 mol/L). The grey-to-white shaded regions (all the plots except for (d)) denote various mechanistic steps (denoted by Roman numerals in the first plot).

# Supporting Section 6. Temporal regression results common to both GEM and noninteracting (Gans) models

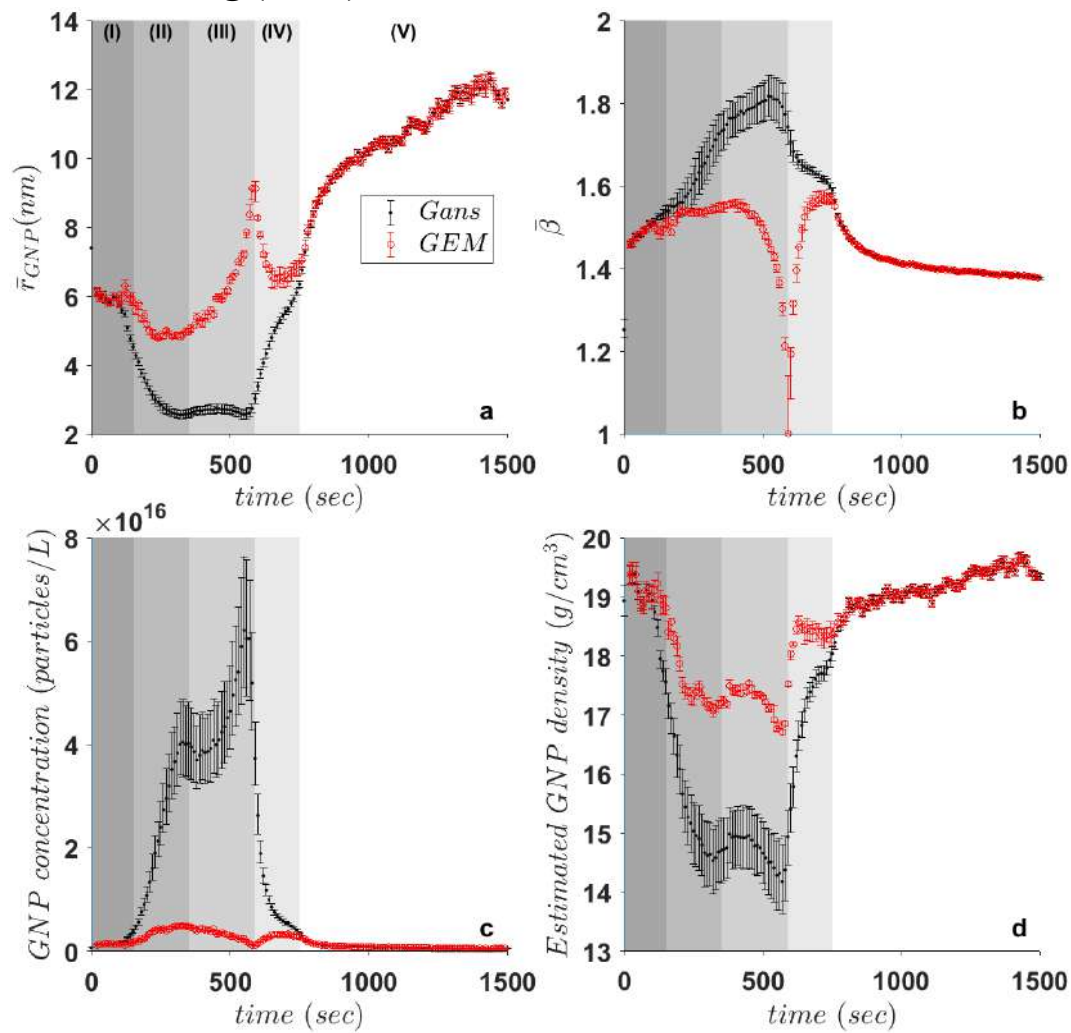

**Supporting Fig. 30.** Temporal evolution of optimal model outputs common between the Gans and the GEM formalisms (15:85-70°C experiment). a, Mean particle radius. b, Mean particle aspect ratio. c, Particle number concentration. d, Estimated particle density. 95% confidence bounds were found by a heuristic search method and the legends are common in all the plots. The grey-to-white shaded regions denote various mechanistic steps (denoted by Roman numerals in the first plot).

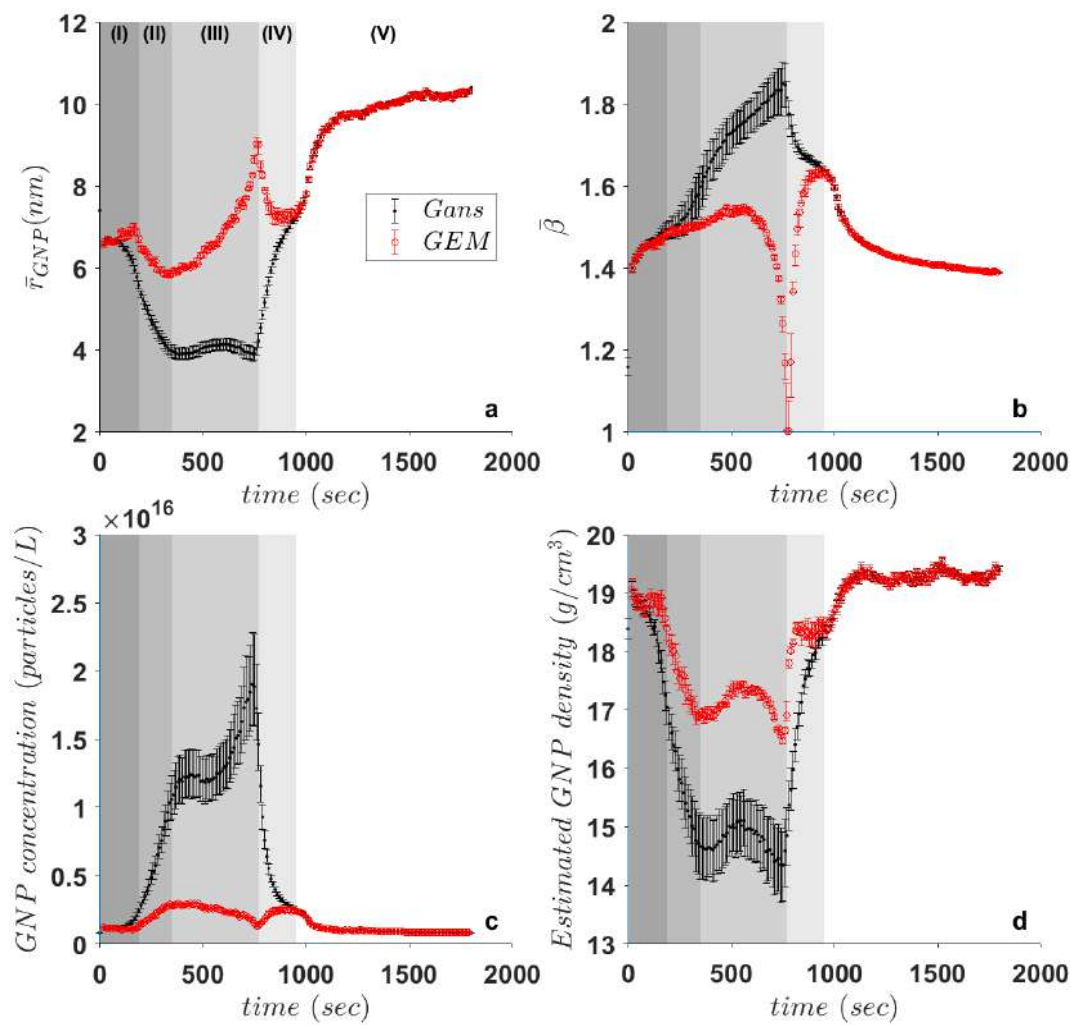

**Supporting Fig. 31.** Temporal evolution of optimal model outputs common between the Gans and the GEM formalisms (15:85-65°C experiment). a, Mean particle radius. b, Mean particle aspect ratio. c, Particle number concentration. d, Estimated particle density. 95% confidence bounds were found by a heuristic search method and the legends are common in all the plots. The grey-to-white shaded regions denote various mechanistic steps (denoted by Roman numerals in the first plot).

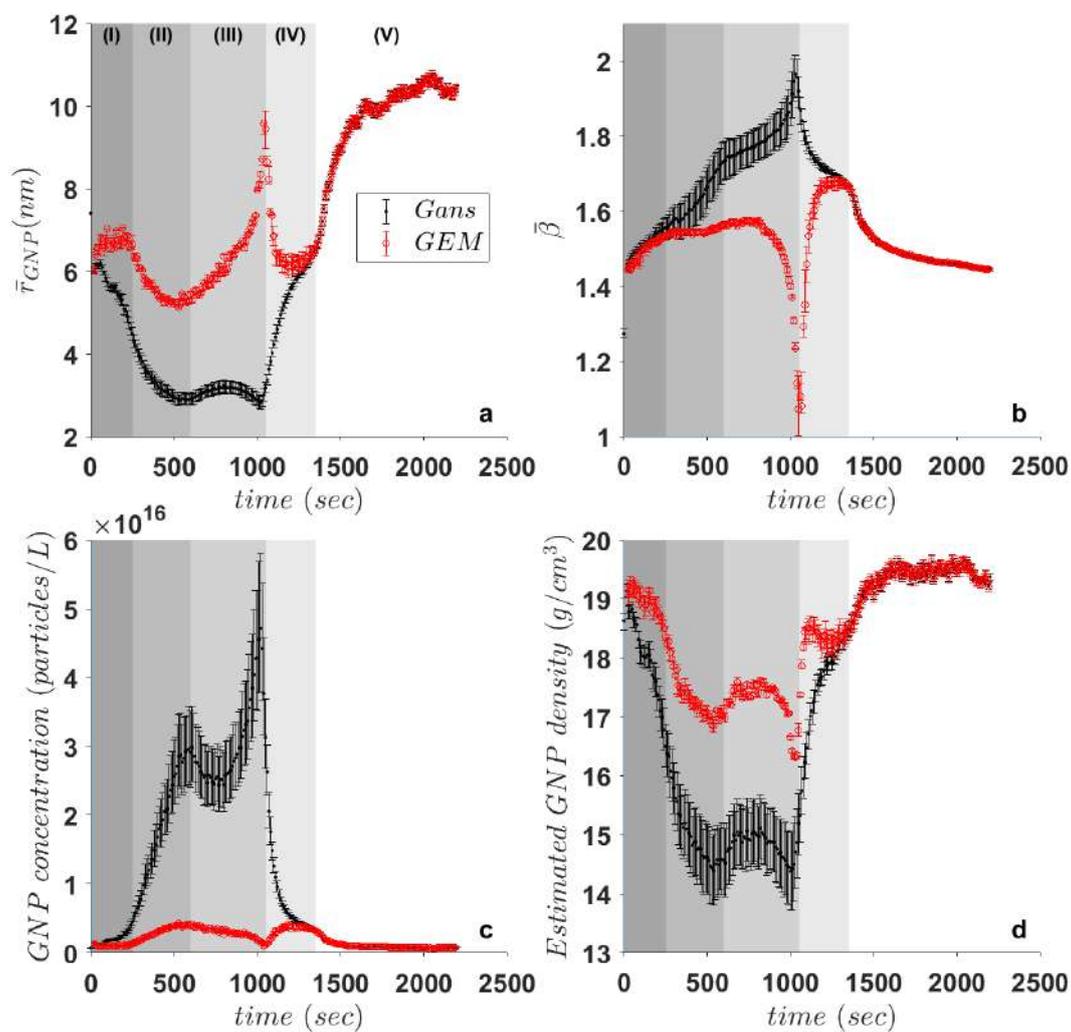

**Supporting Fig. 32.** Temporal evolution of optimal model outputs common between the Gans and the GEM formalisms (15:85-60°C experiment). a, Mean particle radius. b, Mean particle aspect ratio. c, Particle number concentration. d, Estimated particle density. 95% confidence bounds were found by a heuristic search method and the legends are common in all the plots. The grey-to-white shaded regions denote various mechanistic steps (denoted by Roman numerals in the first plot).

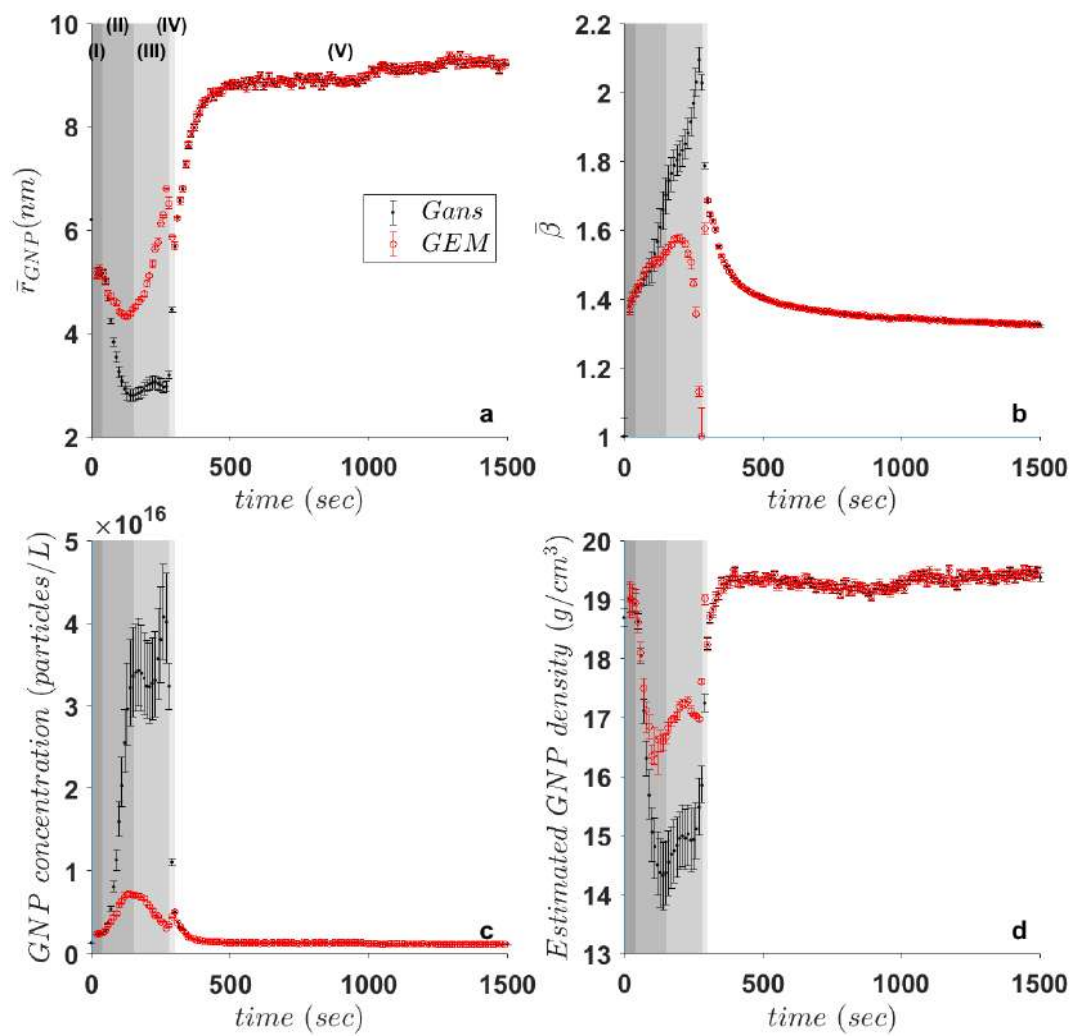

**Supporting Fig. 33.** Temporal evolution of optimal model outputs common between the Gans and the GEM formalisms (25:75-70°C experiment). a, Mean particle radius. b, Mean particle aspect ratio. c, Particle number concentration. d, Estimated particle density. 95% confidence bounds were found by a heuristic search method and the legends are common in all the plots. The grey-to-white shaded regions denote various mechanistic steps (denoted by Roman numerals in the first plot).

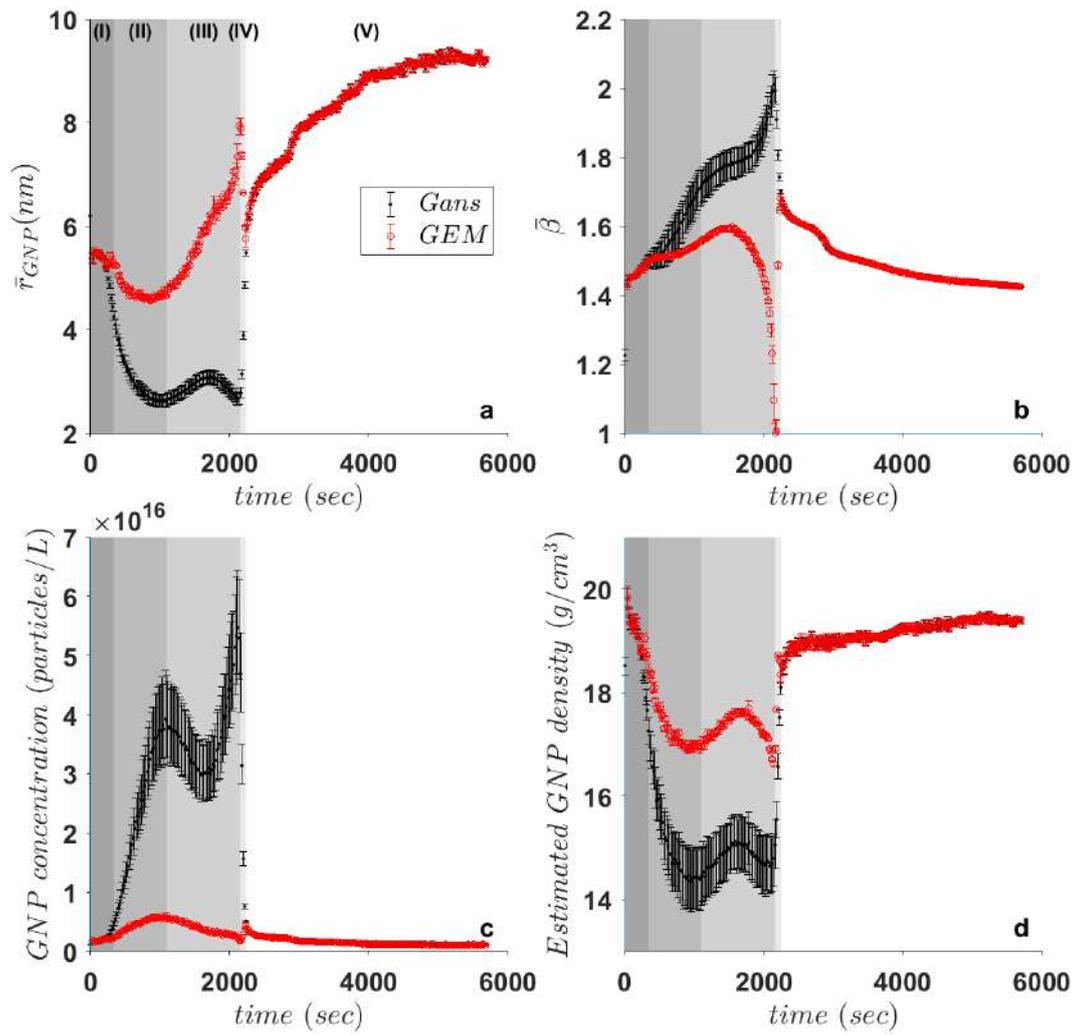

**Supporting Fig. 34.** Temporal evolution of optimal model outputs common between the Gans and the GEM formalisms (25:75-40°C experiment). a, Mean particle radius. b, Mean particle aspect ratio. c, Particle number concentration. d, Estimated particle density. 95% confidence bounds were found by a heuristic search method and the legends are common in all the plots. The grey-to-white shaded regions denote various mechanistic steps (denoted by Roman numerals in the first plot).

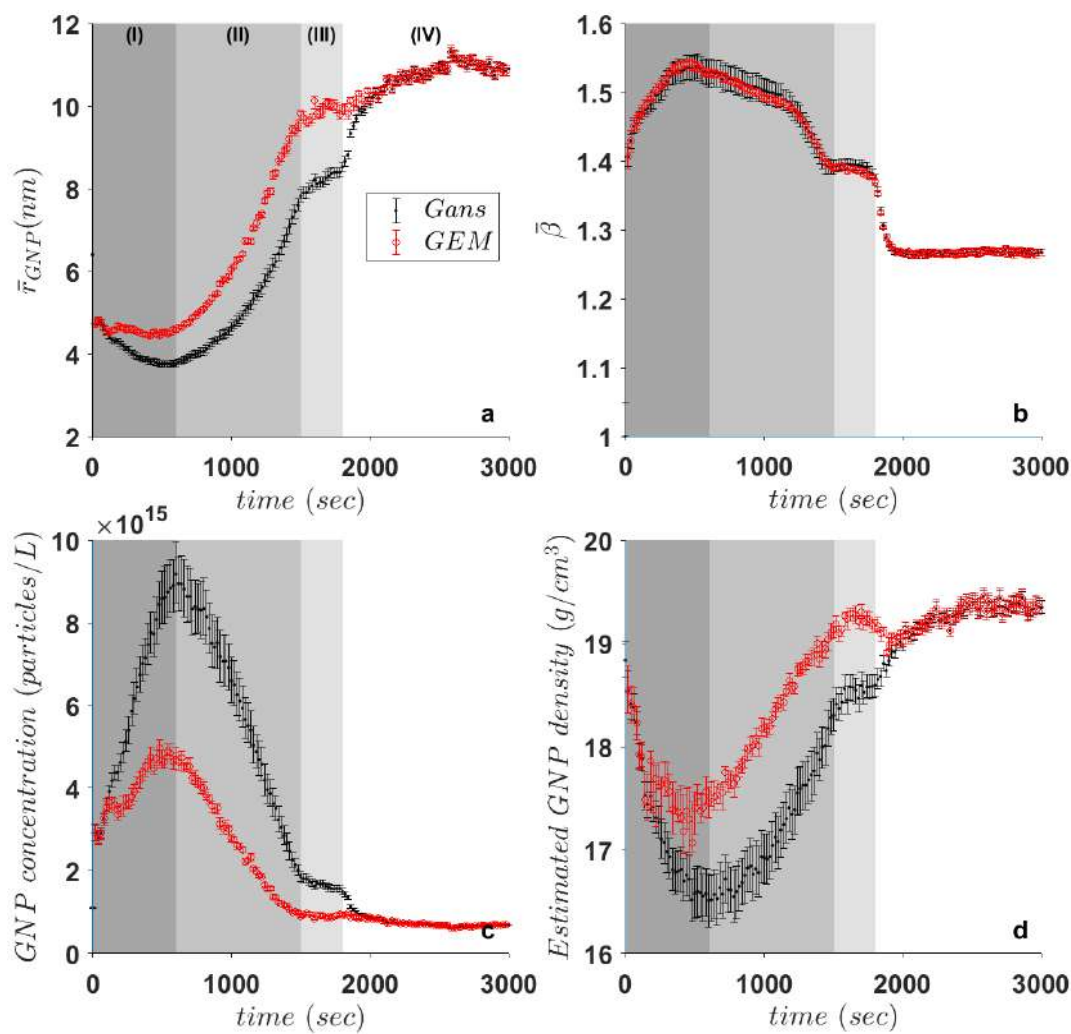

**Supporting Fig. 35.** Temporal evolution of optimal model outputs common between the Gans and the GEM formalisms (0:100-70°C experiment). a, Mean particle radius. b, Mean particle aspect ratio. c, Particle number concentration. d, Estimated particle density. 95% confidence bounds were found by a heuristic search method and the legends are common in all the plots. The grey-to-white shaded regions denote various mechanistic steps (denoted by Roman numerals in the first plot).

## Supporting Section 7. Principal component analysis on temporal spectra

As we discussed in Supporting Section 4, all the datasets exhibit very similar mechanistic behavior (with different time scales). PCA plots reflect this similarity as well (Supporting Figs. 36-39). The most important difference is in the experiment 0:100-70°C, where the electromagnetic interaction effects are far less significant compared to the rest of the experiments. Quantitatively, PC1 to PC3 account for 94.9, 5, and 0.1% of the variance, respectively. Therefore, the contribution of electromagnetic interactions is around 4.5 times less than that in the experiment 15:85-70°C. The corresponding PCA plots also validate this statement. In this respect, conferring Supporting Fig. 40b we see that the score corresponding to PC2 is considerably smaller than in the other datasets (Supporting Figs. 36b-39b).

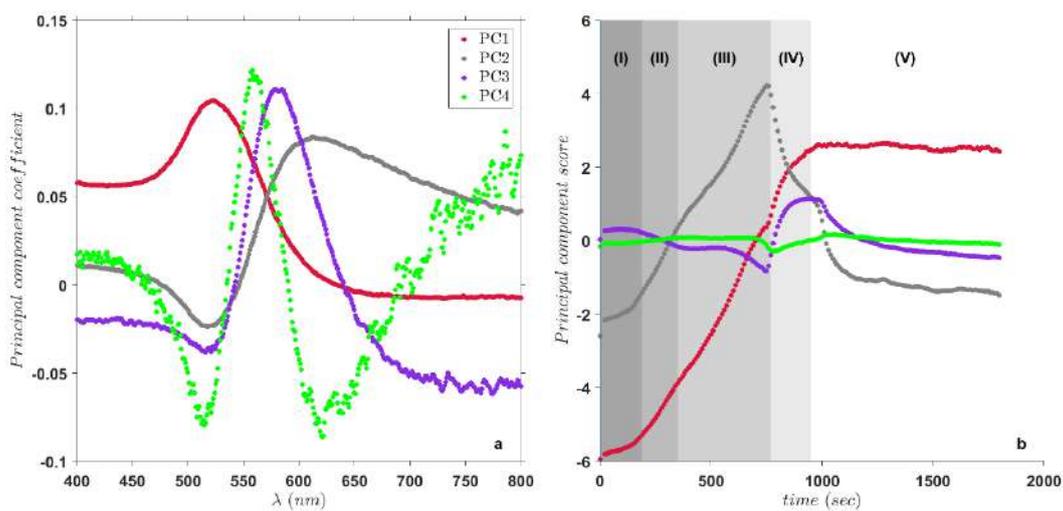

**Supporting Fig. 36.** Principal component analysis on temporal UV-Vis spectra (15:85-65°C experiment). a, PCA coefficients. b, PCA scores. Legends are common in both plots.

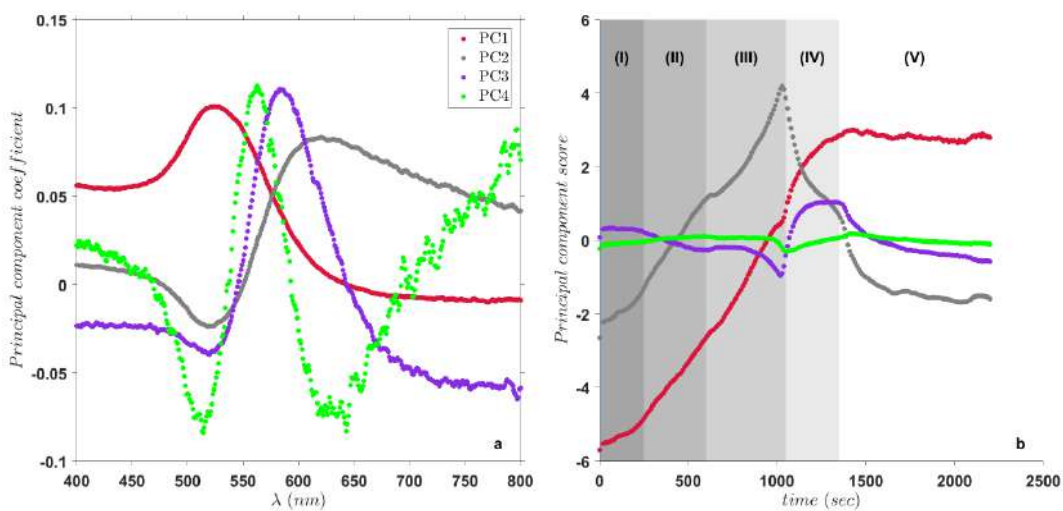

**Supporting Fig. 37.** Principal component analysis on temporal UV-Vis spectra (15:85-60°C experiment). a, PCA coefficients. b, PCA scores. Legends are common in both plots.

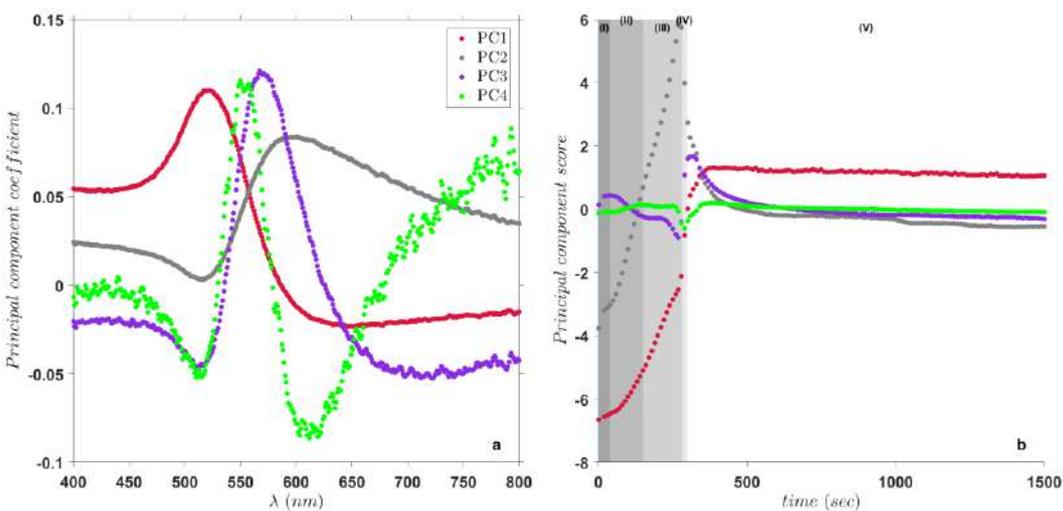

**Supporting Fig. 38.** Principal component analysis on temporal UV-Vis spectra (25:75-70°C experiment). a, PCA coefficients. b, PCA scores. Legends are common in both plots.

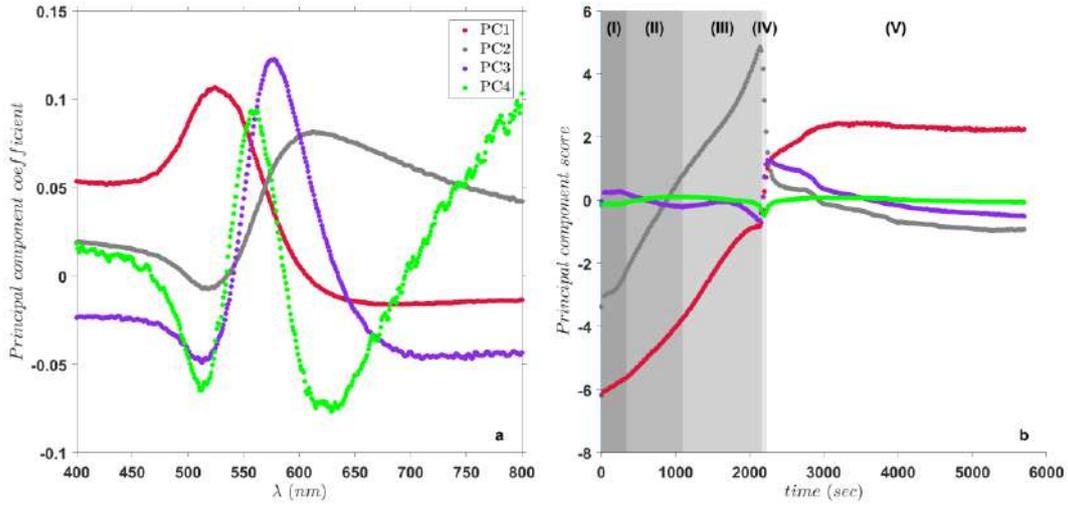

**Supporting Fig. 39.** Principal component analysis on temporal UV-Vis spectra (25:75-40°C experiment). a, PCA coefficients. b, PCA scores. Legends are common in both plots.

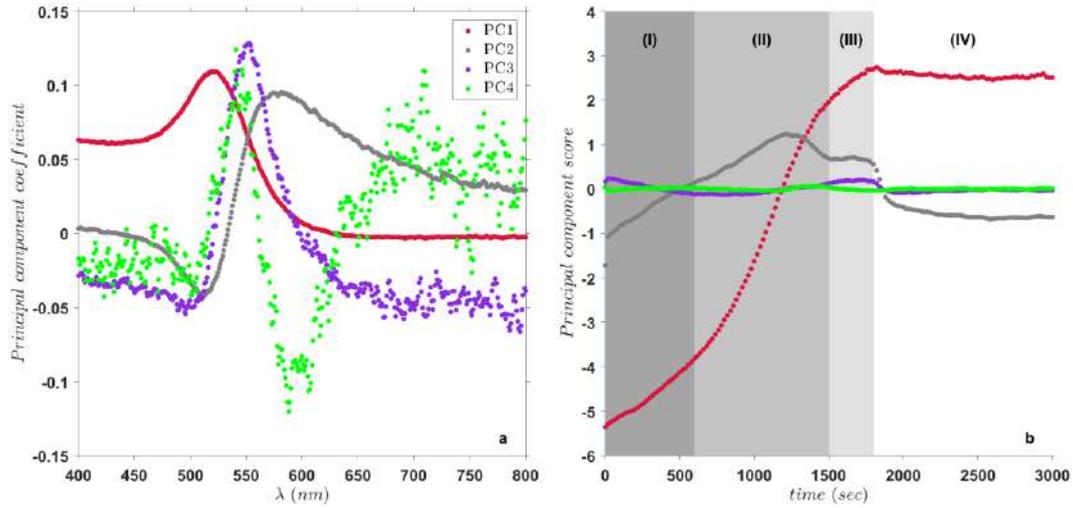

**Supporting Fig. 40.** Principal component analysis on temporal UV-Vis spectra (0:100-70°C experiment). a, PCA coefficients. b, PCA scores. Legends are common in both plots.

# Supporting Section 8. Representative confidence-region plots for the experiment 15:85-70°C

In this section, we have presented representative 95% confidence (likelihood) regions estimated for individual model parameters obtained from regression with the GEM model. The confidence regions are estimated using the approach presented by Schwaab et al.[12,20] This heuristic method of constructing the confidence bounds provides an invaluable and easy-to-use visual tool to assess the behavior of the model at hand. Below, we have provided such plots for the experiment 15:85-70°C at several time points during the seeded growth process. Initially, in the absence of electromagnetic interaction between the particles, only parameters common with the Gans theory ($\bar{r}_{GNP}$, $\bar{\beta}$, and GNP concentration) are well constrained (Supporting Figs. 41 and 42). For these parameters, $fval \equiv 100 \times \chi^2 = 100 \times \sum_i (A_{experimental} - A_{caluclated})^2$ assumes a parabolic shape near the minimum (Supporting Fig. 41a,b,f and 42a,b,f).[56] On the contrary, $x_{SC}$, $f$, and $\bar{r}_{SC}$ adopt very broad confidence regions, which implies model insensitiveness with respect to these parameters (Supporting Figs. 41c,d,e and 42c,d,e). As the interaction effects emerge, parabolic basins of attraction develop for the EMT parameters as well (Supporting Figs. 43-48). Particularly at the onset of interaction, there might be several basins of attraction and/or oddly shaped likelihood regions (Supporting Figs. 44-47). This is a common observation in nonlinear regression problems and similar examples are provided by Schwaab and coworkers.[12] On the other hand, when the interaction effects are strong enough, the likelihood regions become narrow and close to parabolic for all the parameters (Supporting Fig. 48).

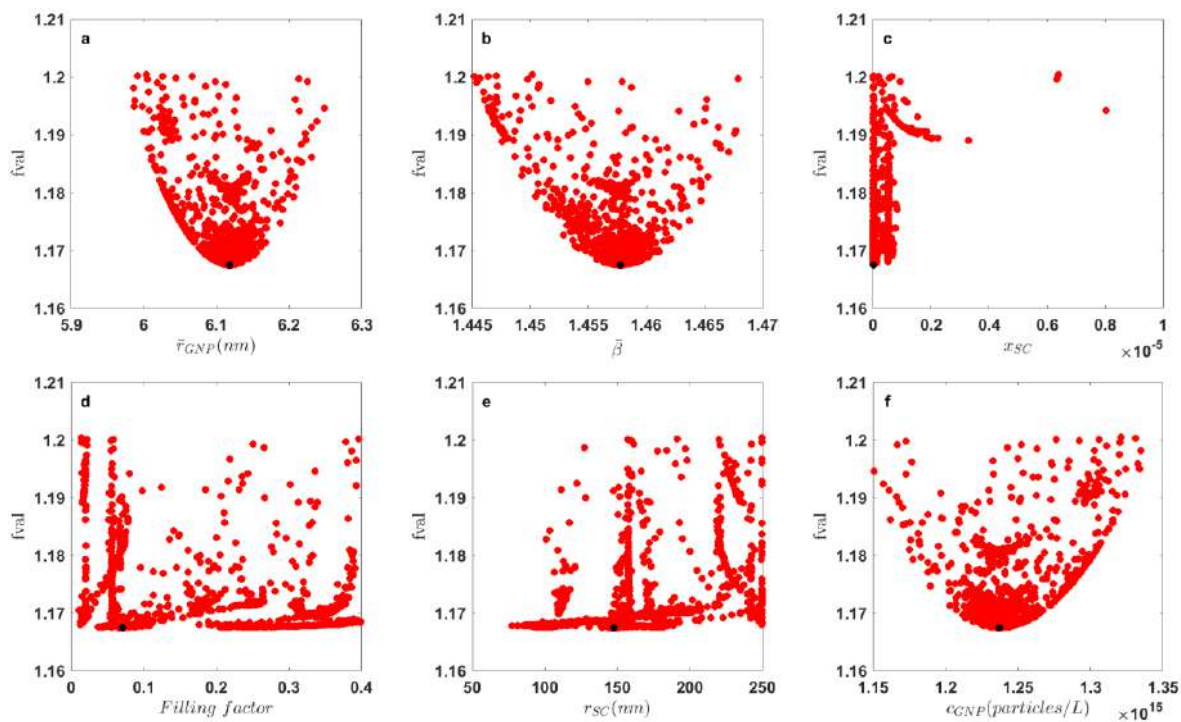

**Supporting Fig. 41.** 95% confidence regions (for individual model parameters) as estimated by the heuristic method of Schwaab *et al.* (15:85-70°C experiment; t = 20 s).

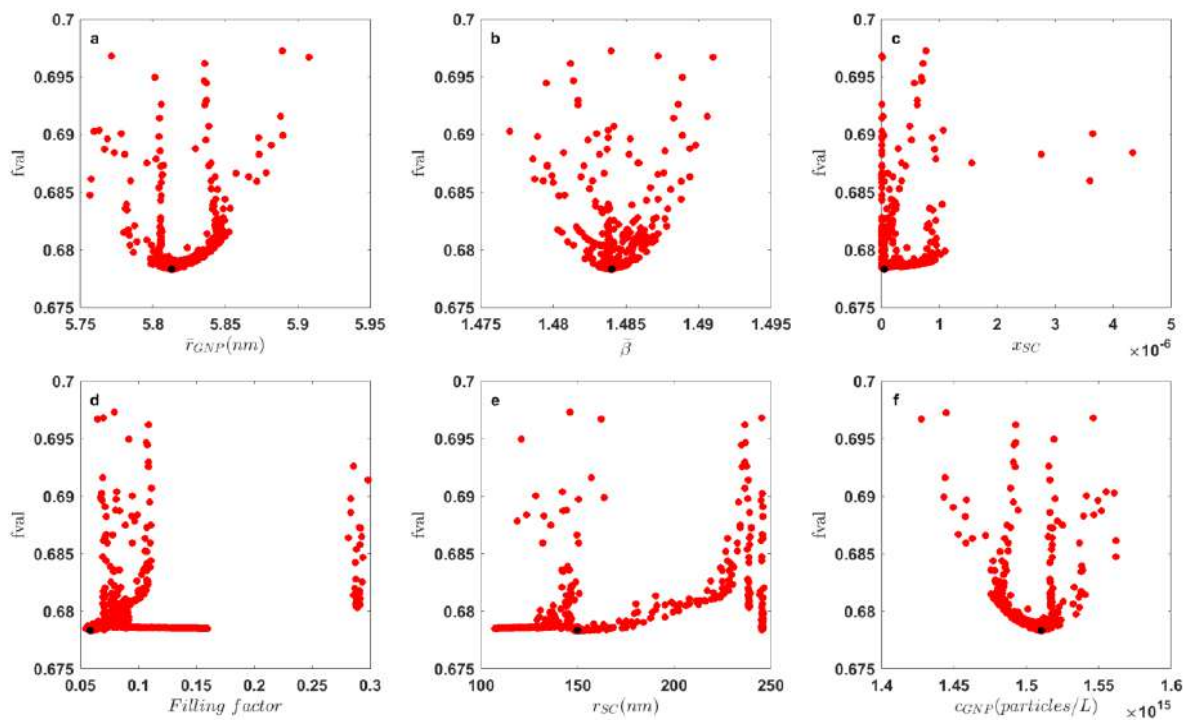

**Supporting Fig. 42.** 95% confidence regions (for individual model parameters) as estimated by the heuristic method of Schwaab *et al.* (15:85-70°C experiment; t = 60 s).

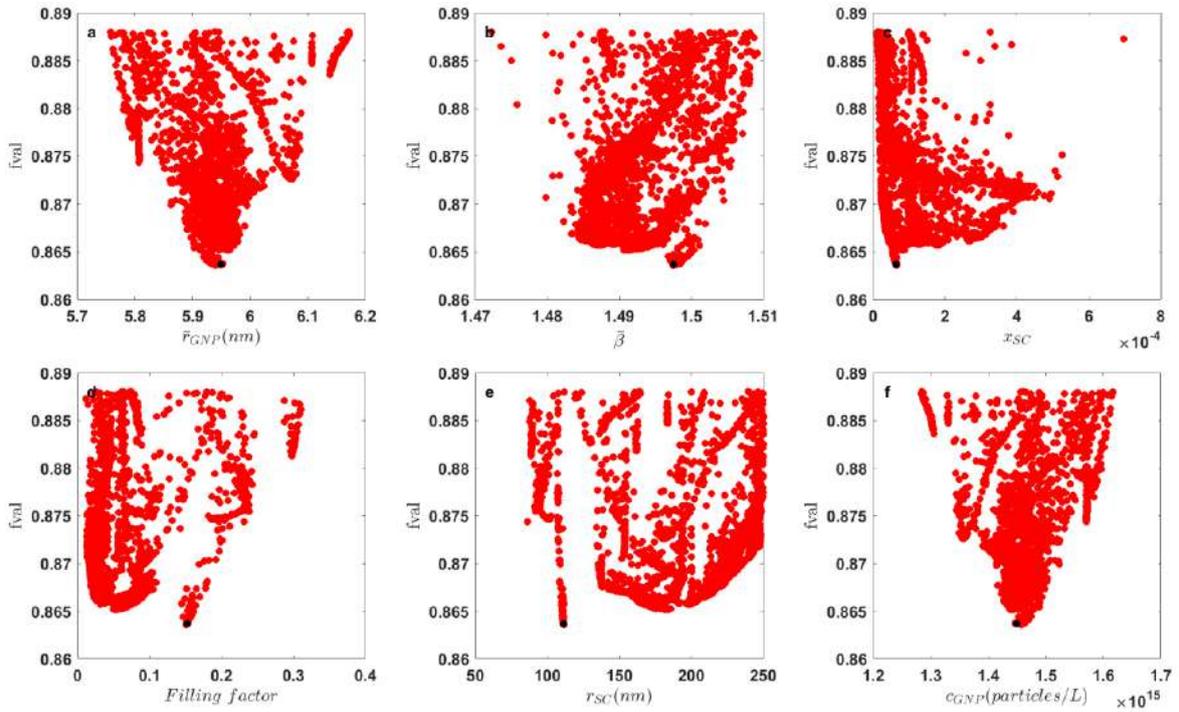

**Supporting Fig. 43.** 95% confidence regions (for individual model parameters) as estimated by the heuristic method of Schwaab *et al.* (15:85-70°C experiment; t = 110 s).

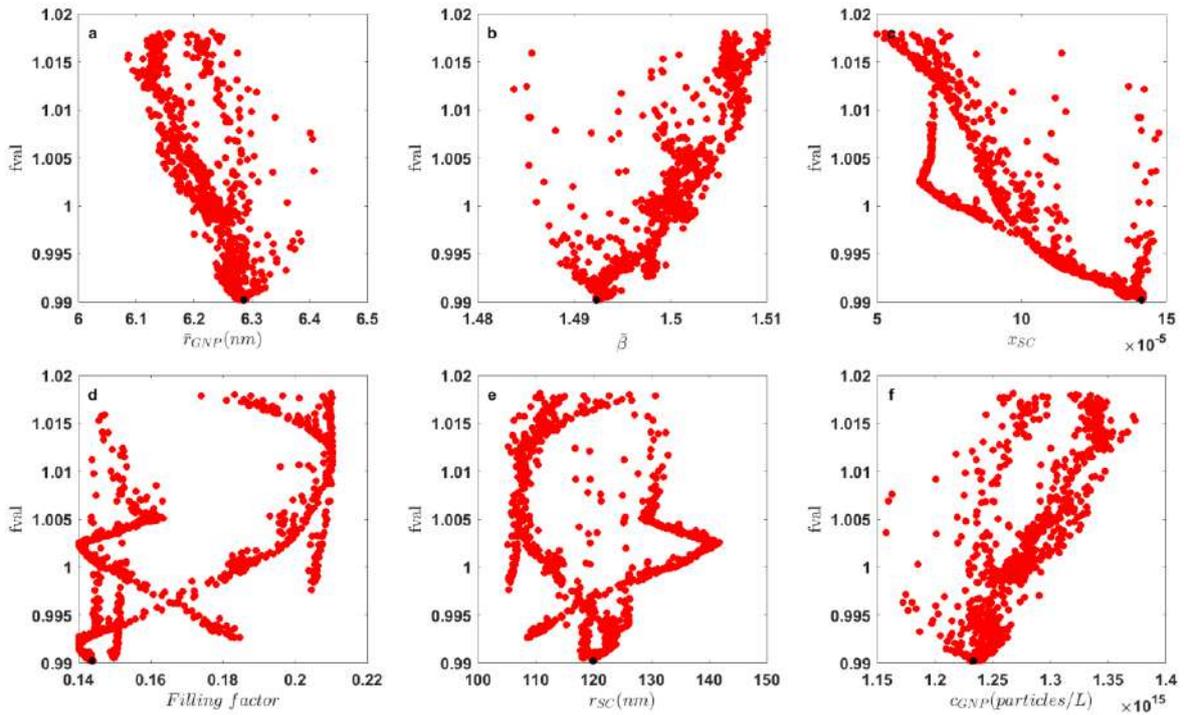

**Supporting Fig. 44.** 95% confidence regions (for individual model parameters) as estimated by the heuristic method of Schwaab *et al.* (15:85-70°C experiment; t = 120 s).

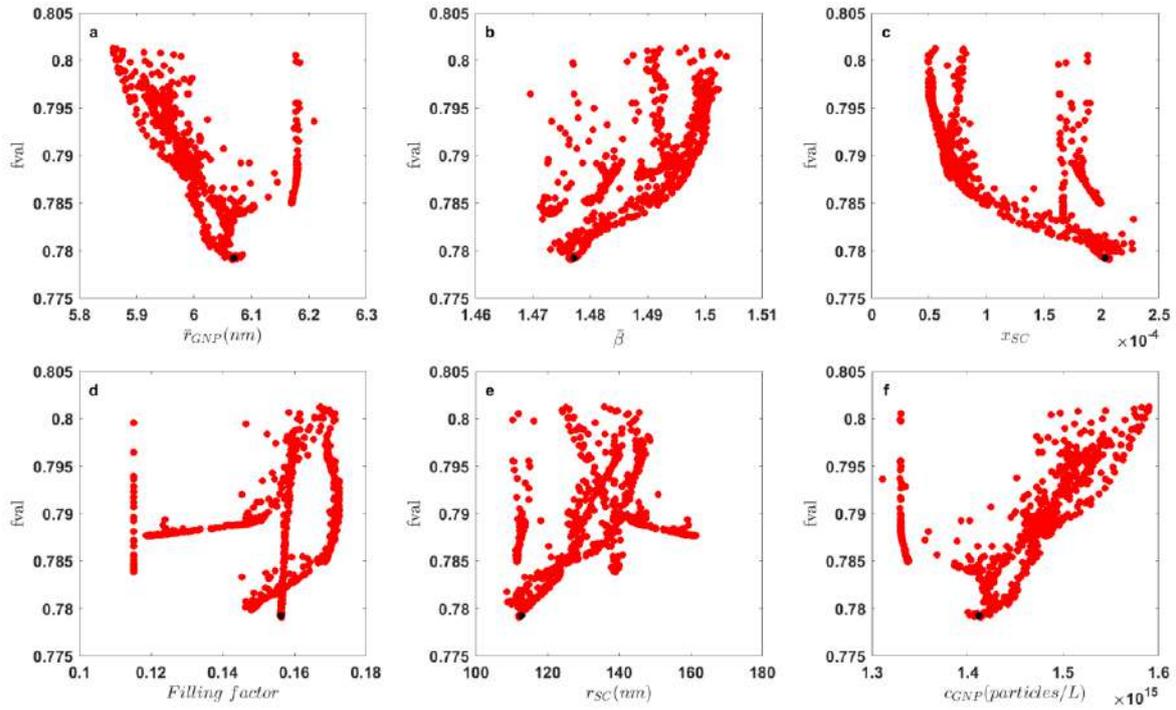

**Supporting Fig. 45.** 95% confidence regions (for individual model parameters) as estimated by the heuristic method of Schwaab *et al.* (15:85-70°C experiment; t = 130 s).

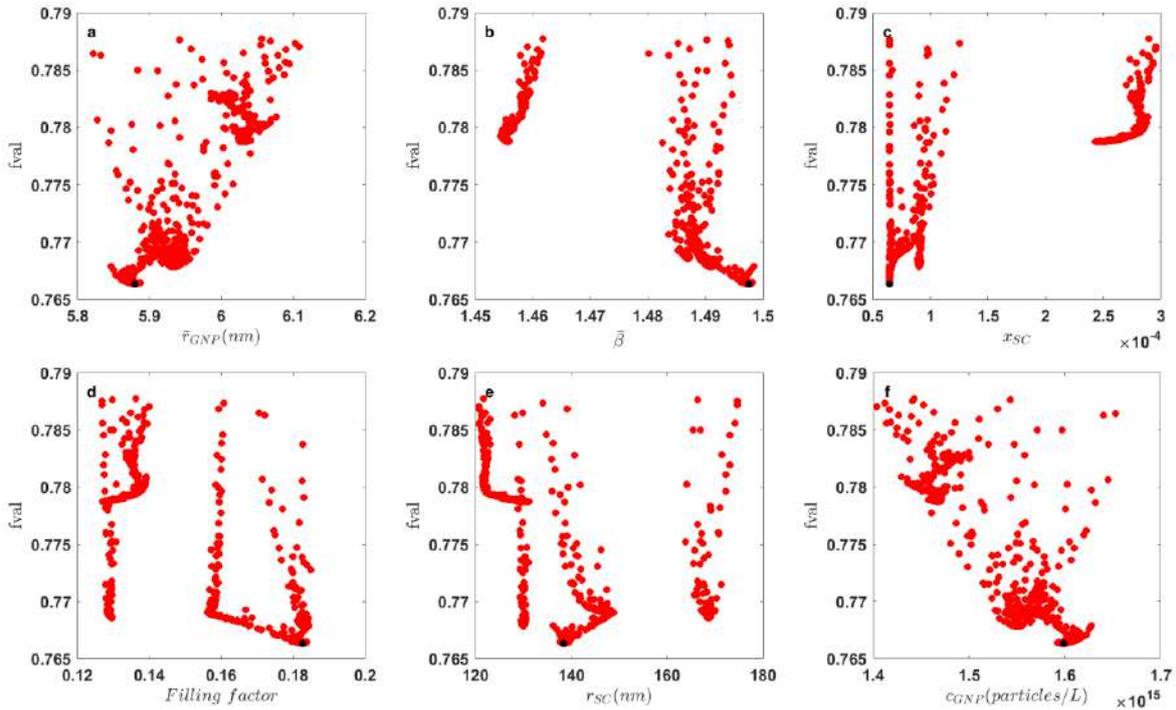

**Supporting Fig. 46.** 95% confidence regions (for individual model parameters) as estimated by the heuristic method of Schwaab *et al.* (15:85-70°C experiment; t = 140 s).

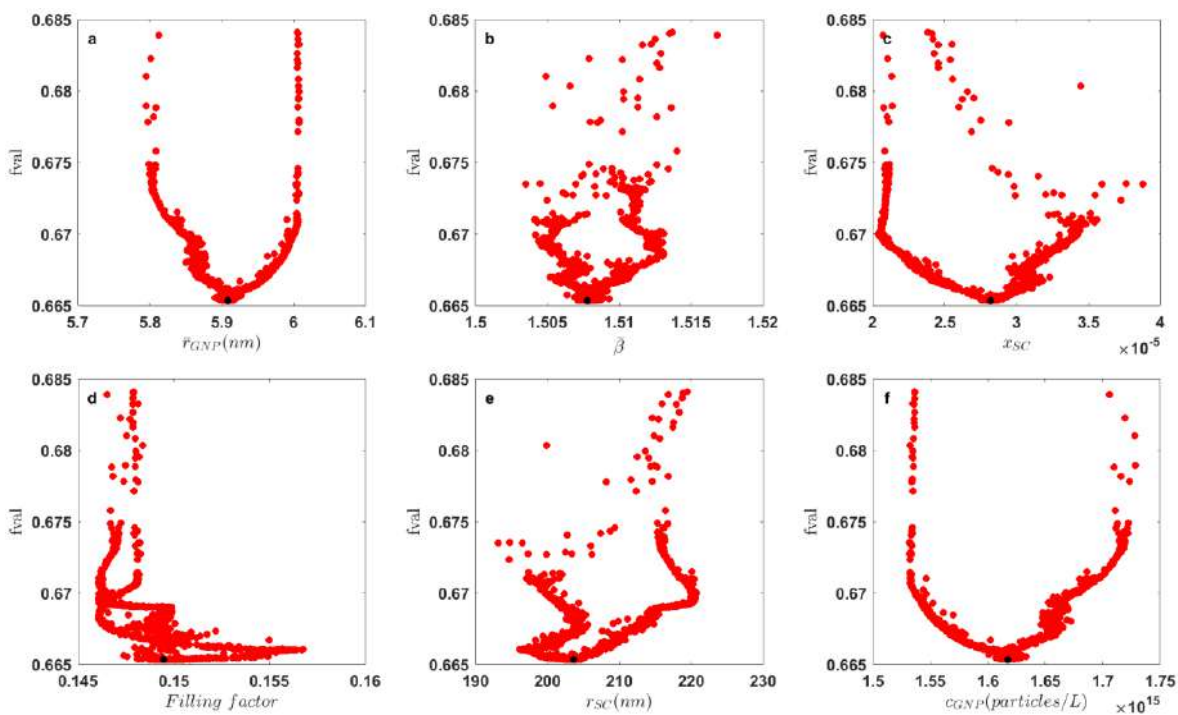

**Supporting Fig. 47.** 95% confidence regions (for individual model parameters) as estimated by the heuristic method of Schwaab *et al*. (15:85-70°C experiment; t = 150 s).

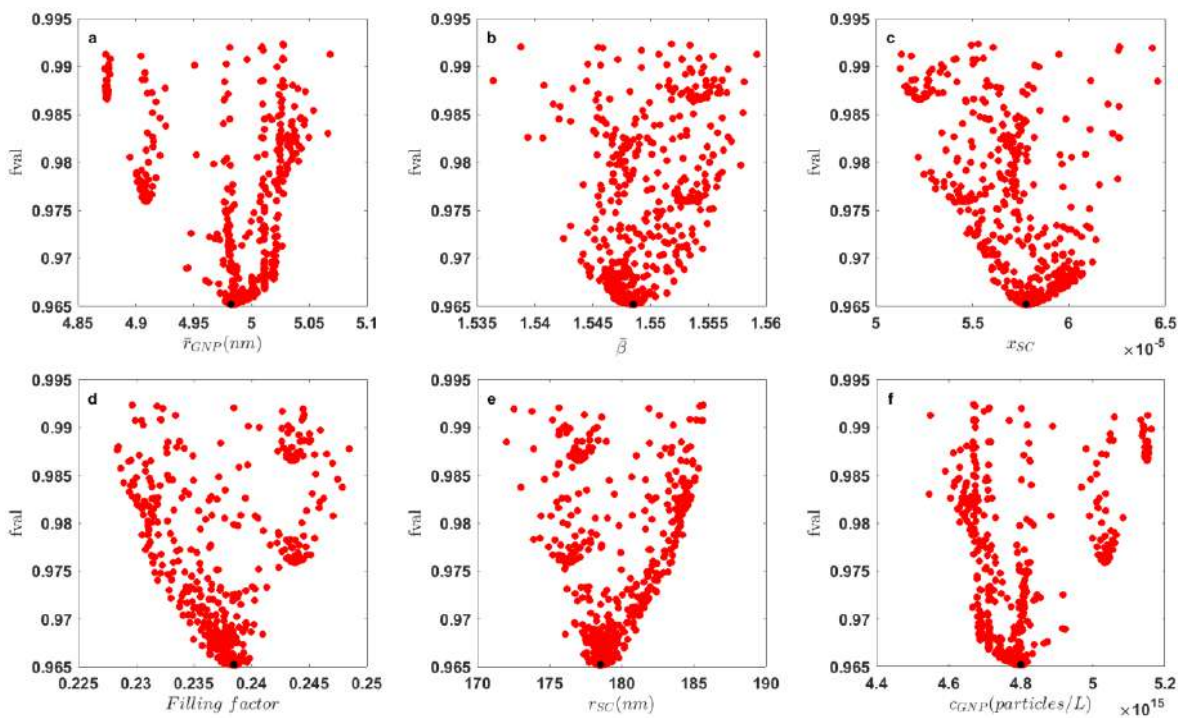

**Supporting Fig. 48.** 95% confidence regions (for individual model parameters) as estimated by the heuristic method of Schwaab *et al*. (15:85-70°C experiment; t = 350 s).

# References

(1)  Alemán, J. V.; Chadwick, A. V.; He, J.; Hess, M.; Horie, K.; Jones, R. G.; Kratochvíl, P.; Meisel, I.; Mita, I.; Moad, G.; Penczek, S.; Stepto, R. F. T. Definitions of Terms Relating to the Structure and Processing of Sols, Gels, Networks, and Inorganic-Organic Hybrid Materials (IUPAC Recommendations 2007). *Pure Appl. Chem.* **2007**, *79,* 1801–1829.

(2)     Carl, N.; Prévost, S.; Fitzgerald, J. P. S.; Karg, M. Salt-Induced Cluster Formation of Gold Nanoparticles Followed by Stopped-Flow SAXS, DLS and Extinction Spectroscopy. *Phys. Chem. Chem. Phys.* **2017**, *19*, 16348–16357.

(3)     Biggs, S.; Mulvaney, P.; Zukoski, C. F.; Grieser, F. Study of Anion Adsorption at the Gold-Aqueous Solution Interface by Atomic Force Microscopy. *J. Am. Chem. Soc.* **1994**, *116*, 9150–9157.

(4)     Chow, M. K.; Zukoski, C. F. Gold Sol Formation Mechanisms: Role of Colloidal Stability. *J. Colloid Interface Sci.* **1994**, *165*, 97–109.

(5)     Hühn, J.; Carrillo-Carrion, C.; Soliman, M. G.; Pfeiffer, C.; Valdeperez, D.; Masood, A.; Chakraborty, I.; Zhu, L.; Gallego, M.; Yue, Z.; Carril, M.; Feliu, N.; Escudero, A.; Alkilany, A. M.; Pelaz, B.; del Pino, P.; Parak, W. J. Selected Standard Protocols for the Synthesis, Phase Transfer, and Characterization of Inorganic Colloidal Nanoparticles. *Chem. Mater.* **2017**, *29*, 399–461.

(6)     Amendola, V.; Meneghetti, M. Size Evaluation of Gold Nanoparticles by UV−Vis Spectroscopy. *J. Phys. Chem. C* **2009**, *113*, 4277–4285.

(7)     Kreibig, U.; Vollmer, M. *Optical Properties of Metal Clusters*; Springer Series in Materials Science; Springer: Berlin, Heidelberg, 1995; Vol. 25; pp 13–186.

(8)     Kreibig, U.; Bour, G.; Hilger, A.; Gartz, M. Optical Properties of Cluster–Matter: Influences of Interfaces. *Phys. status solidi* **1999**, *175*, 351–366.

(9)     Markel, V. A. Introduction to the Maxwell Garnett Approximation: Tutorial. *J. Opt. Soc. Am. A* **2016**, *33*, 1244.

(10)    Oldenhuis, R. MATLAB Minimization Algorithm "Minimize" Version 1.7. Https://Github.Com/Rodyo/FEX-Minimize (date of access: 05/15/2019).

(11)    Messac, A. *Optimization in Practice with MATLAB®: For Engineering Students and Professionals*; Cambridge University Press, 2015; pp 158–199.

(12)    Schwaab, M.; Biscaia, Jr., E. C.; Monteiro, J. L.; Pinto, J. C. Nonlinear Parameter Estimation through Particle Swarm Optimization. *Chem. Eng. Sci.* **2008**, *63*, 1542–1552.

(13)    Hendel, T.; Wuithschick, M.; Kettemann, F.; Birnbaum, A.; Rademann, K.; Polte, J. In Situ Determination of Colloidal Gold Concentrations with UV-Vis Spectroscopy: Limitations and Perspectives. *Anal. Chem.* **2014**, *86*, 11115–11124.

(14)    Song, C.; Wang, P.; Makse, H. A. A Phase Diagram for Jammed Matter. *Nature* **2008**, *453*, 629–632.

(15)    Lingane, J. J. Standard Potentials of Half-Reactions Involving + 1 and + 3 Gold in Chloride Medium. *J. Electroanal. Chem.* **1962**, *4*, 332–342.

(16)    Rodríguez-González, B.; Mulvaney, P.; Liz-Marzán, L. M. An Electrochemical Model for Gold Colloid Formation via Citrate Reduction. *Zeitschrift fur Phys. Chemie* **2007**, *221*, 415–426.

(17)    Cheng, Y.; Tao, J.; Zhu, G.; Soltis, J. A.; Legg, B. A.; Nakouzi, E.; De Yoreo, J. J.; Sushko, M. L.; Liu, J. Near Surface Nucleation and Particle Mediated Growth of Colloidal Au Nanocrystals. *Nanoscale* **2018**, *10*, 11907–11912.

(18)    Loh, N. D.; Sen, S.; Bosman, M.; Tan, S. F.; Zhong, J.; Nijhuis, C. A.; Král, P.; Matsudaira, P.; Mirsaidov, U. Multistep Nucleation of Nanocrystals in Aqueous Solution. *Nat. Chem.* **2016**, *9*, 77.

(19) Ji, X.; Song, X.; Li, J.; Bai, Y.; Yang, W.; Peng, X. Size Control of Gold Nanocrystals in Citrate Reduction: The Third Role of Citrate. *J. Am. Chem. Soc.* **2007**, *129*, 13939–13948.

(20) Bates, D. M.; Watts, D. G. *Nonlinear Regression Analysis and Its Applications*; Bates, D. M., Watts, D. G., Eds.; Wiley Series in Probability and Statistics; John Wiley & Sons, Inc.: Hoboken, NJ, USA, 1988; Vol. 32; pp 200–263.